\documentclass[journal]{IEEEtran}
\usepackage{subfig}
\usepackage{amsthm}
\usepackage{amssymb} 
\usepackage[cmex10]{amsmath}
\usepackage{array}
\usepackage{mdwmath}
\usepackage{mdwtab}
\usepackage{eqparbox}
\usepackage{graphicx}
\usepackage{caption}
\usepackage[ruled]{algorithm2e} 
\usepackage{epstopdf}
\usepackage{algcompatible}
\usepackage[mathscr]{eucal}
\usepackage[section]{placeins}
\usepackage{cite}
\usepackage{mathtools}
\usepackage{color}

\let\labelindent\relax
\usepackage{enumitem}

\newtheorem{theorem}{Theorem}
\newtheorem{define}{Definition}

\newtheorem{lemma}{Lemma}
\newtheorem{myrule}{Remark}

\newcommand{\REV}[1]{{\color{black} #1}}
\newcommand{\REFV}[1]{}

\usepackage{lineno,hyperref}

\begin{document}

\title{A Coverage-Aware Distributed $k$-Connectivity Maintenance Algorithm for Arbitrarily Large $k$ in Mobile Sensor Networks}

\author{Vahid~Khalilpour~Akram,~\IEEEmembership{Member,~IEEE,}
        Orhan~Dagdeviren,~\IEEEmembership{Member,~IEEE,}\\
        and~Bulent~Tavli,~\IEEEmembership{Senior Member,~IEEE}
\thanks{V. K. Akram is with the International Computer Institute, Ege University, Izmir, Turkey. (e-mail: vahid.akram@ege.edu.tr).}
\thanks{O. Dagdeviren is with the International Computer Institute, Ege University, Izmir, Turkey. (e-mail: orhan.dagdeviren@ege.edu.tr).}
\thanks{B. Tavli is with the Electrical and Electronics Engineering Department, TOBB University of Economics and Technology, Ankara, Turkey (e-mail: btavli@etu.edu.tr).}
\thanks{ This work was supported by TUBITAK (The Scientific and Technological Research Council of Turkey) under Project Number 113E470.}
}

\maketitle

\begin{abstract}
Mobile sensor networks (MSNs) have emerged from the interaction between mobile robotics and wireless sensor networks. MSNs can be deployed in harsh environments, where failures in some nodes can partition MSNs into disconnected network segments or reduce the coverage area. A $k$-connected network can tolerate at least $k$-1 arbitrary node failures without losing its connectivity. In this study, we present a coverage-aware distributed $k$-connectivity maintenance (restoration) algorithm that generates minimum-cost movements of active nodes after a node failure to preserve a persistent $k$ value subject to a coverage conservation criterion. The algorithm accepts a coverage conservation ratio (as a trade-off parameter between coverage and movements) and facilitates coverage with the generated movements according to this value. Extensive simulations and testbed experiments reveal that the proposed algorithm restores $k$-connectivity more efficiently than the existing restoration algorithms. Furthermore, our algorithm can be utilized to maintain $k$-connectivity without sacrificing the coverage, significantly.
\end{abstract}

\IEEEpeerreviewmaketitle

\section{Introduction}
\label{Intro}

Mobile Sensor Networks (MSNs) have a wide range of applications~\cite{song2019minimum}. Robust connectivity is a vital and challenging requirement for MSNs. A network is $k$-connected if it remains connected after failures in $k$-1 arbitrary nodes. Preserving the $k$ value after a node failure keeps the fault tolerance of an MSN at the desired level. $k$-connectivity maintenance (restoration) is a process that is initiated after a node failure to restore the $k$ of a network to its primary value (the $k$ value before the failure). Since MSNs are inherently mobile, moving nodes for $k$-connectivity restoration is a convenient solution.

Movement-based connectivity restoration can, potentially, lead to loss of coverage in an MSN to some extent (in the absence of backup nodes). However, maintaining connectivity without the loss of coverage can impose higher movement costs. Therefore, there is an inherent trade-off between movements required for connectivity and network coverage. Nevertheless, connectivity and coverage problems are closely related and should be addressed jointly. Although connectivity restoration is a well-known problem  for $k$=1 or $k$=2 (1 or 2-connectivity restoration), the design, analysis, and implementation of a movement-based coverage-aware distributed $k$-connectivity restoration approach (for arbitrarily large $k$) for MSNs is left unstudied to the best of our knowledge.

To address these challenges, we create a movement-based coverage-aware distributed $k$-connectivity maintenance algorithm for MSNs. Our contributions are as follows:
\begin{enumerate}
\item We propose a novel distributed movement-based algorithm for $k$-connectivity maintenance (for arbitrarily large $k$)  in MSNs.
\item The proposed approach accepts a coverage conservation ratio and adjusts the generated movements to maintain the $k$-connectivity with minimal coverage loss.
\item We provide the theoretical foundations of the proposed algorithm and its resource consumption analysis in terms of bit, space, time, and computational complexities.
\item Through extensive simulation and testbed experiments by employing IRIS nodes and Kobuki robots, we evaluate the performance of the proposed algorithm in terms of movement costs, coverage loss, wallclock times, and transmitted bits. Our findings clearly demonstrate that the proposed algorithm outperforms its counterparts.
\end{enumerate}

The rest of the paper is organized as follows; Section~\ref{relatedwork} provides a concise survey on existing approaches. In Section~\ref{problemformulation}, we formulate the problem and explain the network model. The proposed algorithm is described in detail and theoretically analyzed in Section~\ref{ouralgorithms}. Section~\ref{PerformanceEvaluation} presents the experimental results. The conclusions and future research directions are given in Section~\ref{conclusion}.

\section{Related Work}
\label{relatedwork}
Studies on $k$-connectivity can be organized into detection, deployment, and restoration categories. The $k$-connectivity detection problem is the problem of finding the $k$ value of a network \cite{henzinger2000computing,szczytowski2012DKM,akram2018deck,censor2014distributed,dagdeviren2019design}. In the $k$-connectivity node deployment problem, the aim is to deploy the nodes and arrange their radio ranges in such a way that the resulting topology becomes $k$-connected  \cite{bredin2010deploying,deniz2016adaptive}.

For constant and predefined $k$ values (e.g., $k$=1, 2), there are many studies on connectivity restoration that attempts to reconnect the nodes in a partitioned network by adding relay nodes~\cite{lee2015connectivity,zeng2016fault,almasaeid2009minimum, han2010fault,ranga2015relay,lee2018leef,shriwastav2018round,liu2020restoring,sharma2016distributed}, utilizing reserved nodes~\cite{atay2009mobile,szczytowski2012DKM}, changing the communication range~\cite{wang2016adaptive}, or moving other active nodes~\cite{zhang2018autonomous,akkaya2009distributed,dimarogonas2008decentralized,liu2017survivability, yang2010decentralized,uzun2015distributed,senturk2016towards}. However, there is no distributed solution for the $k$-connectivity restoration problem for arbitrary $k$.

Another important problem in MSNs is maximizing the area covered by the available sensor nodes considering various constraints such as connectivity, energy, and lifetime~\cite{zhang2019coverage,zhang2016approximating}. A detailed review 
of the coverage problem has been provided in~\cite{elhabyan2019coverage}. 
The connected $k$-coverage problem is defined as ensuring the coverage of each point in the network by, at least, $k$ sensor nodes~\cite{zhou2004connected}. In fact, our algorithm is complementary to the $k$-coverage algorithm in~\cite{zhou2004connected} (i.e., the two algorithms can work in cooperation to create a $k$-connected and $k$-covered MSN). 

A central Minimum-Cost $k$-Connectivity Restoration (MCCR) algorithm is proposed for the movement-based general $k$-connectivity restoration problem in \cite{wang2011onmovement}.
MCCR uses a $k$-connectivity test and a maximum weighted matching algorithm to find the best possible movements for $k$-connectivity restoration. After each failure, MCCR creates a set $P$ that includes the positions of all the nodes and a set $V$ that includes the remaining active nodes. It then removes a position from $P$ and  if the remaining positions form a $k$-connected graph, the algorithm creates a bipartite graph over the $P$ and $V$ sets. Next, it applies a maximum matching algorithm to the resulting bipartite graph and repeats this process for all other positions in $P$ selecting a matching with the maximum value as the best movements. The best time complexities of the $k$-connectivity testing and maximum matching are $O(n^2\times k^2)$ \cite{henzinger2000computing} and $O(n^3)$ \cite{kuhn1955hungarian}, respectively. This leads to $O(max\{n^3k^2,n^4\})$ time complexity for MCCR. MCCR is improved by the TAPU algorithm, which replaces the matching algorithm with the shortest path tree algorithm \cite{akram2019tapu}. In this central approach, a shortest path tree rooted by the failed node is established between the nodes and a safe node with the minimum-cost to the root is moved to the position of the failed node. This approach generates exactly the same movements as MCCR, but its running time is lower than that of MCCR. Using central algorithms on large-scale MSNs can be inefficient because it imposes a large amount of message passing and resource consumption to gather the topology information in the sink node. 

A restricted version of the $k$-connectivity restoration problem in heterogeneous mobile networks is presented in~\cite{akram2020distributed}. It is assumed that the network has a limited number of mobile nodes, which simplifies the 
model and reduces the problem complexity by restricting the number of possible solutions. 

The main novelty of the proposed distributed algorithm is that nodes can estimate the remote paths in the general topology and can decide on the required movements based on the local neighborhood information without requring global topology information, in most cases. Indeed, this is the first distributed algorithm that can maintain $k$-connectivity for any arbitrary $k$ value in MSNs. Furthermore, the proposed algorithm is able to balance the cost of the generated movements for $k$-connectivity restoration with the overall coverage loss.

\section{Problem Formulation and Motivation}
\label{problemformulation}

\begin{figure}[!h]
    \centering
        \includegraphics[width=0.48\textwidth]{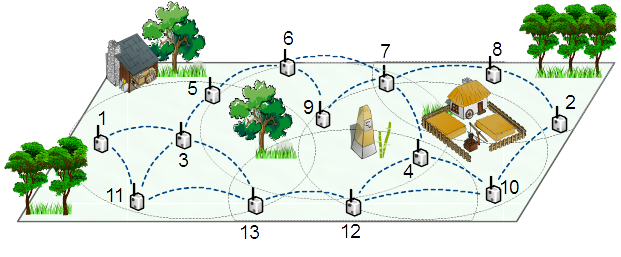}
        \caption{Illustration of an example MSN topology.}
        \label{model}
\end{figure}

We can model an MSN as an undirected graph $G(V,E)$ where $V$ is the set of nodes and $E$ is the set of links (edges). The nodes in MSNs can exchange messages if they are in radio range of each other. Fig. \ref{model} shows a sample 2-connected MSN where the large dashed circles show the radio range of each node. This network remains connected in case of single node failure, yet, failures in two nodes can partition the network.

We assume that the links are bidirectional and stable; consequently, the topology does not change (due to the unstable links) during the algorithm execution. Otherwise, the $k$ value of the network could continuously change making $k$-connectivity restoration extremely complicated. In a $k$-connected network, there are at least $k$ disjoint paths between each pair of nodes. Two paths are disjoint if they share no common nodes except the source and target. In Fig.~\ref{model}, removing any node except nodes 1, 4, and 9 reduces the $k$ value to 1. Thus, we can divide the nodes in an MSN into \textit{Joint} and \textit{Trusted} sets. The failure of any \textit{Joint} node reduces the $k$ value, but failure in a \textit{Trusted} node has no effect on $k$. For example, in Fig.~\ref{model}, nodes 1, 4, and 9 are \textit{Trusted} and the other nodes are \textit{Joint}.

\subsection{Problem Formulation}
\begin{figure}[!h]
\captionsetup[subfloat]{farskip=9pt}
\centering
\subfloat[]{\includegraphics[width=0.2\textwidth]{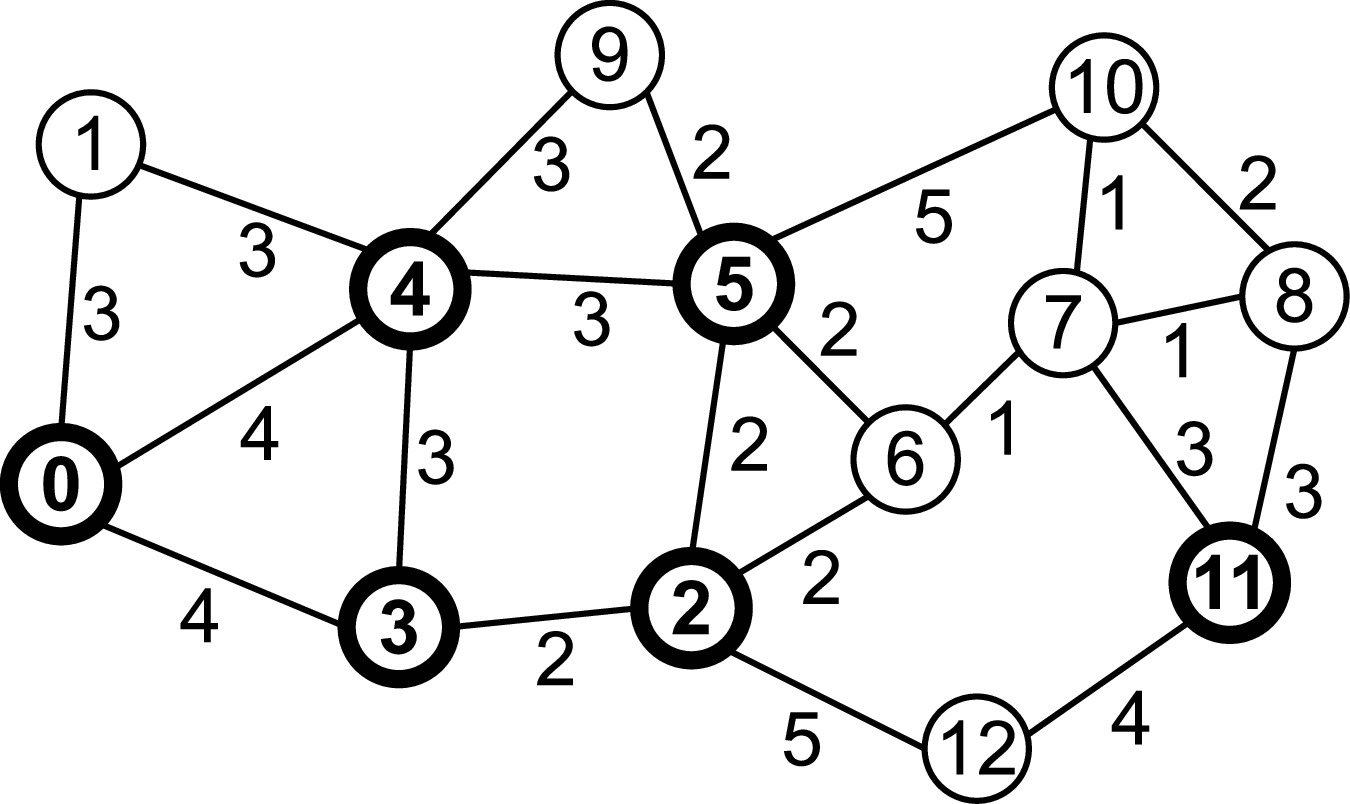}
\label{mainnetwok}}
\subfloat[]{\includegraphics[width=0.14\textwidth]{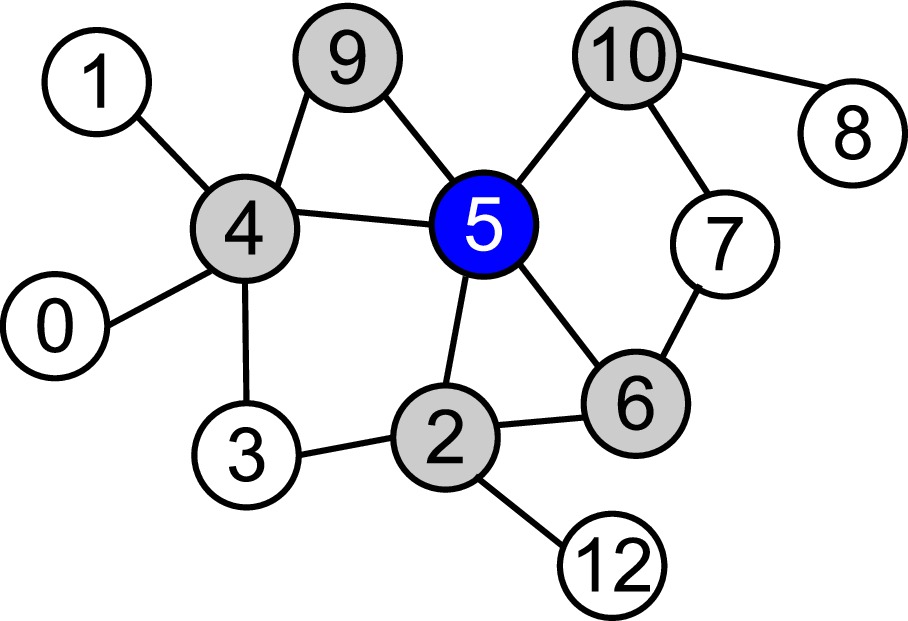}
\label{subgraph5}}
\subfloat[]{\includegraphics[width=0.14\textwidth]{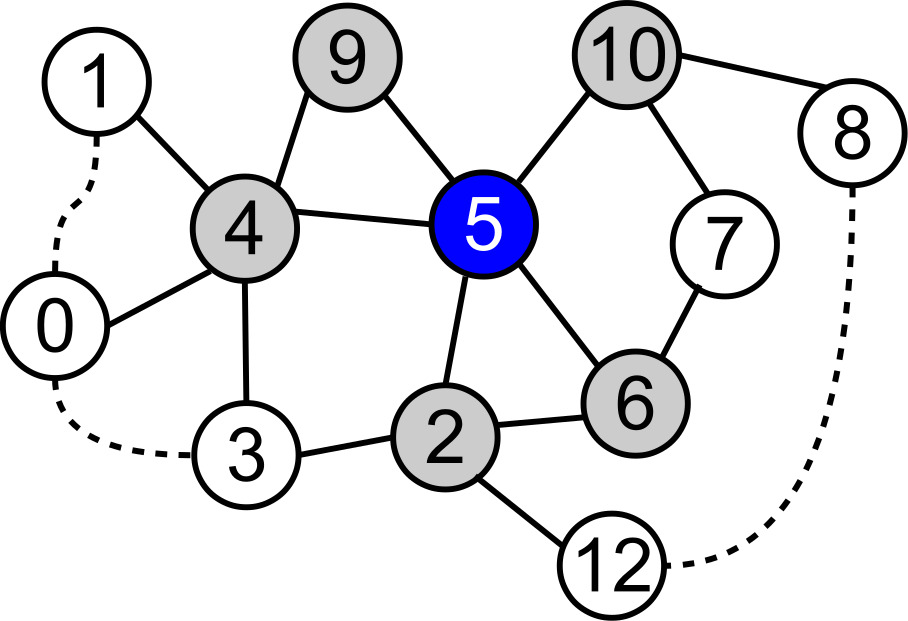}
\label{subgraphk5}}
\caption{a) 2-connected MSN, b) $G_5$, c) $G_5^k$. }
\label{figure:det}
\end{figure}

Throughout the paper, $n=|V|$ is the total number of nodes, $D$ is the network diameter, and $\Gamma_v$ is the set of 1-hop neighbors of node $v$. We denote the maximum and minimum degrees of nodes by $\Delta$ and $\delta$, respectively, the degree of node $v$ by $d_v$, the $k$ value of graph $G$ by $\kappa(G)$, a path between nodes $x$ and $y$ by $\rho(x,y)$, and the number of disjoint paths between nodes $x$ and $y$ in $G$ by $x \overset{G}{\sim} y$. In Fig.~\ref{mainnetwok}, for example, we have $n=13$, $d_1=2$, $\kappa(G)=2$, $\Gamma_3=\{0,2,4\}$, $D=5$, $\Delta=5$, $\delta=2$, $\rho(1,6)=(1,4,5,6)$, and $0\overset{G}{\sim} 8$=2. In this figure, the weights of the edges are the movement costs between the related nodes. We assume that $G_v(V_v,E_v)$ is the 2-hop local subgraph of node $v$ (Fig. \ref{subgraph5} shows $G_{5}$), and $G/v$ indicates graph $G$ without node $v$. In this figure, the gray nodes are \textit{Joint}, and the white nodes are \textit{Trusted} nodes. Removing a \textit{Trusted} node has no effect on $\kappa(G)$, but removing a \textit{Joint} node reduces $\kappa(G)$. Formally, we can define the \textit{Trusted} and \textit{Joint} nodes as follows

\begin{define}
  In a $k$-connected graph $G(V,E)$, node $v\in V$ is \textit{Trusted} if $\kappa(G)=\kappa(G/v)$, otherwise $v$ is \textit{Joint}.
\end{define}

The \textbf{k-connectivity restoration problem} can be defined, formally, as follows: Let $V= \{v_1,v_2,...,v_n\}$ be the set of nodes where $v_i$ has position $i$, $r$ is the radio range, $E=\{(v_i,v_j): v_i, v_j \in V \; and \; ||v_i-v_j||\leq r\}$ is the set of edges, $f:V\times V \rightarrow R^+$ is the movement cost function, and $G(V,E)$ is the topology graph. Given a failed node $v_x$, the $k$-connectivity restoration problem is finding a movement cost function $M:V\rightarrow V$ such that
\begin{equation}
minimize \sum_{v_i\in V/v_x} f(v_i,M(v_i))	
\end{equation}
under the condition that
\begin{equation}
 \kappa(G(V_M,E_M))=\kappa(G(V,E))
\end{equation}
where
\begin{equation}
\begin{split}
V_M=\{M(v_i): \forall v_i\in V/v_x\}.\\
E_M=\{(v_i,v_j): v_i, v_j \in V_M \; and \; ||v_i-v_j||\leq r\}.
\end{split}
\end{equation}

The aim of $k$-connectivity restoration is to find a set of movements with minimal cost to restore the $k$ of the network to its initial value after the failure of a node.

\subsection{Motivation}
Maintaining $k$-connectivity in an MSN brings many direct and indirect benefits. The most important benefit of $k$-connectivity is the inherent \textbf{fault tolerance} built in by design (i.e., a $k$-connected MSN stays connected even if any $k-1$ node combinations are left inoperational). Indeed, the impact of \textbf{bottlenecks and critical nodes} is alleviated in a $k$-connected network provided that the number of such nodes is less than $k$. Since the $k$ value is a lower bound for the \textbf{edge connectivity} of any network, a $k$-connected network stays connected even if $k-1$ links stop functioning.
Since each node in a $k$-connected MSN has at least $k$ \textbf{disjoint paths} to any other node, the \textbf{communication reliability} of such an MSN increases directly with the $k$ value. All the network services in a $k$-connected MSN (e.g., topology management, routing, and backbone formation) can continue uninterruptedly because unless, at least, $k$ nodes are incapacitated, there is, at least, one path to every operational node from any other operational node. The availability of $k$ \textbf{disjoint paths} toward the sink from each node is an invaluable opportunity for balancing the \textbf{energy consumption} throughout the network, which is necessary for \textbf{network lifetime maximization}.

The $k$ value of a network can be utilized for estimating the lower or upper bounds of many other properties of the network. The \textbf{minimum degree} of a $k$-connected network is, at least, $k$ (i.e., each node has, at least, $k$ neighbors). The initial knowledge of the degree of a network is a useful input to \textbf{clustering and partitioning algorithms}. We can always find a \textbf{cycle} of $k$ arbitrary nodes in any $k$-connected network. If all the links in the network have the same capacity, the $k$ value determines the upper bound of the \textbf{network flow} between every pair of nodes. A network with a higher $k$ value, typically, has a higher \textbf{node density} than a network with a lower $k$ value. Therefore, the $k$ value can be utilized as an indicator of the node density. 

\section{The Proposed Approach}
\label{ouralgorithms}
The proposed Localized Imaginary Network Aided Restoration (LINAR) algorithm has two main phases. In the first phase, nodes detect their status (\textit{Trusted} or \textit{Joint}) and support degree and share this information with their 2-hop neighbors. The second phase of the algorithm starts when a node detects the failure of a \textit{Joint} neighbor, which triggers the movements of a single node (or multiple nodes). The second phase is initiated by a neighbor of the failed node and finishes when the $k$ value of the network returns to its initial value. In the second phase, the nodes around the failed/moved nodes make local decisions (regarding to move or not to move) based on their 2-hop local subgraphs. Failure in any node can be restored by the network because each node in the network continuously monitors its neighbors and initiates the restoration process when required.

In the first phase, the nodes find their status using an imaginary $k$-connected local subgraph and broadcast this state to their neighbors. Generally, the 2-hop local subgraph of nodes has a lower $k$ value than the actual $k$ because local subgraphs omit remote paths between the nodes. For example, the presented network in Fig. \ref{mainnetwok} is 2-connected whereas $G_{5}$ is 1-connected. To extract the correct state information from the 2-hop local subgraphs, we first create an imaginary $k$-connected local subgraph, then we find the number of disjoint paths between the 1-hop neighbors of each node.

\subsection{Imaginary $k$-Connected Subgraphs}
\label{imagk}
\label{ruls}

In the proposed algorithm, all nodes estimate some paths existing in the global topology (but they do not exist in their 2-hop local subgraphs) by converting their 2-hop local subgraphs into $k$-connected graphs. For example, in Fig. \ref{subgraph5}, $\kappa(G_5)=1$ whereas the $k$ value of general topology is 2. Using the provided lemmas and theorems, node $v\in V$ can estimate remote paths between its 2-hop neighbors and converts $G_v$ to a $k$-connected graph. If $G(V,E)$ is $k$-connected, then we, definitely, have, at least, $k$ paths between any pair of nodes in $G_v$ even if some of the paths pass through some remote nodes beyond $G_v$. In other words, if the $k$ value of a local graph, say $G_v$, is smaller than the $k$ value of the general graph $G$, then any pair of nodes in $G_v$ that have fewer than $k$ paths in $G_v$ have some remote paths beyond $G_v$ that connect them by the nodes out of $G_v$. Formally, if $\kappa(G_v)<\kappa(G)$, then for any $x,y \in V_v/\Gamma_{v}$ with $i<k$ paths in $G_v$ there are at least $k-i$ remote paths beyond $G_v$ that connect them using remote nodes in $G$. For example, in Fig.~\ref{subgraph5}, nodes 0 and 1 have only one path in $G_5$. However, since the network is 2-connected, node 5 can, safely, conclude that there is at least one more path beyond $G_5$ between nodes 0 and 1. In the proposed algorithm, each node converts its local subgraph to a $k$-connected graph that represents the possible existing remote paths between its 2-hop neighbors. To do so, each node adds imaginary edges to  its subgraph (in its local  memory) to incorporate the possible missing remote paths in $G_v$. An imaginary edge can be defined as follows
\begin{define}
	An \textbf{imaginary edge} $e=(x,y)$ is a hypothetical edge that is added to $G_v$ to incorporate the path $\rho(x,y)=(x,...,y) \in G$ where  $\rho(x,y) \notin G_v$.
\end{define}

Our aim is to increase the $\kappa(G_v)$ to the $\kappa(G)$ in such a way that $v$ has the same state in both graphs. Note that no imaginary edge can be added to the 1-hop neighbors of $v$ in $G_v$ because all neighbors of every $u \in \Gamma_v$  are already present in $G_v$ as stated in Remark~\ref{rule:onehop}. Therefore, each node can add only imaginary edges between its 2-hop neighbors. In~Fig.~\ref{subgraph5}, for example, node 5 can add only edges between the white nodes.
\begin{myrule}
	\label{rule:onehop}
	Node $v$ cannot attach any imaginary edge to any node $u \in \Gamma_v$.
\end{myrule}

The purpose of adding imaginary edges is to incorporate missing remote paths between the nodes in $G_v$ until $G_v$ becomes $k$-connected. Therefore, every added imaginary edge should represent a missing remote path between the nodes. If nodes $a,b \in V_v/\Gamma_v$ already have, at least, $k$ paths in $G_v$, it is possible that they have no other paths beyond $G_v$. Therefore, node $v$ can add only imaginary edges between its 2-hop neighbors if fewer than $k$ paths exist between them in $G_v$ as stated in Remark~\ref{rule:newpath}. 
\begin{myrule}
	\label{rule:newpath}
	Node $v$ can add imaginary edges between  $a,b \in V_v/\Gamma_v$ if $a \overset{G_v}{\sim} b<k$.
\end{myrule}

In the proposed algorithm, each node tries to convert its local subgraph to a $k$-connected graph by creating minimal (ideally zero) extra paths between its 1-hop neighbors. We will prove that this strategy guarantees that the state of  $Joint$ nodes remains $Joint$ in their $k$-connected local subgraphs. Formally, we define the resulting $k$-connected local subgraph of node $v$ (after adding imaginary edges) as follows
\begin{define}
	\label{rule:minpath}
$G^k_v$ is the $k$-connected graph obtained after adding the minimum number of imaginary edges to $G_v$ that minimizes $\sum_{x,y\in \Gamma_v} x \overset{G_v}{\sim} y$.
\end{define}

$\sum_{x,y\in \Gamma_v} x \overset{G_v}{\sim} y$ is the total number of disjoint paths between every 1-hop neighbor of node $v$. Definition~\ref{rule:minpath} implies that node $v$ generates $G_v^k$ such that the number of paths between 1-hop neighbors of node $v$ is minimized. We will prove that if we minimize $\sum_{x,y\in \Gamma_v} x \overset{G_v}{\sim} y$ while adding imaginary edges, a $Trusted$ node $v$ in $G_v^k$ will, definitely, be $Trusted$ in $G$. Fig.~\ref{subgraphk5} shows $G_5^k$. Lemma~\ref{theorem:trusted} proves an important property about $Trusted$ nodes that helps us identify them by finding the number of paths between their 1-hop neighbors. 

\begin{lemma}
	\label{theorem:trusted}
	For any $v \in V$, $G/v$ is $k$-connected if  $\forall$ $\!(x,y) \!\in\! \Gamma_v \!\! :\!  x\!\overset{G/v}{\sim}\!y\!\ge \! k$.
\end{lemma}

Note that the proofs of all the lemmas and theorems presented in the paper can be found in the online supplement. Lemma~\ref{theorem:trusted} implies  that if we have  $i \overset{G/v}{\sim} j \ge k$ for any pair of nodes $i,j \in \Gamma_v$, then $v\in V$ is a $Trusted$ node in $G$. In other words, any $Joint$ node $v \in V$ has at least two neighbors $i,j \in \Gamma_v$ such that $i \overset{G/v}{\sim} j <k$. Consequently, to change the state of $Joint$ node $v\in V$ to $Trusted$, we need to create new paths (not passing over $v$) between  $i$ and $j$. In other words, as long as the number of paths between $i$ and $j$ is not increased the state of a $Joint$ node $v$ does not change to $Trusted$. Lemma~\ref{theorem:samestate} states that as long as the number of paths between the 1-hop neighbors of $v$ does not increase, adding any edge to the graph has no effect on the state of $v$. 

\begin{lemma}
	\label{theorem:samestate}
	For any $e \notin E$ and $v \in V$, $k(G(V,E)/v)=k(G'(V,E\cup e)/v)$ if $\forall (x,y) \in \Gamma_v: x \overset{G}{\sim} y = x \overset{G'}{\sim} y$.
\end{lemma}

Lemma~\ref{theorem:samestate} indicates that if adding an edge to the graph does not increase the number of paths between the 1-hop neighbors of $v$, then node $v$ remains $Trusted$ if it was $Trusted$, and remains $Joint$ if it was $Joint$ before adding the edge. In other words, as long as the number of disjoint paths between the 1-hop neighbors of node $v$ does not increase, adding any edge to the graph has no effect on the $Trusted$ or $Joint$ state of node $v$. Therefore, to maintain the same state for $v$ in both $G$ and $G_v^k$, we need to keep the number of paths between the 1-hop neighbors of $v$ unchanged. However, this is not always possible. In some cases, we have no choice but to increase the number of paths between the 1-hop neighbors of $v$. Using Definition~\ref{rule:minpath}, we convert $G_v$ to a $k$-connected graph by creating the minimum number of new paths between the 1-hop neighbors of $v$. Lemma~\ref{lemma:localpath} states that if we generate $G_v^k$ according to Definition~\ref{rule:minpath}, the number of paths between 1-hop neighbors of $v$ in $G_v^k$ will be less than or, at most, equal to the number of paths between the same nodes in $G$. 

\begin{lemma}
	\label{lemma:localpath}
	For any $k$-connected graph $G(V,E)$, $\forall$ $v\in V$, and $\forall$ $(x,y) \in \Gamma_v$, we have $x \overset{G^\kappa_v}{\sim} y \leq x \overset{G}{\sim} y$.
\end{lemma} 

\begin{figure}[!h]
	\captionsetup[subfloat]{farskip=9pt}
	\centering
	\subfloat[]{\includegraphics[width=0.135\textwidth]{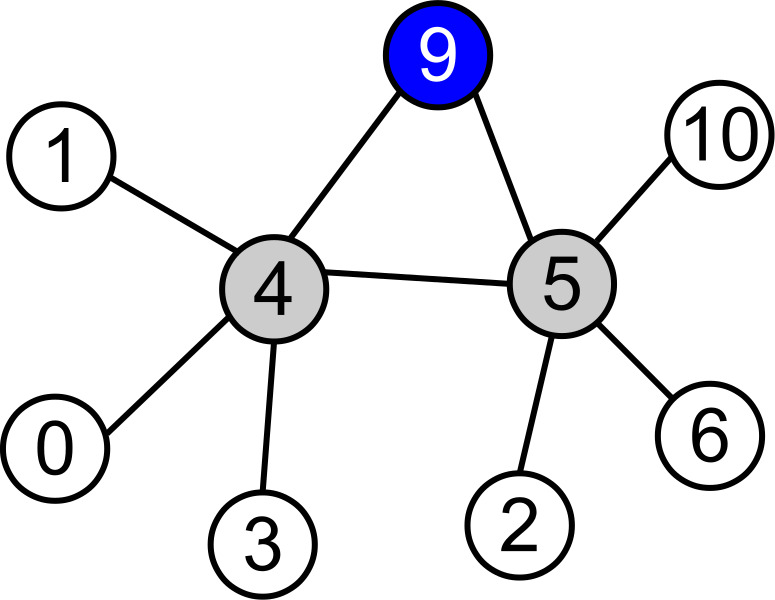}
		\label{sub9}}\hspace*{0.5em}
	\subfloat[]{\includegraphics[width=0.135\textwidth]{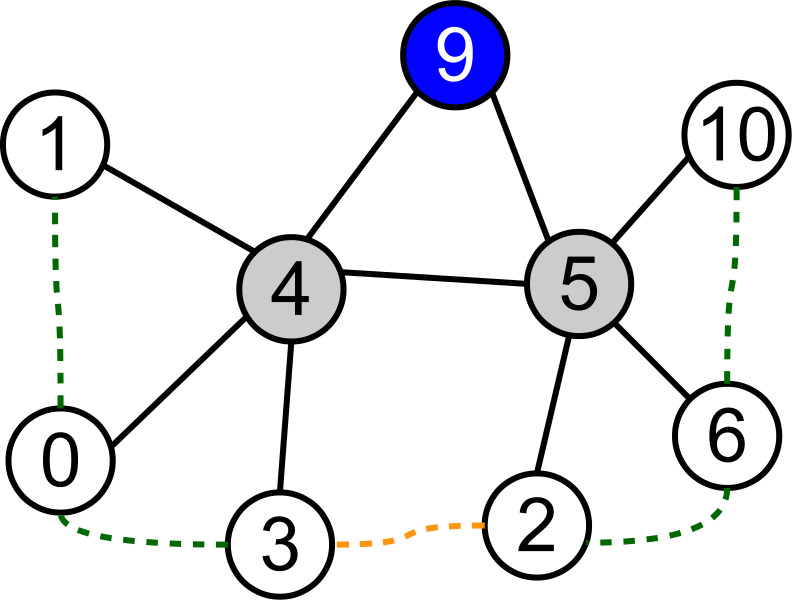}
		\label{sub9h}}\hspace*{0.5em}
	\subfloat[]{\includegraphics[width=0.138\textwidth]{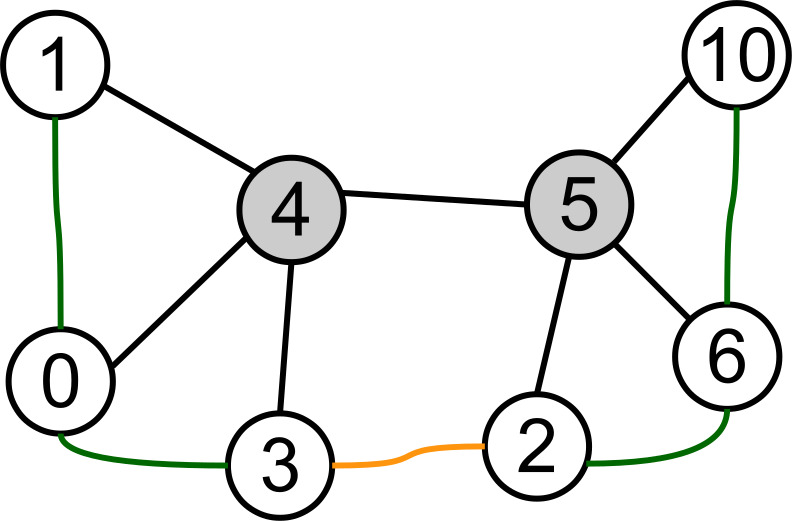}
		\label{sub9k}}\\\vspace*{-0.5em}
	\subfloat[]{\includegraphics[width=0.16\textwidth]{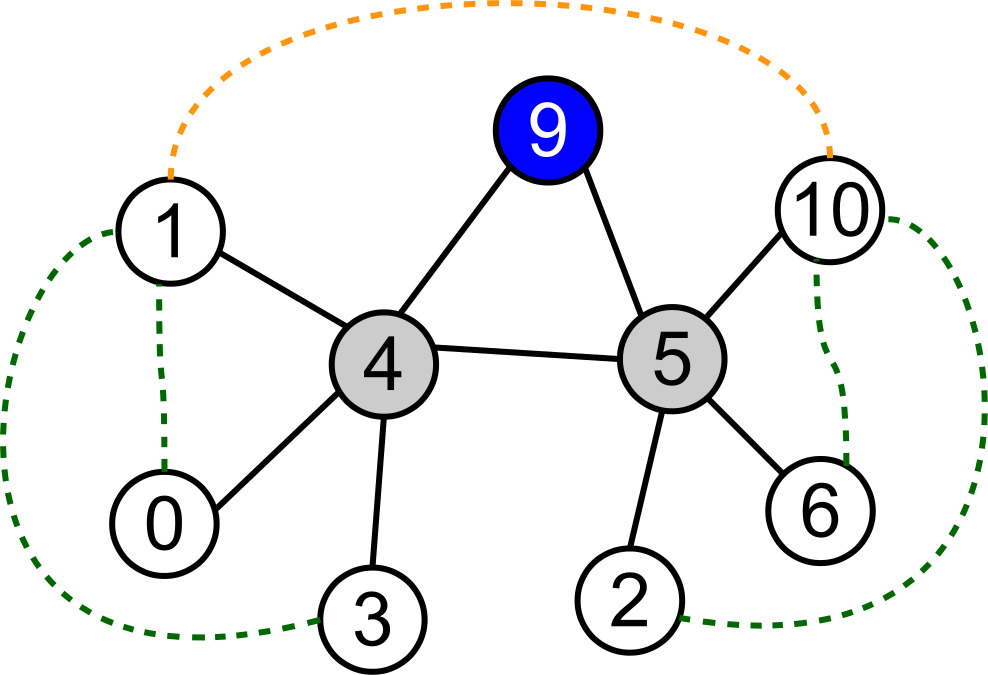}
		\label{sub9h2}}
	\subfloat[]{\includegraphics[width=0.16\textwidth]{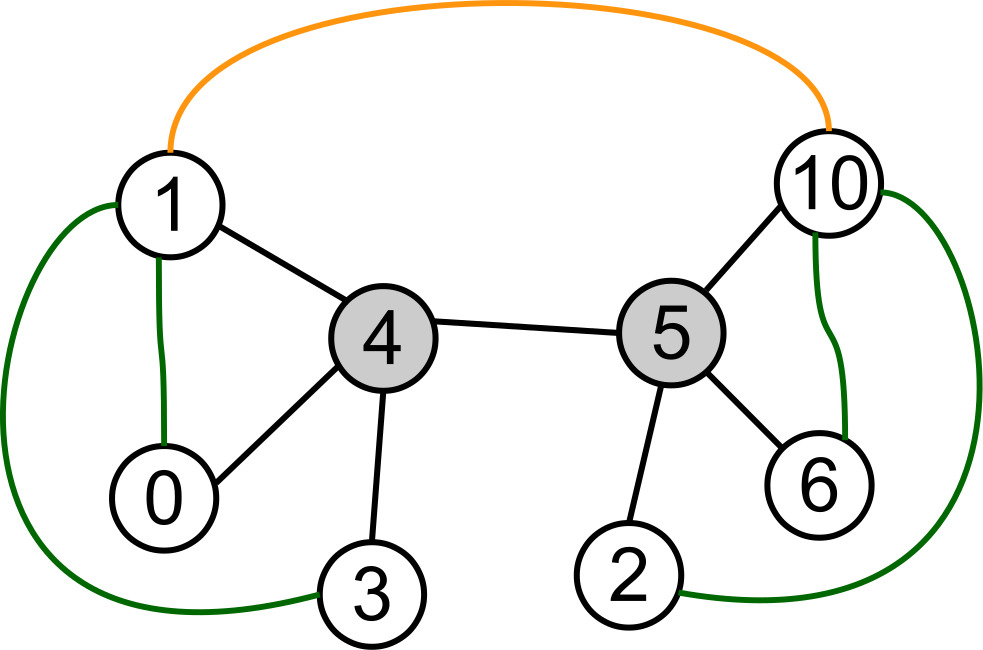}
		\label{sub9k2}}
	\subfloat[]{\includegraphics[width=0.13\textwidth]{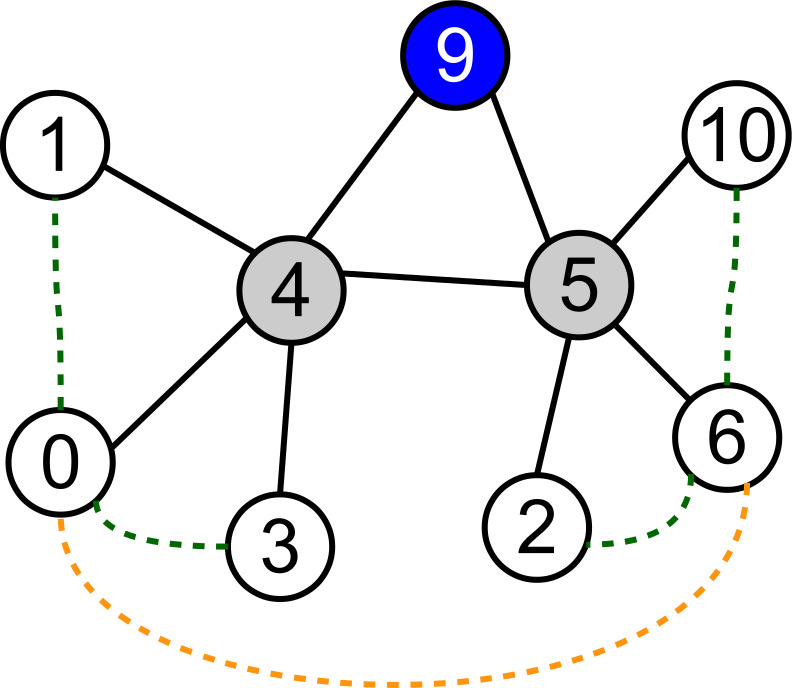}
		\label{sub9h3}}\\\vspace*{-0.5em}
	\subfloat[]{\includegraphics[width=0.13\textwidth]{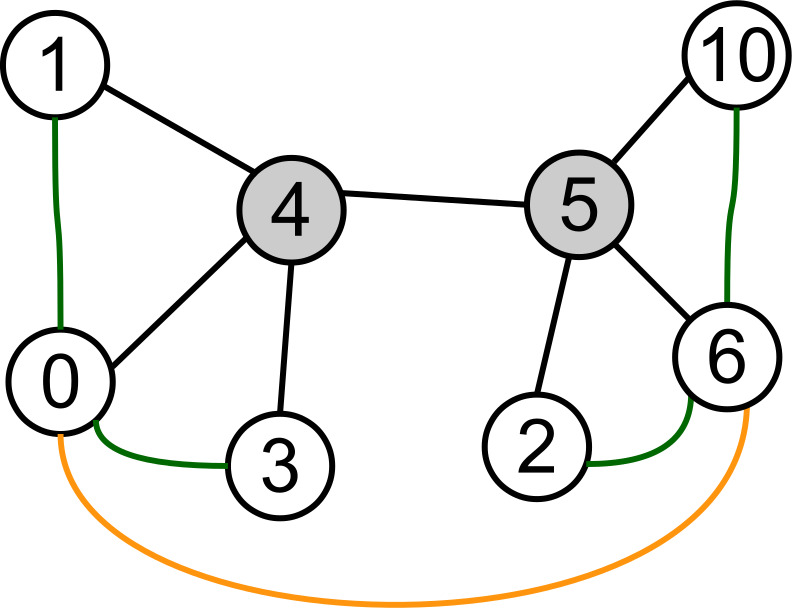}
		\label{sub9k3}}
		\subfloat[]{\includegraphics[width=0.143\textwidth]{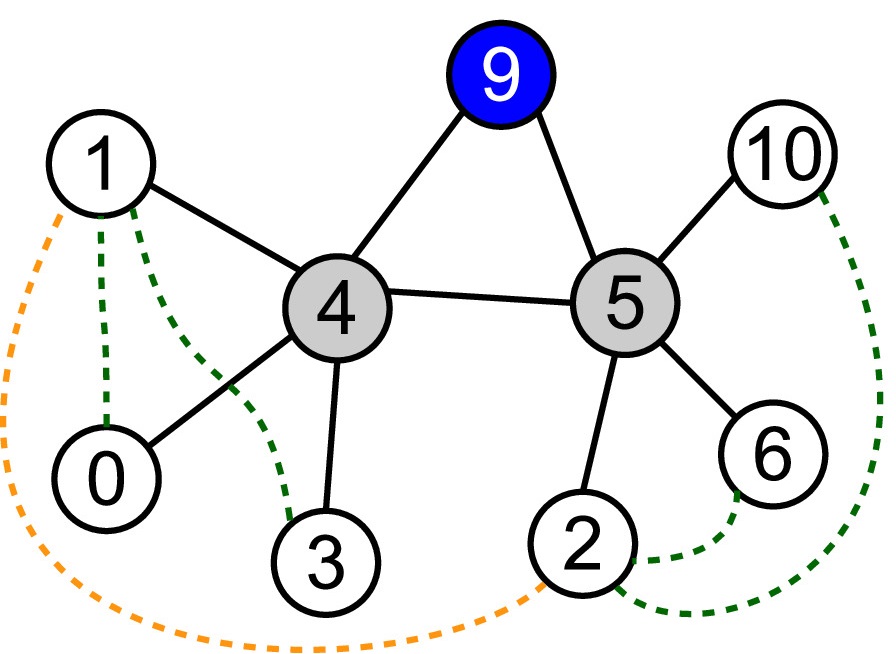}
		\label{sub9h4}}
	\subfloat[]{\includegraphics[width=0.15\textwidth]{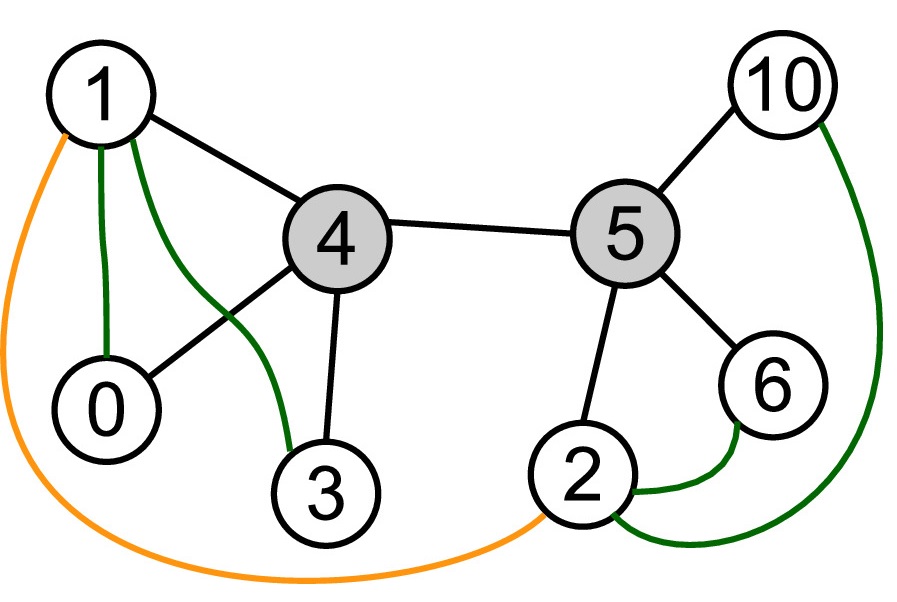}
		\label{sub9k4}}
	\caption{a) $G_{9}$: Local subgraph of node 9; b), d), f), and h) $G_{9}^k$: 2-connected local subgraphs of node 9; c), e), g), and i)  $G_{9}^k/9$: removing node 9 from $G_{9}^k$. }
	\label{figure:det}
\end{figure}

Lemma~\ref{lemma:localpath} implies that for any added imaginary edge to $G_v^k$, we, indeed, have an actual path in $G$. There can be multiple imaginary edge sets to create $G^\kappa_v$ that minimize $\sum_{x,y\in \Gamma_v} x \overset{G_v}{\sim} y$, yet, all of them satisfy $x \overset{G^\kappa_v}{\sim} y \leq x \overset{G}{\sim} y$. For example, Fig.~\ref{sub9} shows the local subgraph of node~9 and Fig.~\ref{sub9h} shows the same graph after adding imaginary edges. Adding the first four imaginary edges (green edges) to $G_9$ does not increase the number of edges between its 1-hop neighbors. However, to make $G_9$ 2-connected we have no choice but to add one more edge (the orange edge) which increases the number of disjoint paths between the 1-hop neighbors of node~9. Fig.~\ref{sub9k} shows the graph obtained after removing node 9 from its 2-connected local subgraph. Since this graph is 2-connected, node 9 is a $Trusted$ node in $G_9^2$. Node $v$ has more than one choice to generate $G_v^k$ according to Definition~\ref{rule:minpath}. In fact, Figs.~\ref{sub9h2}, \ref{sub9h3}, and \ref{sub9h4} can be used to create $G_9^2$. Figs.~\ref{sub9k2}, \ref{sub9k3}, and \ref{sub9k4} show that after adding these edges and removing node 9, all three $G_9^2$'s remain 2-connected as in the case depicted in Fig.~\ref{sub9k}. 

To generate $G_v^k$, node~$v$, first, obtains $E'_v$, which is the set of all possible edges between the 2-hop neighbors of node~$v$ that do not exist in $E_v$, by performing a linear search among its 2-hop neighbors. Next, any edge $(a,b)$ from $E'_v$ is removed if $a \overset{G_v}{\sim} b \geq k$ (Remark \ref{rule:newpath}) so that only the possible candidate edges between the 2-hop neighbors of $v$  with fewer than $k$ paths remain in $E'_v$. Then, one of the subsets $S_v\subseteq E'_v$ that gives the minimum number of total paths between every pair $x,y\in \Gamma_u$ is selected (Definition~\ref{lemma:localpath}.) and $G_v^k$ is obtained by merging $S_v$ and $G_v$. Theorem~\ref{theorem:imaginary_correctness}, directly, implies that if a node is $Trusted$ in $G_v^k$ then it is $Trusted$ in $G$. 
\begin{theorem}
	\label{theorem:imaginary_correctness}
	If $G^\kappa_v/v$ is $k$-connected, then $G/v$ is $k$-connected.
\end{theorem}

\begin{figure}[!htbp]
\captionsetup[subfloat]{farskip=9pt}
\centering
\subfloat[]{\includegraphics[width=0.14\textwidth]{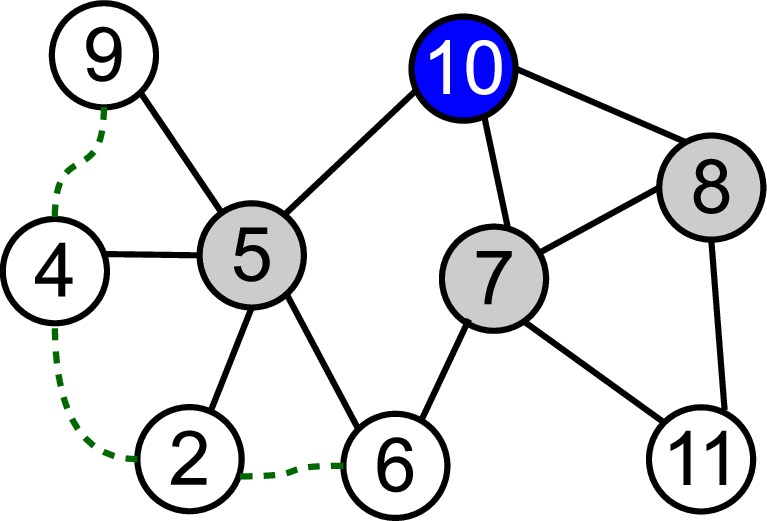}
\label{sub10h}}
\hfil
\subfloat[]{\includegraphics[width=0.14\textwidth]{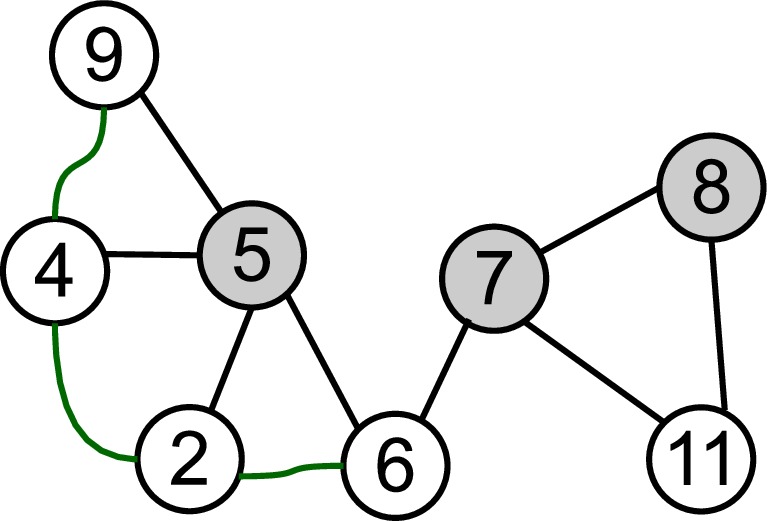}
\label{sub10k}}
\hfil
\subfloat[]{\includegraphics[width=0.17\textwidth]{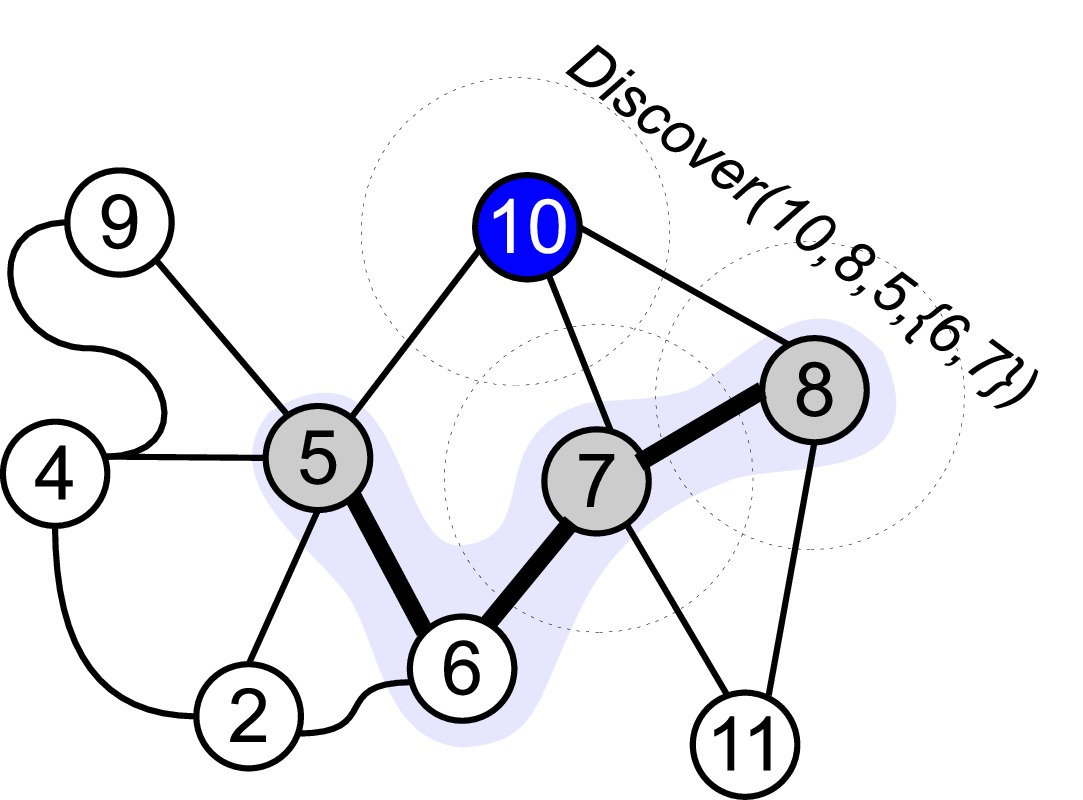}
\label{figure:node104}}
\hfil
\subfloat[]{\includegraphics[width=0.14\textwidth]{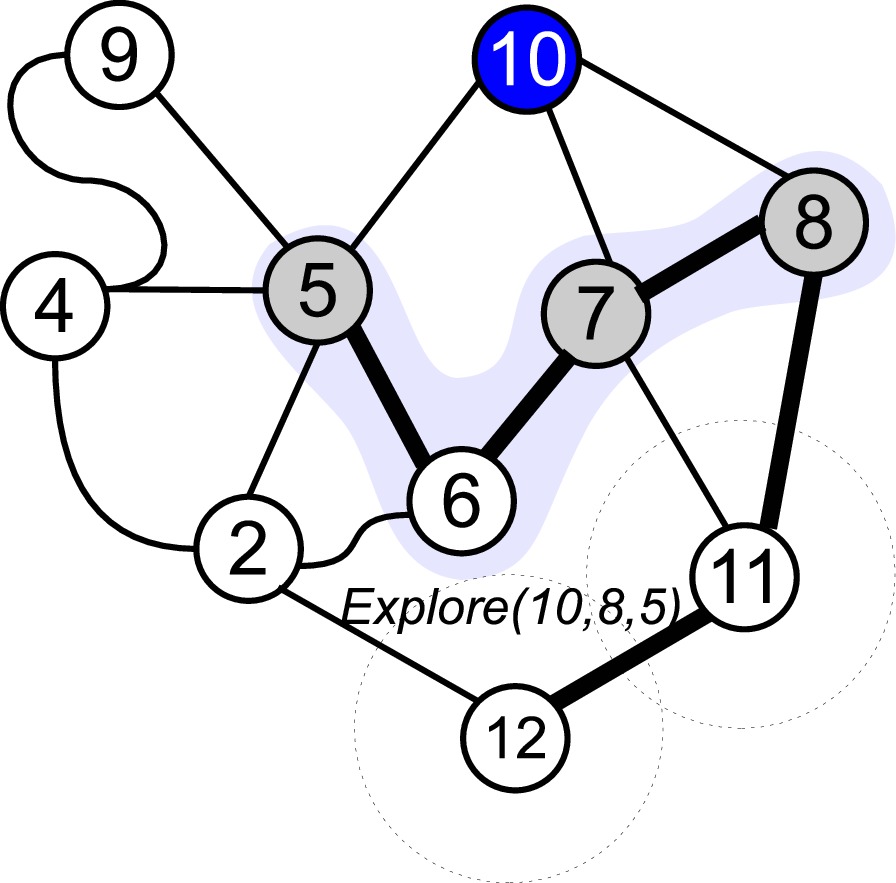}
\label{figure:node105}}
\hfil
\subfloat[]{\includegraphics[width=0.14\textwidth]{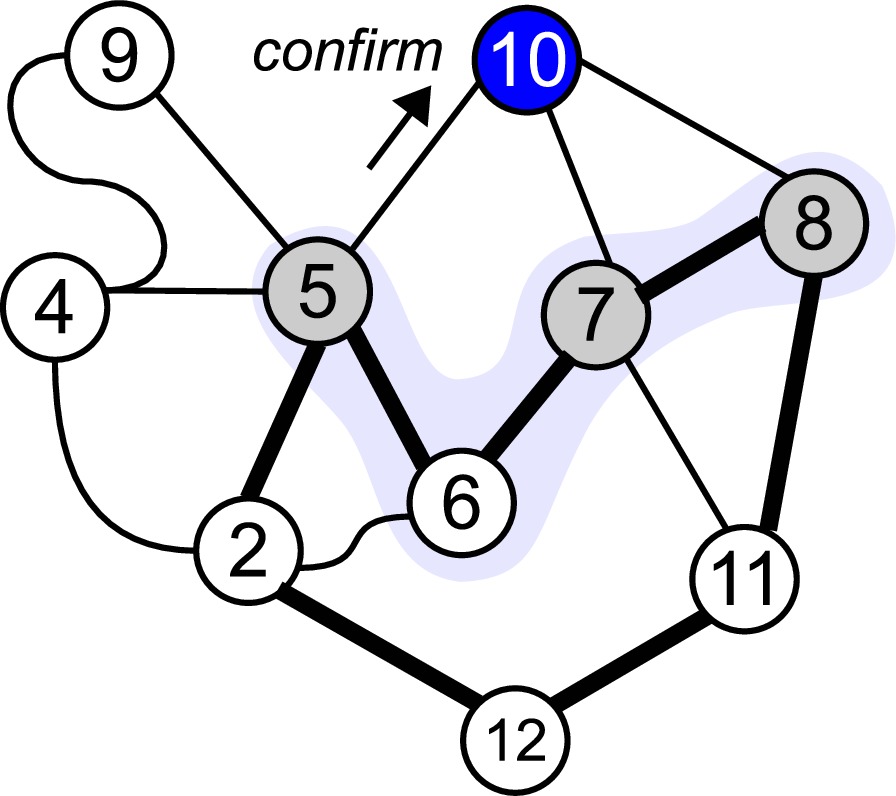}
\label{figure:node106}}
\hfil
\subfloat[]{\includegraphics[width=0.15\textwidth]{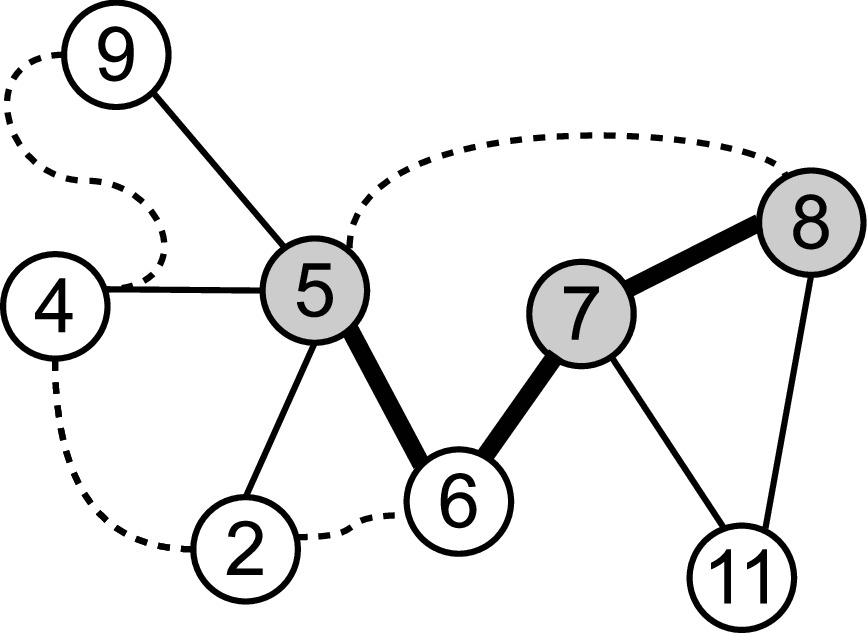}
\label{figure:node107}}
\hfil
\caption{a) $G_{10}^k$: 2-connected local subgraph of node 10, c) $G_{10}^k/10$: removing node 10 from $G_{10}^k$, d) broadcasted $Discover$ messages, e) broadcasted $Explore$ messages, e) sent $Confirm$ message, f) resulting $G_{10}^k$.}
\label{figure:node10}
\end{figure}

If node $v$ is $Joint$ in $G$, then it will, definitely, be $Joint$ in $G_v^k$. However, it is possible for node $v$ to be $Joint$ in $G_v^k$ while it is $Trusted$ in $G$.  For example, Fig.~\ref{sub10h} shows the local subgraph of node 10 after adding imaginary edges ($G$ is illustrated in Fig.~\ref{mainnetwok}). Adding three imaginary edges to $G_{10}$ makes it 2-connected without increasing the number of edges between its 1-hop neighbors. Fig.~\ref{sub10k} shows the graph obtained after removing node 10 from its 2-connected local subgraph. Since this graph is 1-connected, the algorithm considers node 10 as a $Joint$ node. However, node 10 is a $Trusted$ node in $G$. Therefore, to avoid such mislabeling, we confirm the status of a detected $Joint$ node (by using only $G_v^k$) by performing a global path search between $x,y\in \Gamma_v$ with $x\!\overset{G_v/v}{\sim}\!y\!<k$. 

\begin{lemma}
	\label{lemma_min}
	For any $v \in V$, $\kappa(G/v)<\kappa(G)$ if  $\exists u \in \Gamma_v \; s.t \; d_u=k$.
\end{lemma}

We can avoid a global path search for some of the $Joint$ nodes based on the degree of their neighbors. Lemma~\ref{lemma_min} states that if node $v$ has a neighbor $u \in \Gamma_v$  such that $d_u=k$, then $v$ is \textit{Joint} because removing $v$ reduces the $k$ value of the graph. All neighbors of $u$ in $G$ already exist in $G_v$. If $d_u=k$ in $G_u$, then $d_u=k$ in $G$, which means that $v$ is a \textit{Joint} node because in $G/v$ node $u$ has at most $k$-1 neighbors. Hence, a simple rule for detecting some of the $Joint$ nodes can be expressed as
\begin{equation}
\label{jointrule1}
if \; \exists \ u \; \in \; \Gamma_v \; \; s.t \; \;  d_u \, = \, k \;\, in \;\, G_v^k \; \rightarrow \;  v \text{ is } \textit{Joint.}
\end{equation}

For example, in Fig.~\ref{subgraphk5}, the degree of node $9\in \Gamma_5$ is 2; therefore, node 5 can confirm that it is \textit{Joint} without initiating any path search. If node $v$ is $Joint$ in $G_v^k$ and does not satisfy \REFV{@R1C4}\REV{Eq.~\ref{jointrule1}}, then we should search for a global path between a pair of its 1-hop neighbors. \REFV{@R1C3}\REV{ For example,  Fig.~\ref{sub10h} and Fig.~\ref{sub10k} show $G_{10}^k$ and $G_{10}^k/10$, respectively. The degrees of all the gray nodes are higher than 2 and there is only one disjoint path between nodes 2 and 11 in $G_{10}^k/10$.} Therefore, node 10 cannot determine its status using Theorem~\ref{theorem:trusted} and \REFV{@R1C4}\REV{Eq.~\ref{jointrule1}}. The nodes that satisfy Theorem~\ref{theorem:trusted} or \REFV{@R1C4}\REV{Eq.~\ref{jointrule1}} find their status and terminate the identification phase. For the other nodes we start a global path search. If node $v$ does not satisfy Theorem~\ref{theorem:trusted} and \REFV{@R1C4}\REV{Eq.~\ref{jointrule1}} then we have  $x,y \in G_v^k/v $ such that $x \overset{G_v/v}{\sim} y = k-1$. If we can find a path $p(x,y)$ between $x$ and $y$ such that $p(x,y)$ includes none of the nodes from the existing paths in $G_v^k$, then we can add an imaginary edge between $x$ and $y$ in $G_v^k/v$. If the resulting graph satisfies Theorem~\ref{theorem:trusted}, then node $v$ is \textit{Trusted}. Otherwise we repeat the same procedure for the other node pairs that have fewer than $k$ disjoint paths in $G_v^k/v$. If we cannot find any such paths, then $v$ is a \textit{Joint} node. For example, in $G_{10}/10$ (Fig. \ref{sub10k}), a local path $p(8,5)=(8,7,6,5)$ already exists between nodes~8~and~5 and the graph becomes 2-connected if we find another path between them. After adding such an imaginary edge, node 10 can mark itself as \textit{Trusted} because in the updated $G_{10}^k/10$ there are at least two disjoint paths between every $(x,y) \in \Gamma_{10}$.

Let $v$ be an ambiguous node and $w,u \in \Gamma_v$ be the nodes between which $v$ is interested in finding a disjoint path. The first step is finding disjoint paths between $u$ and $w$ in $G_v$, which can be achieved by using a $k$-connectivity detection algorithm \cite{henzinger2000computing}. The nodes used in local disjoint paths should be excluded from the global search. Let $A$ be the set of nodes used in the local disjoint paths between $u$ and $w$. For example, in \REFV{@R1C3}\REV{ Fig.~\ref{sub10k}}, we have $A=\{7,6\}$. In the proposed method, node $v$ broadcasts a $Discover(v,u,w,A)$ message to start a global path search in $G$. In \REFV{@R1C3}\REV{Fig. \ref{figure:node104}}, for example, node 10 broadcasts a $Discover(10,8,5,\{6,7\})$ message to start a path search between nodes 8 and 5 without the participation of nodes~6~and~7. Except node $w$, other 1-hop neighbors of $v$ rebroadcast the $Discover(v,u,w,A)$ \REFV{@R1C3}\REV{(Fig. \ref{figure:node104})} and in this way all nodes in $G_v$ receive the $Discover$ message. $Discover$ messages are not broadcasted by the 2-hop neighbors of the source node. In fact, only the 1-hop neighbors of the source node rebroadcast $Discover$ messages, which leads to at most $\Delta$ $Discover$ messages for each path search initiative. Each node that receives a $Discover(v,u,w,A)$ message from node $u$ broadcasts an $Explore$ message if it is not a member of $A\cup w$. In \REFV{@R1C3}\REV{Fig. \ref{figure:node105}}, node 11  broadcasts an $Explore$ message after receiving a $Discover$ message from node 8 because $11 \notin \{6,7,5\}$. After receiving the $Discover$ message, nodes 6 and 7 ignore all upcoming $Explore$ messages because $A$ covers both nodes. Each $Explore$ message contains the ID of the initiator, source, and target nodes and is rebroadcasted by the other nodes. Nodes 12 and 2 rebroadcast the received $Explore$ message (\REFV{@R1C3}\REV{Figs. \ref{figure:node105}}) and node 5 sends a \textit{Confirm} message to node 10 to inform that the search was successful (\REFV{@R1C3}\REV{Figs. \ref{figure:node106}}). In this way, the resulting graph becomes $k$-connected (\REFV{@R1C3}\REV{Figs. \ref{figure:node107}}).

The nodes that cannot find their status using their 2-hop local subgraphs start the aforementioned global path search process to determine their status. No synchronization is required for exchanging the \textit{Discover}, \textit{Explore}, and \textit{Confirm} messages, hence, the global path search can be conducted asynchronously between the nodes. The global path search operations are performed in the first (setup) phase of the algorithm when the nodes attempt to detect their status. 

Using the outcomes of Lemma~\ref{lemma_min} and Theorem~\ref{theorem:imaginary_correctness}, a subset of the nodes can determine their correct status based on the local information. The other nodes that cannot determine their status using the established theoretical framework start a global path search process and use global information to determine their status. Therefore, all nodes always find their correct status.

\subsection{Trade-off Between Mobility And Coverage}
\label{covvsmov}
Maintaining system connectivity while preserving system coverage are the two indispensable essential requirements.
Hence, many studies on the connectivity restoration problem emphasize the relationship between the connectivity and coverage problems and usually solve the restoration problem by introducing additional mobile nodes~\cite{bai2008complete,yun2010optimal,tian2005connectivity,zhu2012survey,li2009connectivity,sengupta2013multi,al2017optimal}.

To maintain the maximal coverage after a node failure without adding new nodes, a neighbor with a lower individual effect on the covered area can be moved to the position of the failed node. Obviously, this strategy can reduce the $k$ value and can even destroy the connectivity of the network. Another approach is to search the entire network, find a node that has the lowest individual coverage, and move it to the position of the failed node. This approach can generate costly movements. To obtain a balanced and fair method, we used a $\beta \in [0,1]$ parameter in the proposed algorithm that defines the desired coverage conservation ratio. $\beta=0$ means that coverage is not important at all and $\beta=1$ means that coverage preservation has the highest priority. The proposed approach always restores the $k$ value either by generating minimal movements and ignoring the coverage loss (i.e., lower $\beta$ values) or by minimizing the coverage loss and sacrificing the minimization of the movements (i.e., higher $\beta$ values).

\begin{figure}[!htbp]
	\captionsetup[subfloat]{farskip=9pt}
	\centering
	\subfloat[]{\includegraphics[width=0.165\textwidth]{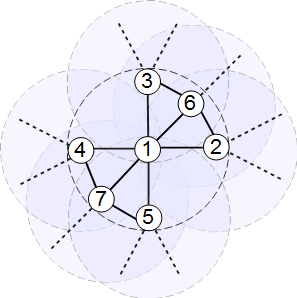}
		\label{cov0}}
	\hfil
	\subfloat[]{\includegraphics[width=0.165\textwidth]{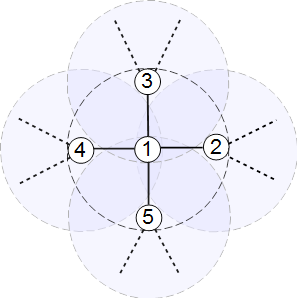}
		\label{cov1}}
	\hfil
	\subfloat[]{\includegraphics[width=0.13\textwidth]{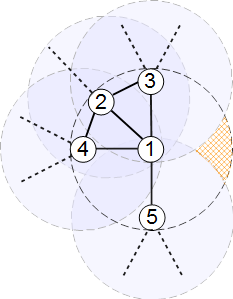}
		\label{cov2}}
	\hfil
	\subfloat[]{\includegraphics[width=0.14\textwidth]{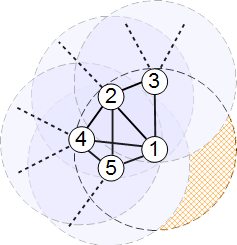}
		\label{cov3}}
	\hfil
	\subfloat[]{\includegraphics[width=0.13\textwidth]{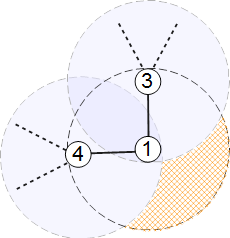}
		\label{cov4}}
	\hfil
	\subfloat[]{\includegraphics[width=0.13\textwidth]{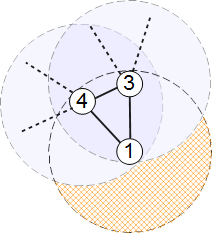}
		\label{cov5}}
	\hfil
	\caption{ The bluish texture indicates the intersecting coverage areas of different nodes and the yellowish texture indicates the coverage loss if node 1 moves away with a) $sup_1=14$, b) $sup_1$=8, c) $sup_1$=6, d) $sup_1$=4, e) $sup_1$=2, and f) $sup_1$=1.}
	\label{cov}
\end{figure}

We use a local-neighborhood-based heuristic to estimate the coverage loss after any event that leads to a decreased $k$. We define a support degree value for each node that determines the number of supporting neighbors that have intersecting coverage areas with the node's coverage area. Formally, $sup(v)$ is a value that is used to estimate the likelihood of the coverage loss due to the departure or failure of node $v$. The support degree is a relative value that allows us to compare the nodes based on their individual covered area. A higher support degree of a node means that we lose a smaller covered area after moving or losing that node in comparison to the other nodes. We can model an MSN as a unit disk graph and our heuristic for calculating the support degree is based on the following two facts about unit disk graphs
 \begin{enumerate}
 	\item The nodes with a higher number of neighbors have a lower individual covered area and can have a higher support degree,
 	\item The nodes that have a higher number of independent neighbors (the neighbors that have no direct link between themselves) have a lower individual covered area and can have a higher support degree.
 \end{enumerate}
For example, in Fig.~\ref{cov0}, node 1 has six neighbors and the area it covers is completely covered by other nodes. 
In Fig.~\ref{cov5}, node 1 has only two connected neighbors and about half of the area it covers is not covered by other nodes. Based on the above facts, we propose the following equation to calculate a support degree for each node
\begin{multline}
\label{coveragerul}
Sup(u,G(V,E))\!=\!\dfrac{1}{k}\times\!\sum_{v\in \Gamma_u}^{}\!d_u- |\{(v,w)\in E : w\in \Gamma_u\}|
\\\REFV{@R1C2}\REV{ if \quad k(G/u)=k(G) \qquad \qquad \qquad }
\end{multline}
In \REFV{@R1C4}\REV{Eq.~\ref{coveragerul}}, for each node $u$ and one of its neighbors $v\in \Gamma_u$, we find the difference between the degree of $u$ and the number of common neighbors of $u$ and $v$, which indicates the number of neighbors of $u$ that are not directly connected to $v$. We calculate the sum of these values for all the neighbors of $v$ and divide it by $k$ to have the same base line for all nodes. Using \REFV{@R1C4}\REV{Eq.~\ref{coveragerul}}, nodes with lower individual covered areas obtain higher support degrees. Fig.~\ref{cov} illustrates the relationship between the neighbors and the covered area of node~1 where we assume that the general network is 2-connected (the nodes are a part of a 2-connected network). In Fig.~\ref{cov0}, for example, node~1 has six neighbors and $d_1=6$. Therefore, for node 1, we have $sup_1=1/2 \times ((6-1)+(6-2)+(6-1)+(6-1)+(6-2)+(6-1))=14$.  
Similarly, in Figs.~\ref{cov1},~\ref{cov2},~\ref{cov3},~\ref{cov4}, and~\ref{cov5}, the support values of node 1 are 8, 6, 4, 2, and 1, respectively. 

\REFV{@R1C2}\REV{Note that, only the support degree of $Trusted$ nodes are used in the restoration phase. Since the $Joint$ nodes are not selected for moving, their support degree has no effect on the moving strategy.}

As illustrated in Fig.~\ref{cov}, it is more likely that for a larger support value, the individual covered area is lower, hence, the coverage loss after moving is also lower.
Note that the support value of a node provides only an estimation about the coverage loss after failure. In the proposed algorithm, node $v$ moves to the position of node $u$ under one of the following conditions
 \begin{enumerate}
	\item $u$ is a failed \textit{Joint} node and $v$ has the lowest coverage-aware moving cost to $u$ among the other neighbors,
	\item $u$ moved to the location of another node, $(sup(u)-sup(v))\times\beta> 1$ and $v$ has the lowest coverage-aware moving cost to $u$ among the other neighbors.
\end{enumerate}
$(sup(u)-sup(v))\times\beta> 1$ implies that moving $v$ to the position of $u$ increases the coverage ratio with respect to the $\beta$ value. To calculate the coverage-aware moving cost of node $v$ to the position of node $u$, we divide the movement cost between nodes $u$ and $v$ by $1+(sup(v)\times \beta)$. Thus, higher $\beta$ and $sup$ values, effectively, reduce the moving costs and lead to the selection of the nodes with lower individual covered areas even if they have relatively higher moving costs. The desired $\beta$ value can be determined by the user based on the requirements of the application.

\subsection{The Steps of the Proposed Algorithm}
\label{algorithmsec}
The steps of the proposed algorithm (LINAR) are presented in Algorithm~\ref{alg:linar}.The algorithm accepts the $k$ value and the coverage conservation ratio $\beta$ as an input parameter. Each node maintains the following set of local variables
\begin{itemize}
	\item $I_u$: Available information about other nodes.
	\item $\Gamma_u$: The 1-hop neighbor list of node $u$.
	\item $V_u$: The nodes in the 2-hop local subgraph of node $u$.
	\item $E_u$: The edges in the 2-hop local subgraph of node $u$.
	\item $stat_u$: Status (\textit{Trusted} or \textit{Joint}) of node $u$.
	\item $T_u$: Target nodes of $u$ in the  path search.
	\item $R_u$: The received messages in node $u$.
\end{itemize}

Each node starts the algorithm by broadcasting a $Start$ message (line 2) that includes its coordinates (position). When node $u$ receives a $Start$ message from node $w$ (line 3), it adds $w$ to $\Gamma_u$ and the position of $w$ to $I_u[w].pos$ (line 4). Each node calls the $CreateGraph$ procedure after $ts$ time units from the first $Start$ message reception (line 5). The $ts$ delay should be long enough to allow sending all the $Start$ messages. In the $CreateGraph$ procedure, each node creates its 1-hop local subgraph and broadcasts its 1-hop neighbor list and their position information in an $Ngb$ message (lines 7-8). When node $u$ receives an $Ngb$ message from node $w$ (line 9), it updates $I_u$, $V_u$, and $E_u$ with the received lists and calls the $DetectState(u)$ procedure if it receives $Ngb$ messages from all $v \in \Gamma_u$ (lines 9-12).

\begin{algorithm} [!h]
\SetAlgoNoLine

\begin{algorithmic}[1] \caption{LINAR ($k$, $\beta$)} \label{alg:linar}\scriptsize

\STATE\textbf{Initially:}
\item[] $I_u\leftarrow [\,]$.  \quad //information  of 2-hop neighbors in $u$
\item[] $\Gamma_u\leftarrow \O$. \quad  //1-hop neighbor list of $u$
\item[] $V_u \leftarrow \{u\}, \; E_u\leftarrow \O$. \quad //nodes and edges sets of local subgraph
\item[] $stat_u \leftarrow joint$, $sup_u \leftarrow 0$. //status and num. of independent neighbors
\item[] $T_u\leftarrow \O$, $R_u\leftarrow \O$. \quad //target nodes and Received messages set
\item[]
\item[] \textit{// Initialization Phase}
\item[]
\STATE Node $u$ starts the algorithm by \textbf{broadcasting} \textit{\textbf{Start}}($pos_{u}$).
\item[]

\STATE\emph{When $u$ receives \textbf{Start}$(pos_w)$ from $w$:}
\STATE \quad $\Gamma_u \leftarrow \Gamma_u \cup \{w\}$, $I_u[w].pos \leftarrow p_w$.
\STATE \quad \textbf{if} $|\Gamma_u| = 1$ \textbf{then} \textbf{call} \emph{CreateGraph} after \emph{ts} time units.
\item[]

\STATE \emph{procedure \textbf{CreateGraph}:}
\STATE \quad$\forall w\in \Gamma_u:  V_u\leftarrow V_v\cup \{w\}$ \emph{and} $E_u\leftarrow E_u\cup \{(w,u)\}$.
\STATE \quad \textbf{broadcast} \emph{\textbf{Ngb}}($\Gamma_u,I_u$).
\item[]
\STATE \emph{When $u$ receives \textbf{Ngb}($\Gamma_w,I_w$) from $w$:}
\STATE \quad $\forall v \in I_w : I_u[v] \leftarrow I_w[v]$.
\STATE \quad$\forall v\in \Gamma_w: V_u\leftarrow V_u\cup \{v\}$ \textbf{and} $E_u\leftarrow E_u\cup \{(v,w)\}$.
\STATE \quad \textbf{if} $\forall v\in \Gamma_u:$ $v$ has sent $Ngb$ \textbf{then} call \emph{DetectState(u)}.

\item[]
\STATE \emph{procedure \textbf{DetectState(u)}:}
\STATE  \quad $G_u \leftarrow (V_v,E_v)$, $sup_u\leftarrow Sup(u,G_u)$.
\STATE  \quad \textbf{if} $\forall v \in \Gamma_u$: $d_v>k$ \textbf{then}
\STATE  \qquad $E'_u\leftarrow \{e=(i,j)\notin E_u : \, i,j\in V_u/\Gamma_u \,\, and\,\, i \overset{G_u}{\sim} j < k\}$.
\STATE  \quad\quad $S_u\!\leftarrow\!\{S\!\!\subseteq\! E'\!:\!\kappa(G'(V_u,S\!\cup\!E_u))\!=\!k \, and \!\!\!\!\!\sum\limits_{\substack{x,y\in \Gamma_u}}\!\!\!\!x\!\overset{G'}{\sim}\!y\,is\, min.\}$.

\vspace*{-0.6em}
\STATE  \quad\quad $G_u'\!\leftarrow\!(V_u,E_u\! \cup\! S_u\!)/u$.

\STATE  \qquad \textbf{if} $\kappa(G'_u)=k$ \textbf{then} $stat_u \leftarrow trusted$.
\STATE  \qquad \textbf{else} \textbf{while} $\kappa(G'_u)<k$ \textbf{do}
\STATE  \qquad \quad  $A \leftarrow dpath(G'_u,v,z)$  \textbf{where} $(v,z)\in \Gamma_u$ \textbf{and}  $v \overset{G'_u/u}{\sim} z < k$.
\STATE  \qquad \quad $G'_u\leftarrow (V_u,E_u\cup(v,z))$,  $T_u\leftarrow T_u\cup \{z\}$.
\STATE  \qquad \quad \textbf{broadcast} \emph{\textbf{Discover}}($u,v,z,A$).
\STATE  \quad \textbf{broadcast} \emph{\textbf{Stat}($stat_u$,$sup_u$)}.

\item[]
\STATE\emph{When $u$ receives \textbf{Discover}$(x,v,z,A)$ from $w$:}
\STATE \quad \textbf{if} $u \ne z$ \textbf{and} $Discover(x,y,z) \notin R_u$ \textbf{then}
\STATE \qquad $R_u \leftarrow R_u \cup Discover(x,y,z)$.
\STATE \qquad \textbf{if} $x = w$ \textbf{then} \textbf{broadcast} \emph{\textbf{Discover}}($x,v,z,A$).
\STATE \qquad  \textbf{if} $u \in A$ \textbf{then} $R_u \leftarrow R_u \cup Explore(x,y,z)$.
\STATE \qquad \textbf{else} \textbf{if} $v \in \Gamma_u$ \textbf{then} \textbf{broadcast} \emph{\textbf{Explore}}($x,v,z$) after \textbf{\emph{ts}} time unit.

\item[]
\STATE\emph{When $u$ receives \textbf{Explore}$(x,v,z)$ from $w$:}
\STATE \quad \textbf{if} $Explore(x,y,z) \notin R_u$ \textbf{then}
\STATE \qquad  $R_u \leftarrow R_u \cup Explor(x,y,z)$.
\STATE \qquad \textbf{if} $z = u$ \textbf{then} \textbf{send} \emph{\textbf{Confirm}} to $x$.
\STATE \qquad \textbf{else} \textbf{broadcast} \emph{\textbf{Explore}}($x,v,z$).

\item[]
\STATE\emph{When $u$ receives \textbf{Confirm} from $w$ :}
\STATE \quad $T_u \leftarrow T_u/w$.
\STATE \quad \textbf{if} $T_u = \O$ \textbf{then} $stat_u \leftarrow trusted$,  \textbf{broadcast} \emph{\textbf{Stat}}($state_u, supp_u$).

\item[]
\item[] \textit{// Recovery Phase}
\item[]
\STATE\emph{When $u$ receives \textbf{Stat}$(s,i)$ from $w$:} $I_u[w].stat \leftarrow s$,  $I_u[w].sup \leftarrow i$.
\STATE\emph{When $u$ \textbf{detects failure} in $w$, $I_u[w].stat$ =joint:} call \textit{update}($w$).
\STATE\emph{When $u$ \textbf{detects move} of $w$, $(I_u[w].sup\!-\!sup_u)\!\times\!\beta\!\!>\!\! 1$:} call \textit{update}($w$).

\item[]
\STATE \emph{procedure \textbf{update}(w):}
\STATE \, \textbf{if} $stat_u $=\textit{Trusted} \textbf{then}
\STATE \quad \, \textbf{if} $\nexists (w,v)\!\in\!E_u$ s.t ($I_u[v].stat$= \textit{trusted} \textbf{and} \textit{LowerCost}$(v,\!w)=$\textit{true}) 
\STATE \qquad \quad \textbf{then} \textbf{move} to position $I_u[w].pos$.
\STATE \, \textbf{else if} $I_u[w].stat$=\textit{joint} \textbf{then}
\STATE \qquad \quad \textbf{if} $\nexists(w,v)$ \textbf{s.t} ($I_u[v].stat$= \textit{Trusted} \textbf{or} \textit{LowerCost}$(v,\!w)$=\textit{true}) 
\STATE \qquad \qquad \textbf{then} \textbf{move} to position $I_u[w].pos$.

\item[]
\STATE \emph{procedure \textbf{LowerCost(v,w)}:}
\STATE \quad $c_v\leftarrow cost(I_u[v].pos,I_u[w].pos)/(1+ (I_u[v].sup*\beta))$.
\STATE \quad $c_u\leftarrow cost(pos_u,I_u[w].pos)/(1+ (sup_u*\beta))$.
\STATE \quad \textbf{if} $c_v<c_u$ \textbf{or} ($c_v=c_u$ \textbf{and} $v<u$) \textbf{then} \textbf{return} \emph{true}.
\STATE \quad \textbf{else return} \emph{false}.

\end{algorithmic}

\end{algorithm}

In the $DetectState$ procedure, node $u$ creates its 2-hop local subgraph $G_u$ using the sets $V_u$ and $E_u$. Node~$u$ then calculates its support degree using $G_u$ and the provided method in \REFV{@R1C4}\REV{Eq.~\ref{coveragerul}} (line 14). Node $u$ ignores the remaining commands in the $DetectState$ procedure (line 15) and broadcasts its default status value (which is \textit{Joint}) if it has at least one neighbor $v$ with $d_v=k$ (based on Lemma~\ref{lemma_min}). Otherwise, node $u$ stores the set of all possible imaginary edges based on rules \ref{rule:onehop} and \ref{rule:newpath} in $E'_u$ (line 16). Node~$u$ then selects one of the subsets of $E'_u$ that minimizes $\sum_{x,y\in \Gamma_v} x \overset{G_v}{\sim} y$ (Definition~\ref{rule:minpath}) and stores this subset in $S_u$ (line 17). 

After creating the $k$-connected local graph $G_u^k$ using $S_u$, node $u$ stores the $G_u^k/u$ in $G'_u$ (line 18) and sets its state to $Trusted$ if $G'_u$ is $k$-connected (line 19). Otherwise, $u$ selects two nodes $\{v,z\} \in \Gamma_u$ with the minimum degree such that $v \overset{G'_u/u}{\sim} z < k$, detects the nodes that are already used in $k-1$ disjoint paths between $v$ and $z$ in $G'_u$, and stores them in set $A$ (line 21). Afterward, $u$ adds an edge between $v$ and $z$, and adds $z$ to the search target list $T_u$ (line 22). Then, $u$ starts a global path search process by broadcasting a $Discover(u,v,z,A)$ message (line 23). Node $u$ sends a new $Discover$ message for each pair of selected nodes from $G'_u$ until $G'_u$ becomes $k$-connected. Finally, node $u$ broadcasts its default ($Joint$) or updated ($Trusted$) state and its support degree (line 24).

When node $u$ receives an unvisited $Discover(x,v,z,A)$ message from node $w$, it adds the message to the received messages set $R_u$ and rebroadcasts the $Discover$ if the sender is the initiator of the search process (lines 25-27). If $u \in A$, then it adds an $Explore(x,y,z)$ to $R_u$ to ignore the upcoming $Explore$ messages (line 29). If $v \in \Gamma_u$, then node $u$ broadcasts an $Explore$ message after $ts$ time units (line 30). The $ts$ delay is to ensure that all nodes receive $Discover$ messages before receiving $Explore$ messages. The first piece of data in each $Explore$ message is the ID of the initiator node. The second and third data pieces are the IDs of the source and target nodes in the path. When node $u$ receives an unvisited $Explore(x,v,z)$ from node $w$, it adds the message to $R_u$ (lines 31-33). If $u$ is the target of the search, it sends a $Confirm$ message to the initiator node (line 34). Otherwise, node $u$ rebroadcasts the received $Explore$ message (line 35).

When node $u$ receives a $Confirm$ message from $w$, it removes $w$ from the target list and changes its state to \textit{Trusted} if the target list becomes empty (lines 36-38). The nodes that change their status broadcast a $Stat$ message including their new states and support values (line 38). When node $u$ receives a $Stat(s,i)$ message from its neighbor $w$, it updates the state and support values of $w$ in its local set $I_u$ (line 39). We assume that the nodes periodically broadcast a beacon packet (e.g., one packet per minute) to indicate that they are still alive. Therefore, when a node does not receive a beacon from a neighbor for a certain number of periods (e.g., five periods), it assumes that the neighbor is unavailable (e.g., it stopped working or moved away).  When node $u$ detects the failure of a \textit{Joint} node $w$, it calls the \textit{update} procedure to move to the failed node's position if there is no better candidate than itself (line 40). Similarly, node $u$ calls the \textit{update} procedure if it detects the movement of a neighbor node $w$ with $(I_u[w].sup-sup_u)\times\beta>1$ (line 41).

The \textit{update} procedure (line 42) of node $u$ accepts the ID of a target node $w$ and moves node $u$ to the position of $w$ if there is no better candidate than $u$. If $u$ is a trusted node and $w$ has no trusted neighbors with a lower movement cost than $u$, then $u$ moves to the position of $w$ (lines 44-45). Otherwise, if $u$ is a joint node and $w$ has no trusted neighbors or joint neighbors with a lower moving cost, than $u$ then $u$ moves to the position of $w$ (lines 46-48).


The $LowerCost$ procedure of node $u$ accepts the IDs of two other nodes ($v$ and $w$) and returns true if the coverage-aware movement cost of $u$ to $w$ is lower than the coverage-aware movement cost of $v$ to $w$ or if their movement costs are equal but the ID of $u$ is smaller than the ID of $v$ (as elaborated in Section~\ref{covvsmov}). Otherwise, this procedure returns false (lines 49-53). 
The movement cost of $u$ to $w$ is calculated in a similar manner. The procedure returns true if $c_v<c_u$ or ($c_v=c_u$ and $v<u$).

\subsection{Complexity Analysis}
\label{complexsec}
In this subsection, we provide theorems on the bit, time, space, and computational complexities of LINAR.

\begin{theorem}
\label{theorem:receivedComplexity}
The bit complexities of LINAR are $O(n^2 \times log_2n)$ and $\Omega(n \times k \times log_2n)$ for the worst and best cases, respectively.
\end{theorem}

\begin{theorem}
\label{theorem:spaceComplexity}
The space complexity of LINAR is $O(\Delta^4+ n\times \Delta)$.
\end{theorem}

\begin{theorem}
\label{theorem:timeComplexity}
The time complexity of LINAR is $O(D)$.
\end{theorem}

\begin{theorem}
\label{theorem:computationalComplexity}
The computational complexity of LINAR is $O(n^2 \times \Delta)$.
\end{theorem}

\section{Performance Evaluation}
\label{PerformanceEvaluation}

We implemented LINAR, MCCR, and TAPU on testbed environments along with a \textit{Localized}, a \textit{Greedy}, and a \textit{Basic} central algorithm to compare their performances with LINAR. In the \textit{Localized} approach, the sink node selects a neighbor of the failed node with the smallest degree and moves it to the position of the failed node and repeats this process until the resulting topology becomes \emph{k}-connected. The sink node sends a message including the target position to each node that should move to a new position. In the \textit{Greedy} algorithm, the sink node selects the nearest neighbor (the neighbor with the minimum moving cost) of the failed node and moves it to the position of the failed node (by sending a message similar to the \textit{Localized} approach) and repeats this process until the resulting network becomes \emph{k}-connected. In the \textit{Basic} algorithm, we assume that there are a sufficient number of redundant nodes and after each failure, one of these nodes is sent to the position of the failed node.

\begin{figure}[!htbp]
\captionsetup[subfloat]{farskip=9pt}
\centering
\subfloat[]{\includegraphics[width=0.1\textwidth]{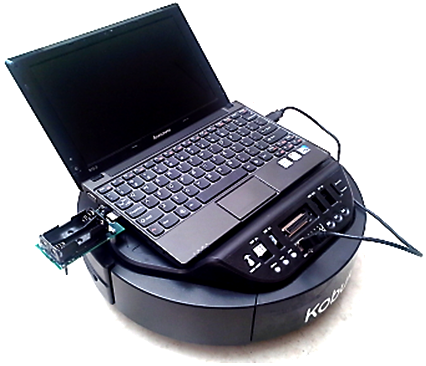}
\label{mobile}}
\subfloat[]{\includegraphics[width=0.35\textwidth]{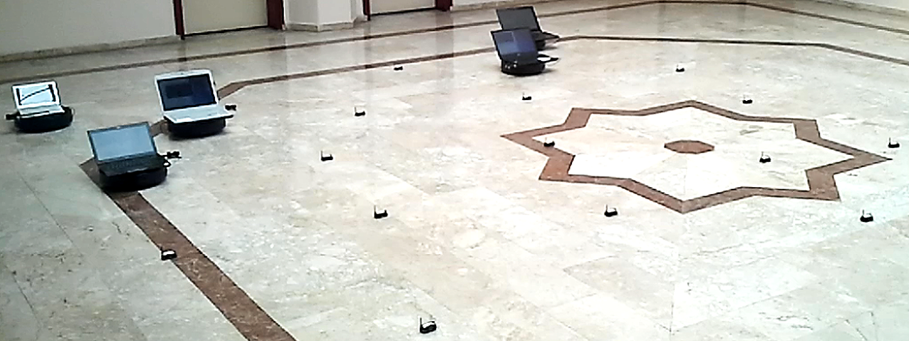}
\label{testbed}}
\hfil
\\
\subfloat[]{\includegraphics[width=0.35\textwidth]{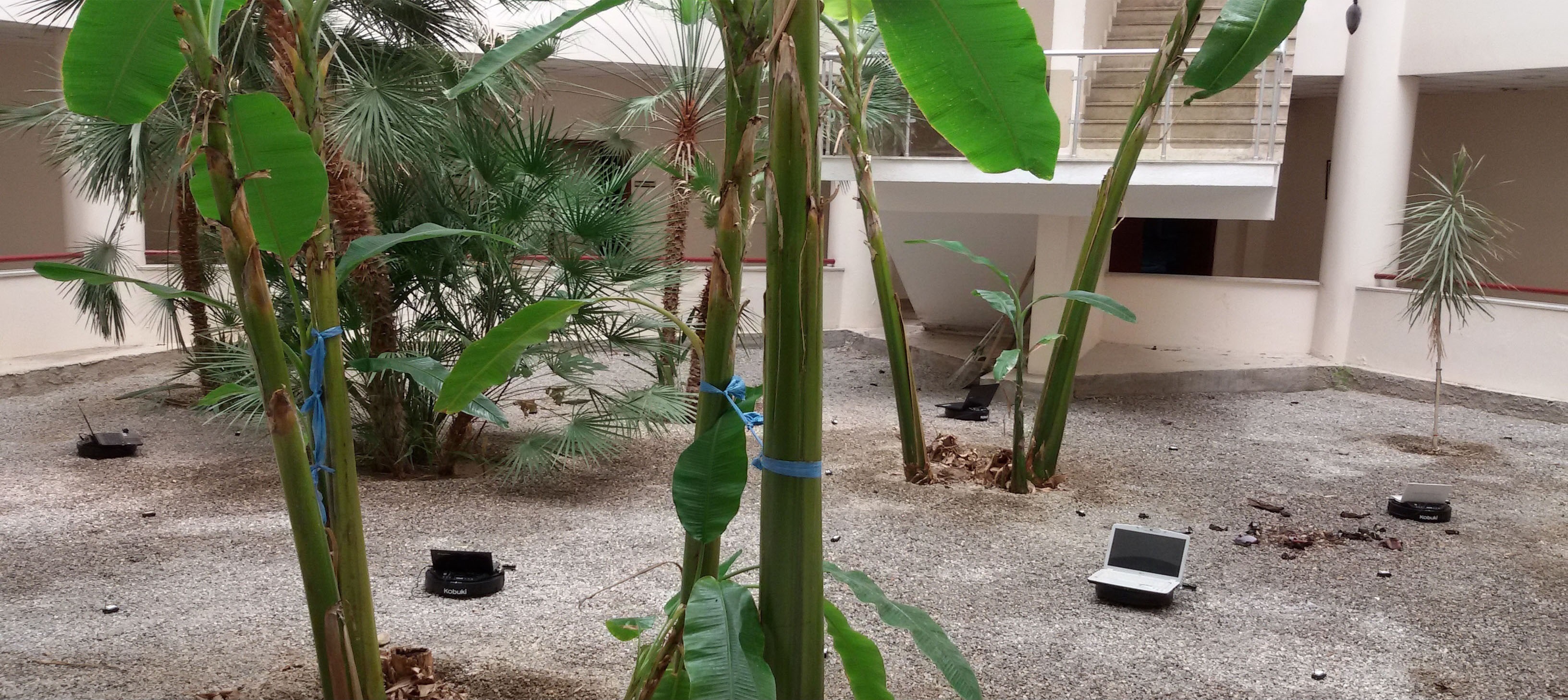}
\label{orman}}
\\
\subfloat[]{\includegraphics[width=0.35\textwidth]{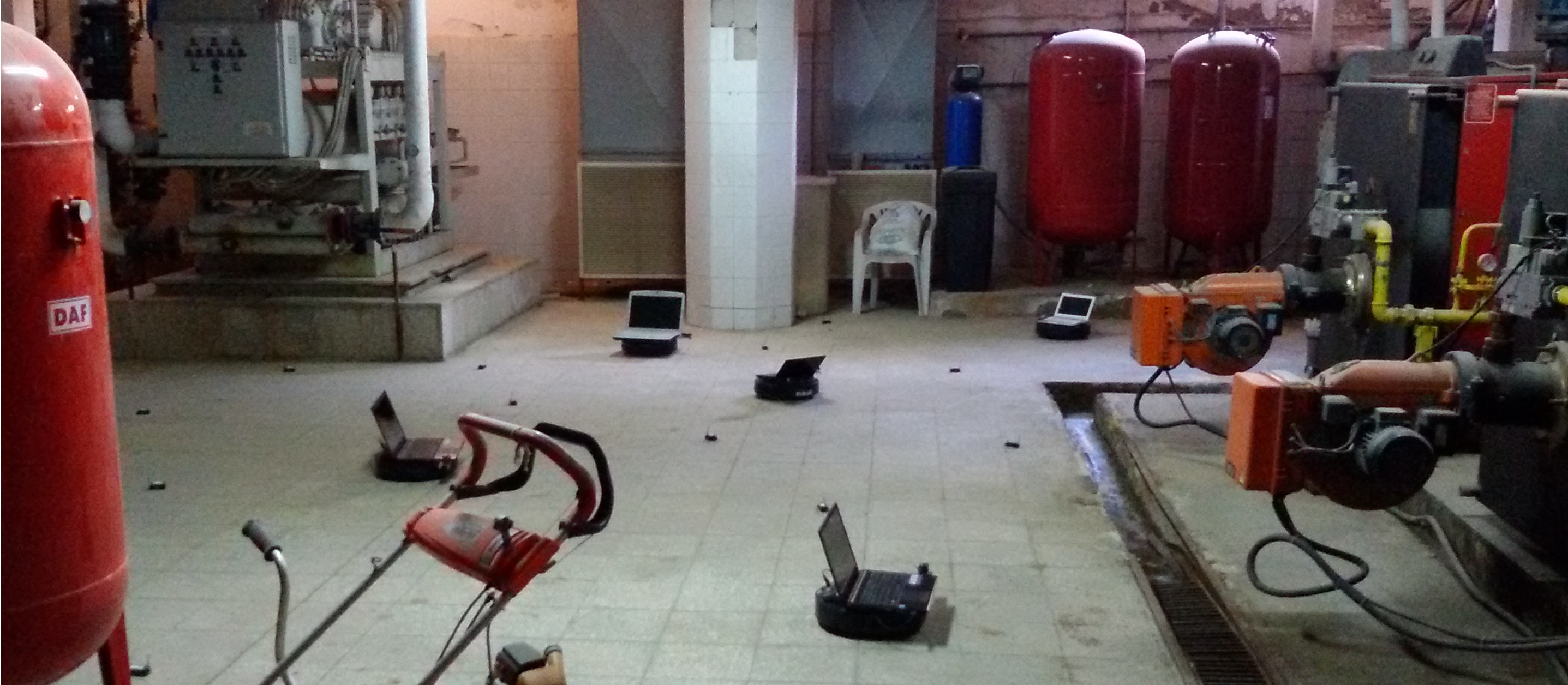}
\label{gas}}
\hfil
\caption{a) A mobile node utilized in experiments b) hall testbed, c) garden testbed, d) powerhouse testbed.}
\label{realexper}
\end{figure}

\begin{figure}[!h]
\captionsetup[subfloat]{farskip=9pt}
\centering
\subfloat[]{\includegraphics[width=0.23\textwidth]{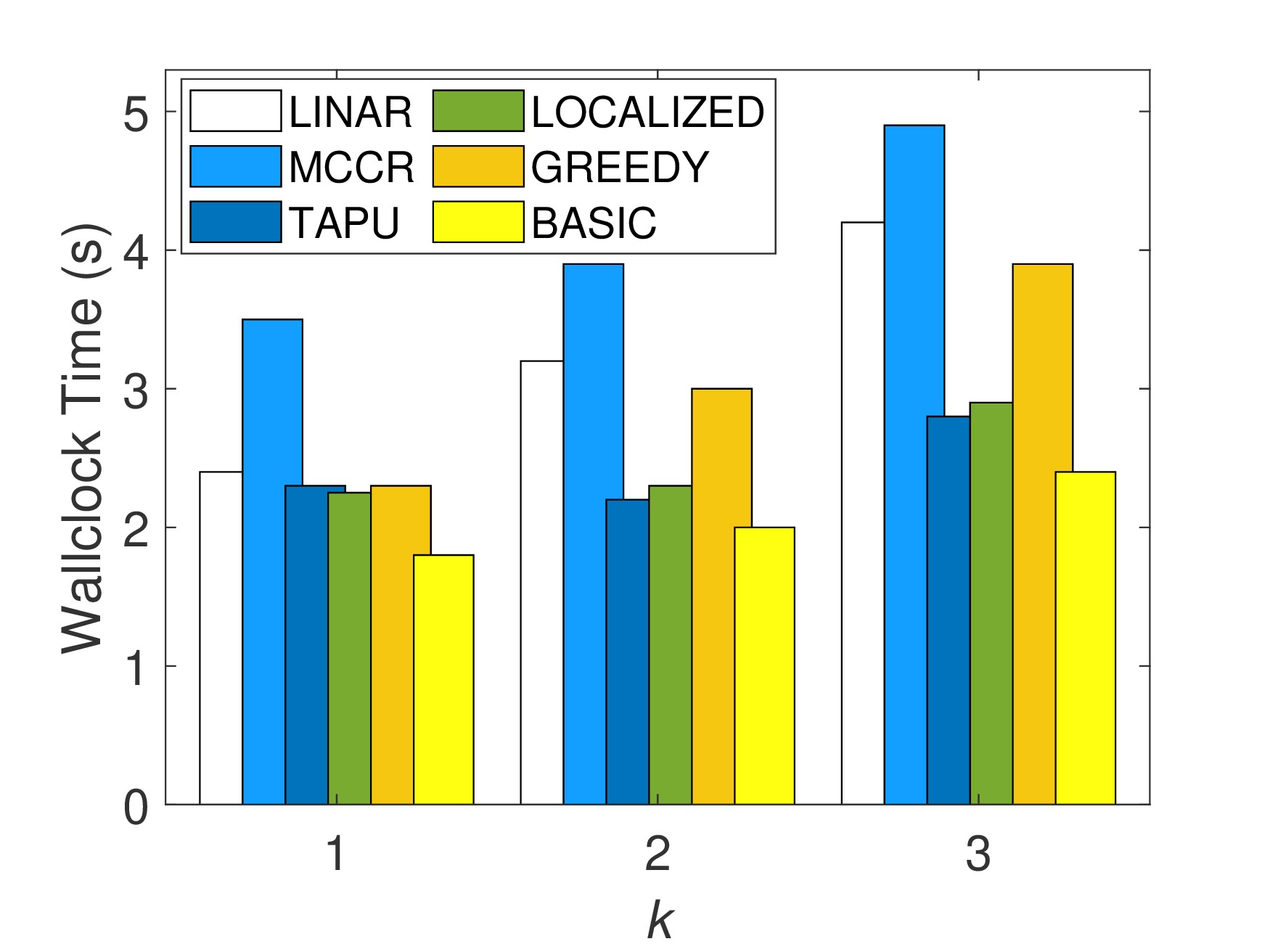}
\label{real_time}}
\hfil
\subfloat[]{\includegraphics[width=0.23\textwidth]{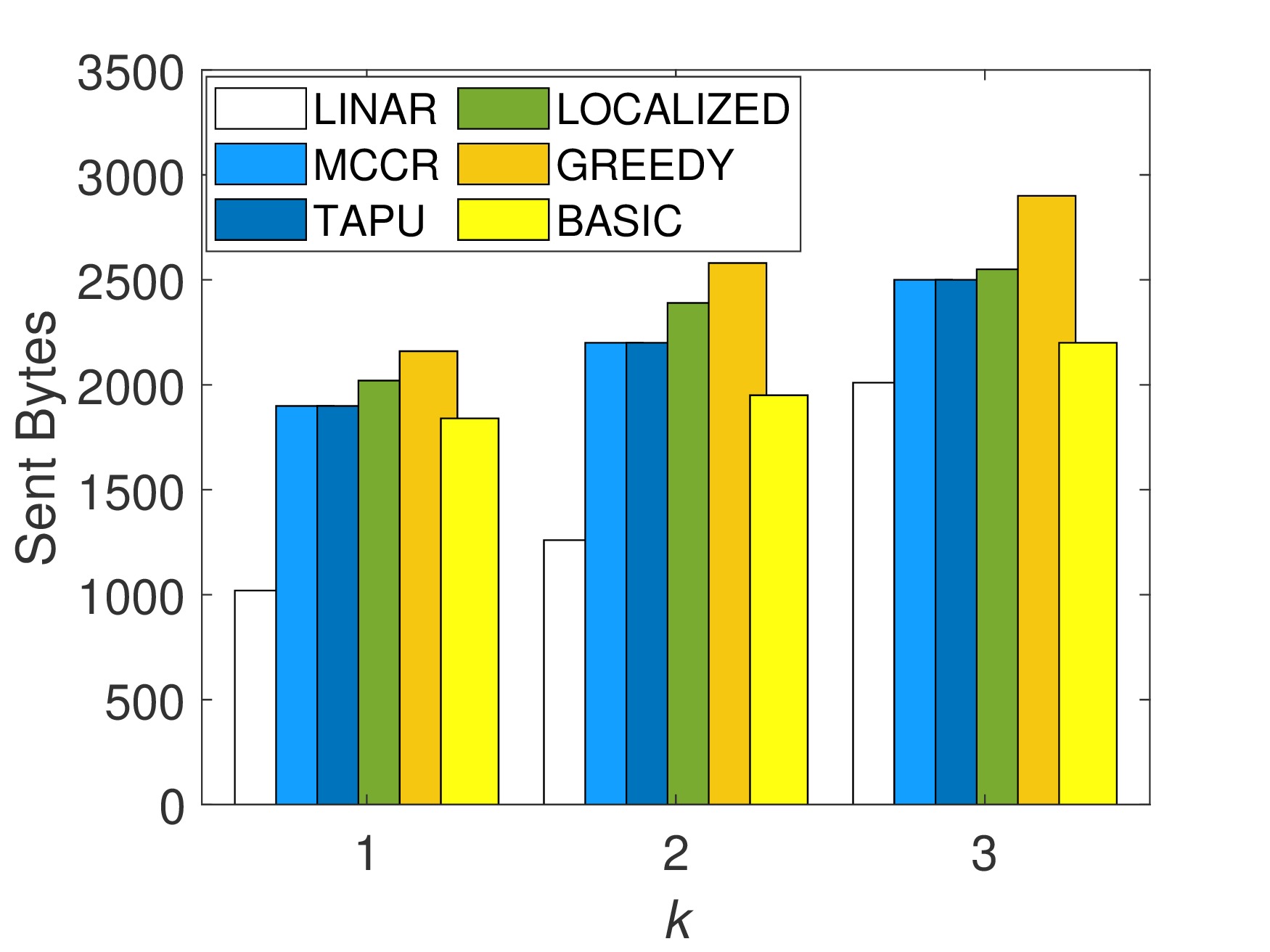}
\label{real_sent}}
\hfil\\
\subfloat[]{\includegraphics[width=0.25\textwidth]{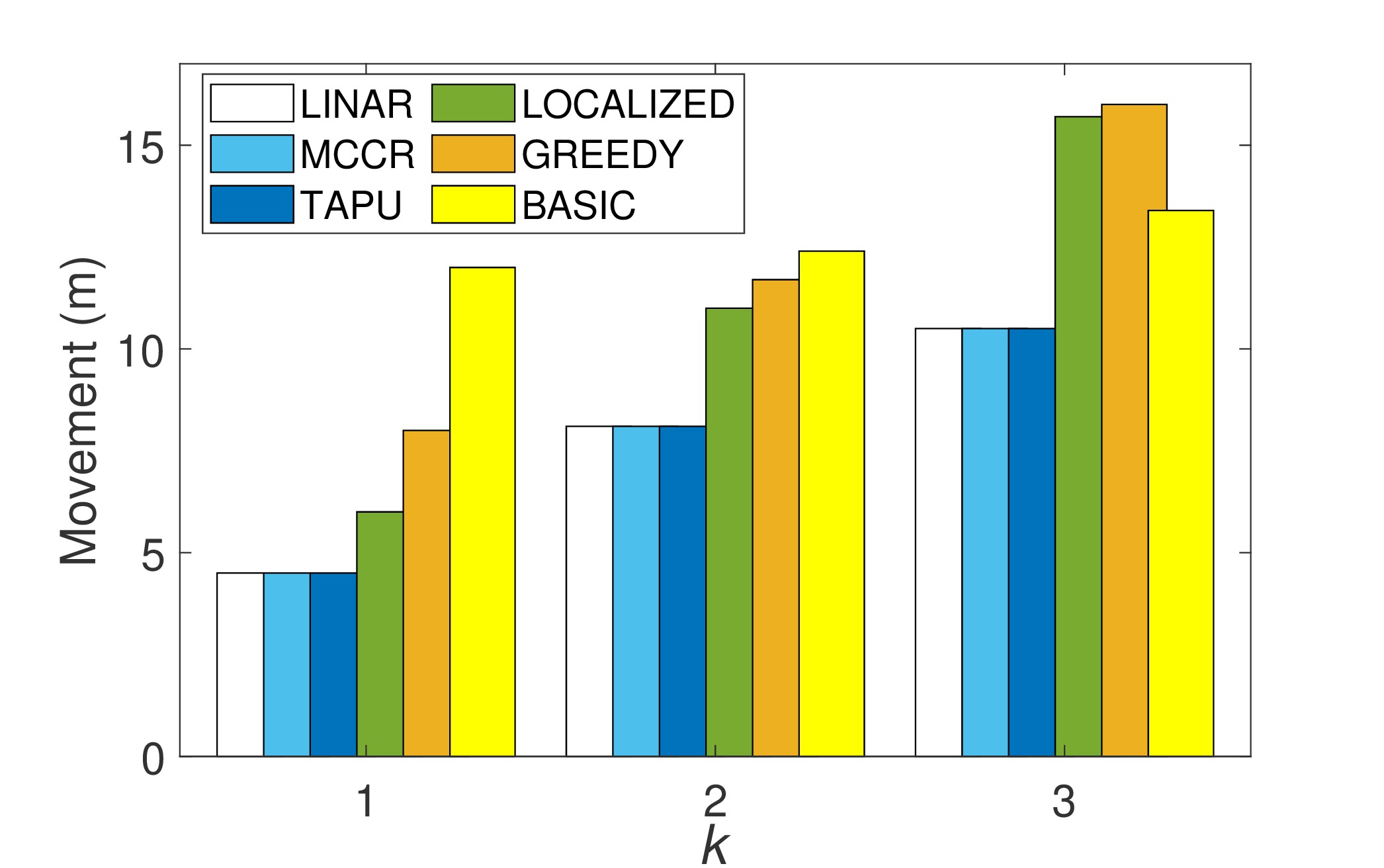}
\label{real_move}}
\hfil\\
\caption{a) Wallclock times b) sent bytes c) movement costs.}
\label{threeCentral}
\end{figure}

All nodes broadcast a beacon packet every 2 s to inform that they are alive. A node is marked as failed by its neighbors if it does not broadcast a beacon for 10 s. In central algorithms, all nodes send their neighbor lists to the sink node and the sink node creates a graph for the entire network topology, furthermore, the nodes that detect a failure send the failed node ID to the sink and wait for the movement messages from the sink. In the experiments, we counted the total sent bytes of all messages  except the beacon packets. We used Crossbow IRIS motes \cite{IRIS} and Kobuki \cite{yujinrobot} robots to create MSNs. 
We integrated an IRIS mote with a Kobuki robot using a small laptop to create a mobile node (Fig. \ref{mobile}). We established 15 random networks in three different environments (five networks for each type of terrain) and set the communication range of all nodes to 1.5 m to establish the desired topologies. We established MSNs with 5 mobile and 15 static nodes in a hall with marble flooring (Fig. \ref{testbed}), a garden with a soil surface (Fig. \ref{orman}), and a powerhouse with mosaic flooring (Fig. \ref{gas}). Because of the limited number of available Kobuki robots and IRIS nodes, we established networks with $k$=1, 2, and 3 with $\beta=0$ in all experiments. However, we performed simulations with different $\beta$ and $k$ values. In all experiments and simulations, we set the moving cost function of nodes to the Euclidean distance between their source and target locations.

Fig. \ref{real_time} shows the average wallclock times of the algorithms in the testbed experiments. The moving times of nodes from the source to the target position are not included in the wallclock times. The wallclock times of MCCR are higher than those of all other algorithms and the \textit{Basic} algorithm has the best performance. The wallclock times of LINAR are lower than those of MCCR. Fig. \ref{real_sent} shows the sent bytes of algorithms after failures in two nodes in different topologies. The sent bytes of LINAR are significantly lower than those of the other algorithms because in this algorithm the topology information is not sent to a single node. For $k$=1, the total sent bytes of LINAR are at least 47.1\% lower than those of the other algorithms. For $k$=3, this ratio is approximately 8.3\%. Fig. \ref{real_move} shows the average movement costs of the implemented algorithms after failures in two \textit{Joint} nodes. LINAR, MCCR, and TAPU generate movements with lower costs than the \textit{Greedy}, \textit{Basic}, and \textit{Localized} algorithms in all $k$ values.

\begin{figure*}[!htbp]
	\captionsetup[subfloat]{farskip=9pt}
	\centering
	\subfloat[]{\includegraphics[width=0.24\textwidth]{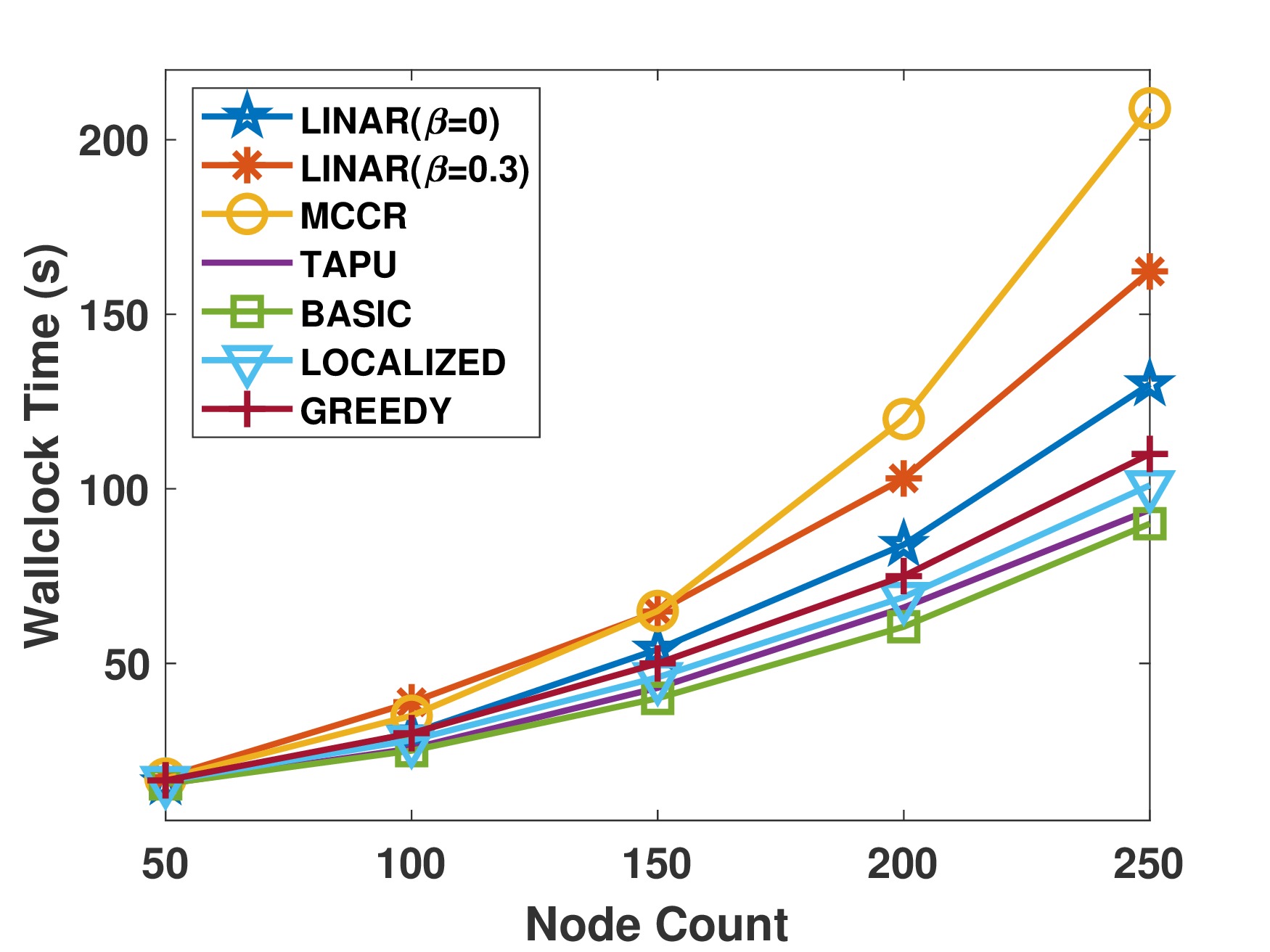}
		\label{time_n_all}}
	\hfil
	\subfloat[]{\includegraphics[width=0.24\textwidth]{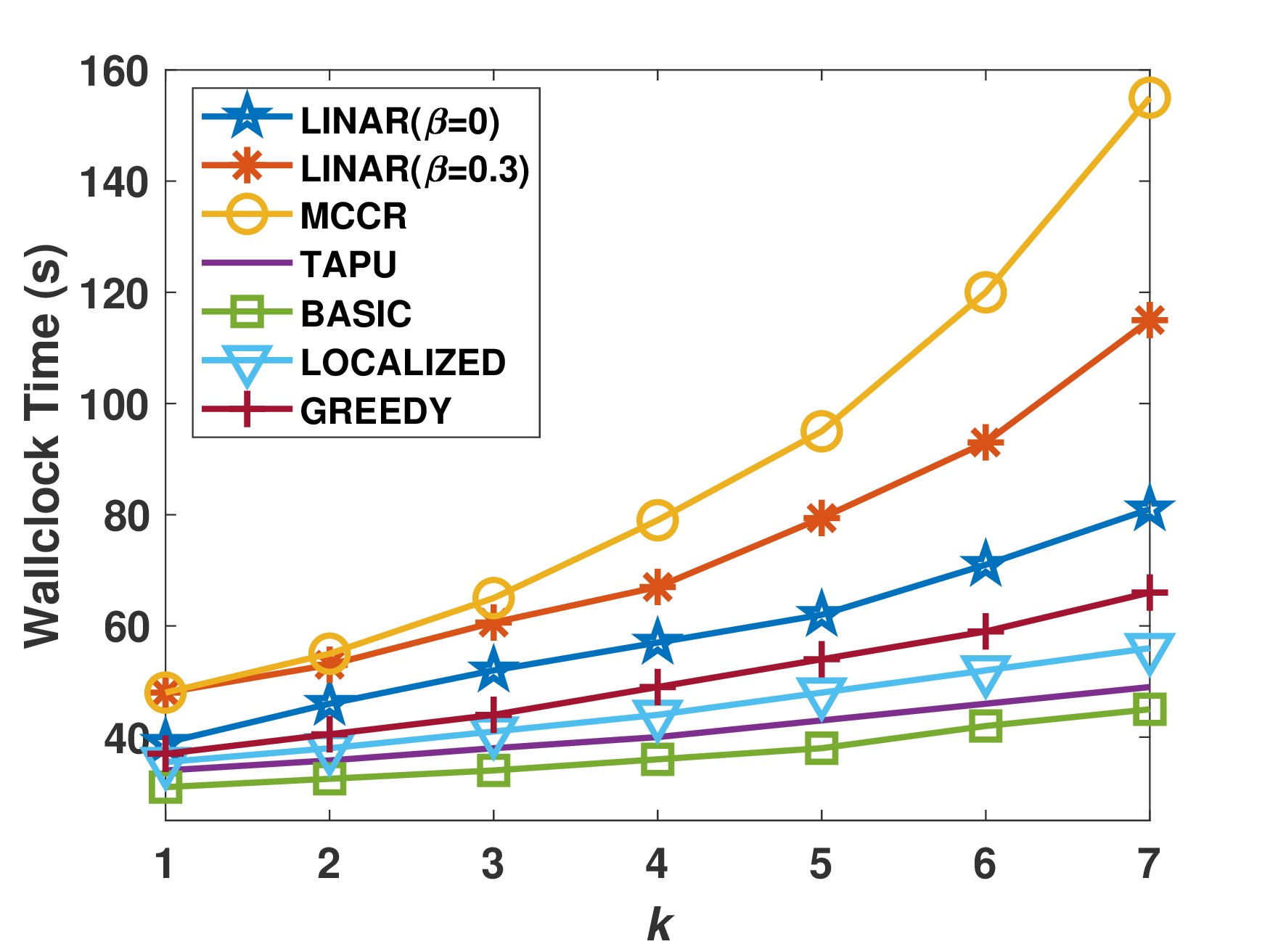}
		\label{time_k_all}}
	\hfil
	\subfloat[]{\includegraphics[width=0.24\textwidth]{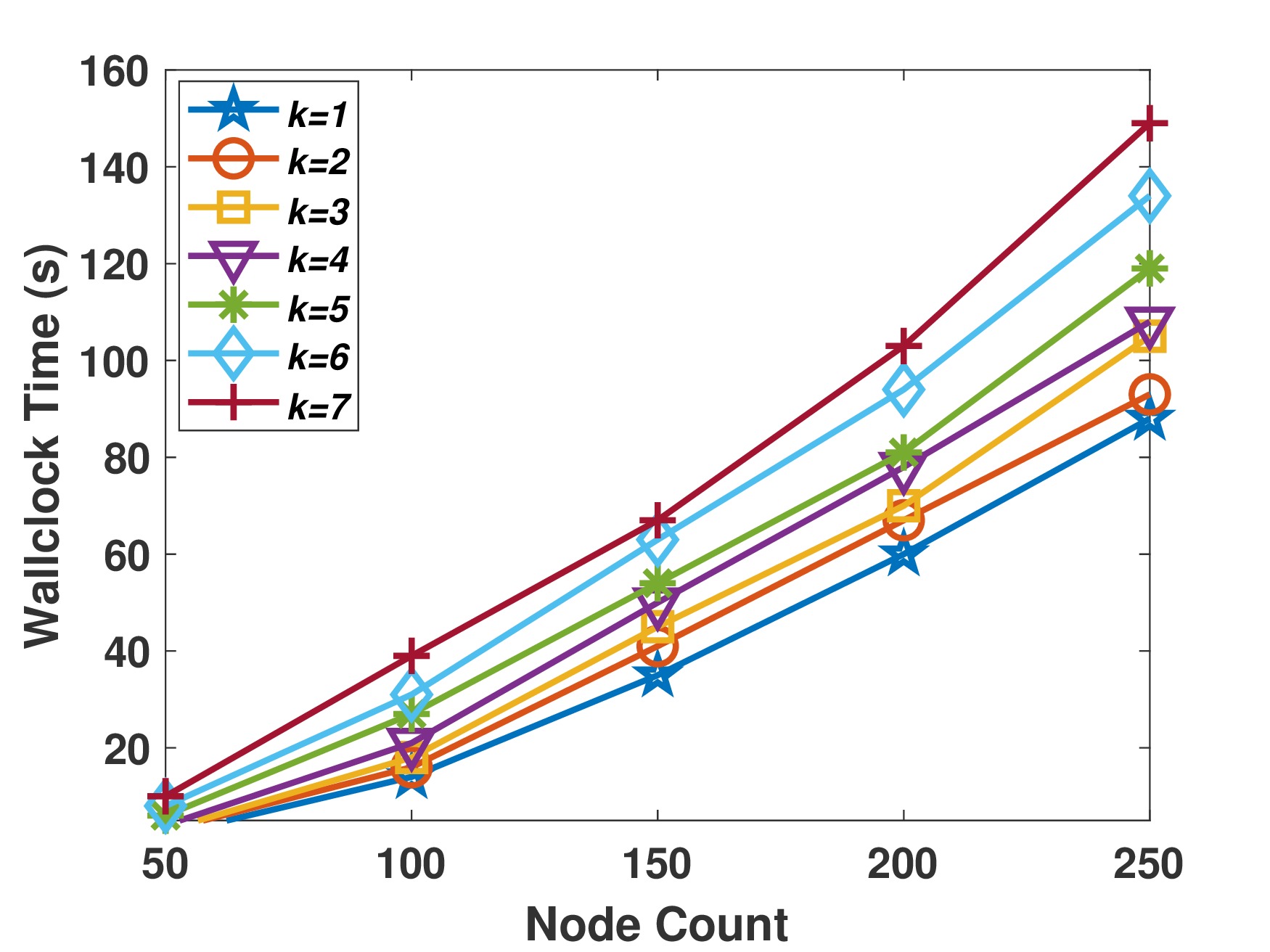}
		\label{time_n_linar}}
	\hfil
	\subfloat[]{\includegraphics[width=0.25\textwidth]{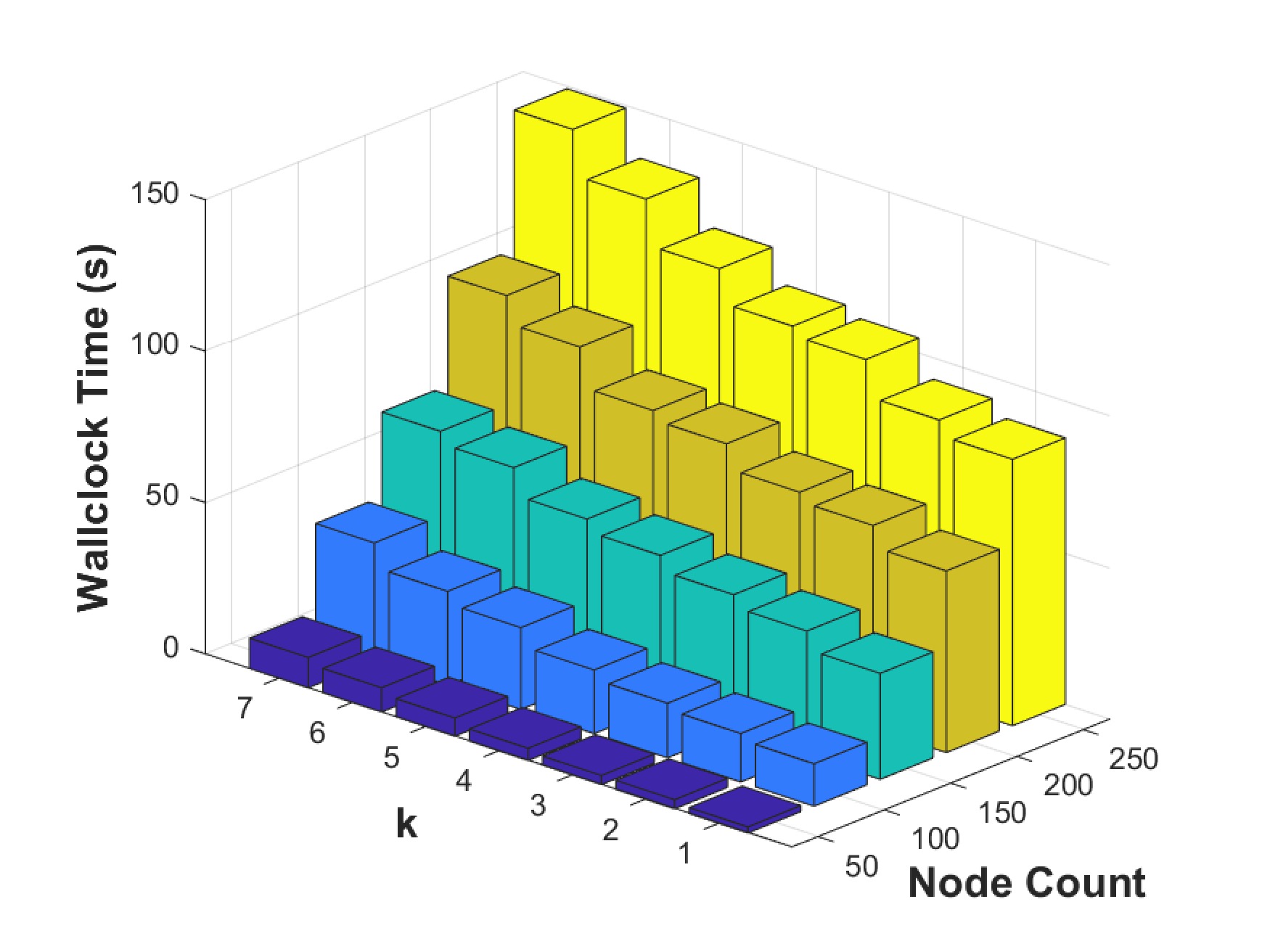}
		\label{time_nk_linar}}
	\hfil
	\caption{Wallclock times of: a) algorithms against node count, b) algorithms against $k$, c) LINAR ($\beta=0$) against node count, d) LINAR ($\beta=0$) against node count and $k$.}
	\label{time}
\end{figure*}

\begin{figure*}[!h]
	\captionsetup[subfloat]{farskip=9pt}
	\centering
	\subfloat[]{\includegraphics[width=0.24\textwidth]{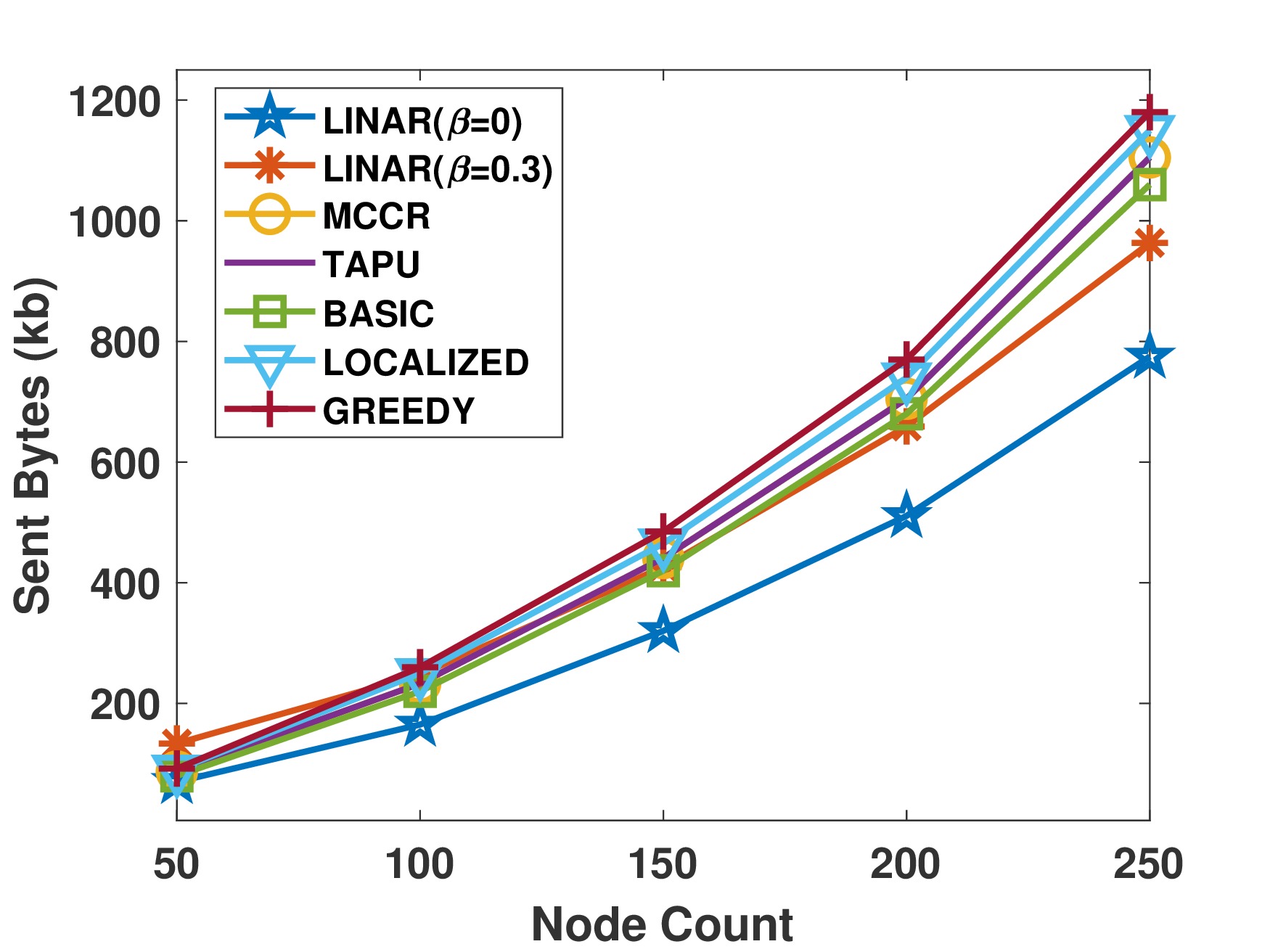}
		\label{sentbytes_n_all}}\hfill
	\subfloat[]{\includegraphics[width=0.24\textwidth]{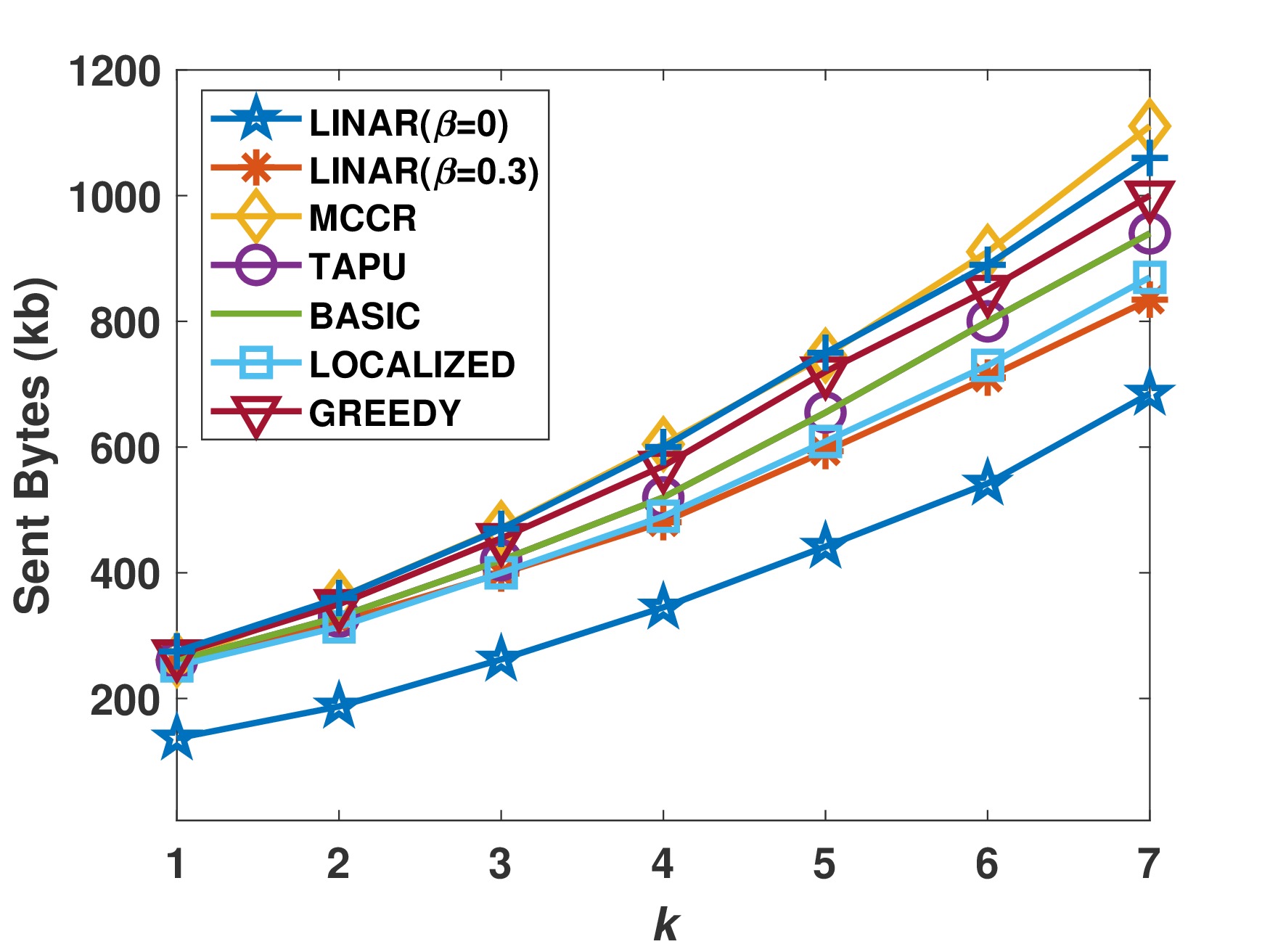}
		\label{sentbytes_k_all}}\hfill
	\subfloat[]{\includegraphics[width=0.24\textwidth]{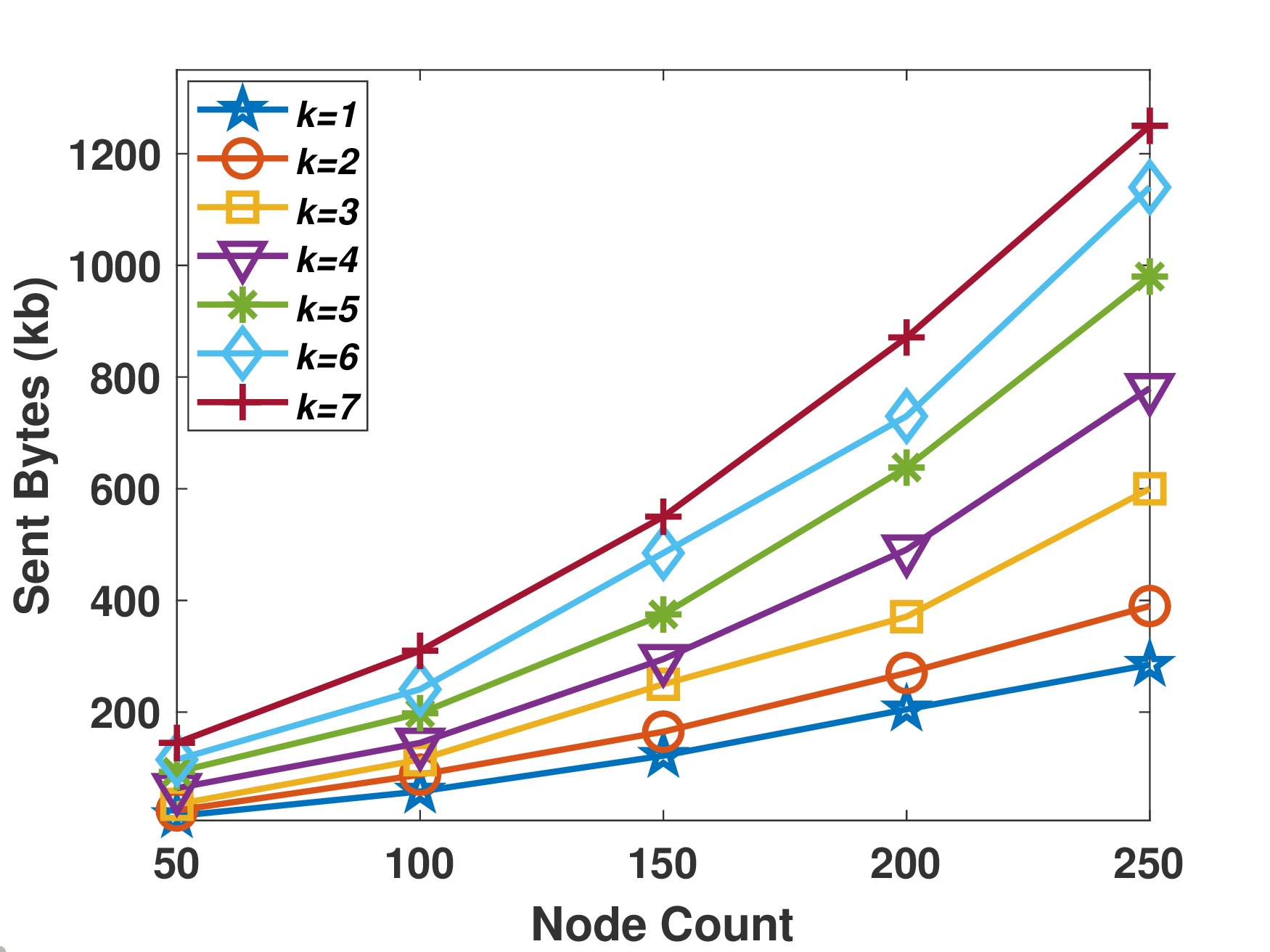}
		\label{sentbytes_n_linar}}\hfill
	\subfloat[]{\includegraphics[width=0.25\textwidth]{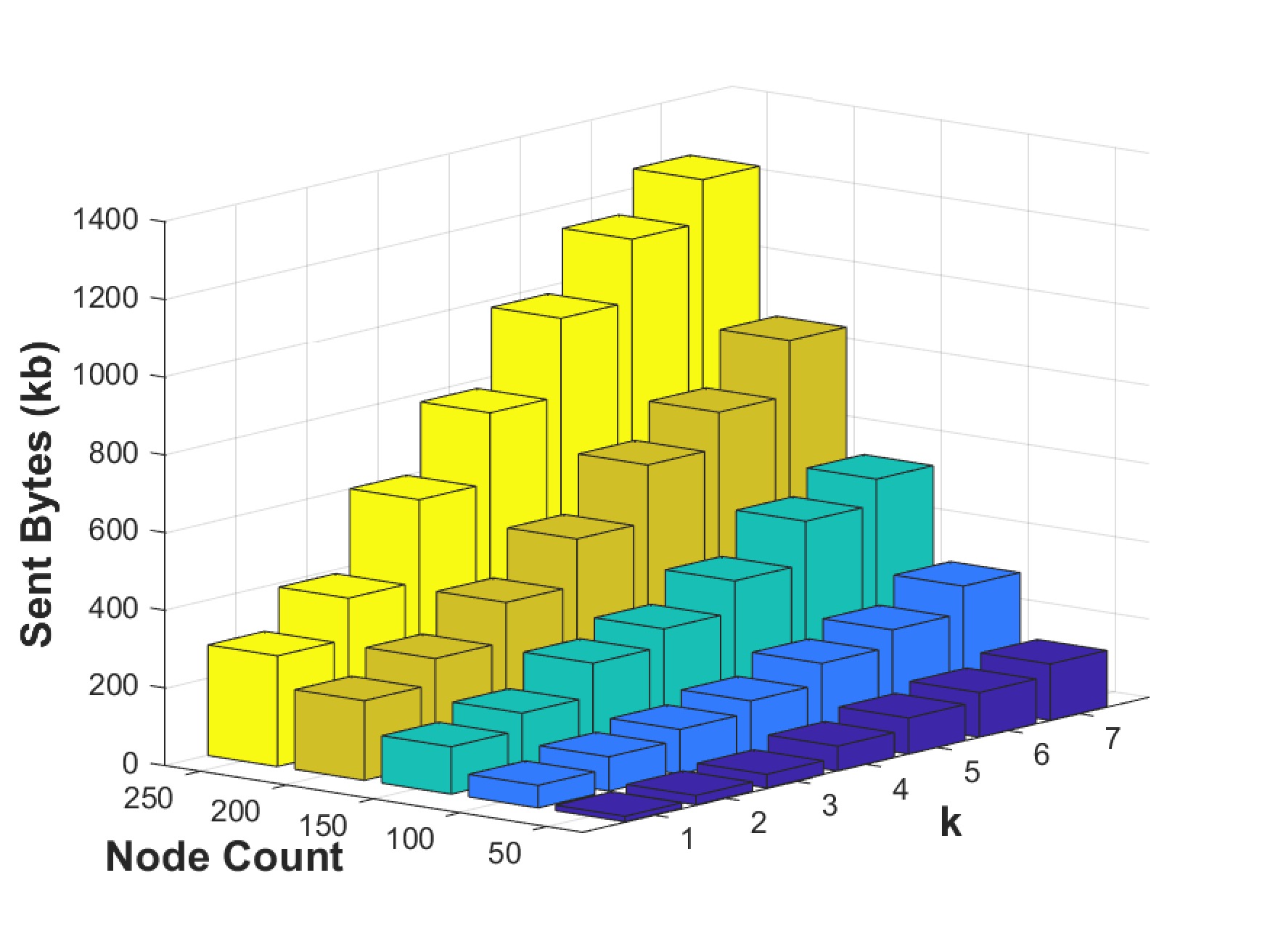}
		\label{sentbytes_nk_linar}}\hfill
	\caption{Average sent bytes of: a) algorithms against node count, b) algorithms against $k$, c) LINAR ($\beta=0$) against node count, d) LINAR ($\beta=0$) against node count and $k$.}
	\label{sentbytes}
\end{figure*}

\begin{figure*}[!h]
\captionsetup[subfloat]{farskip=9pt}
\centering
\subfloat[]{\includegraphics[width=0.24\textwidth]{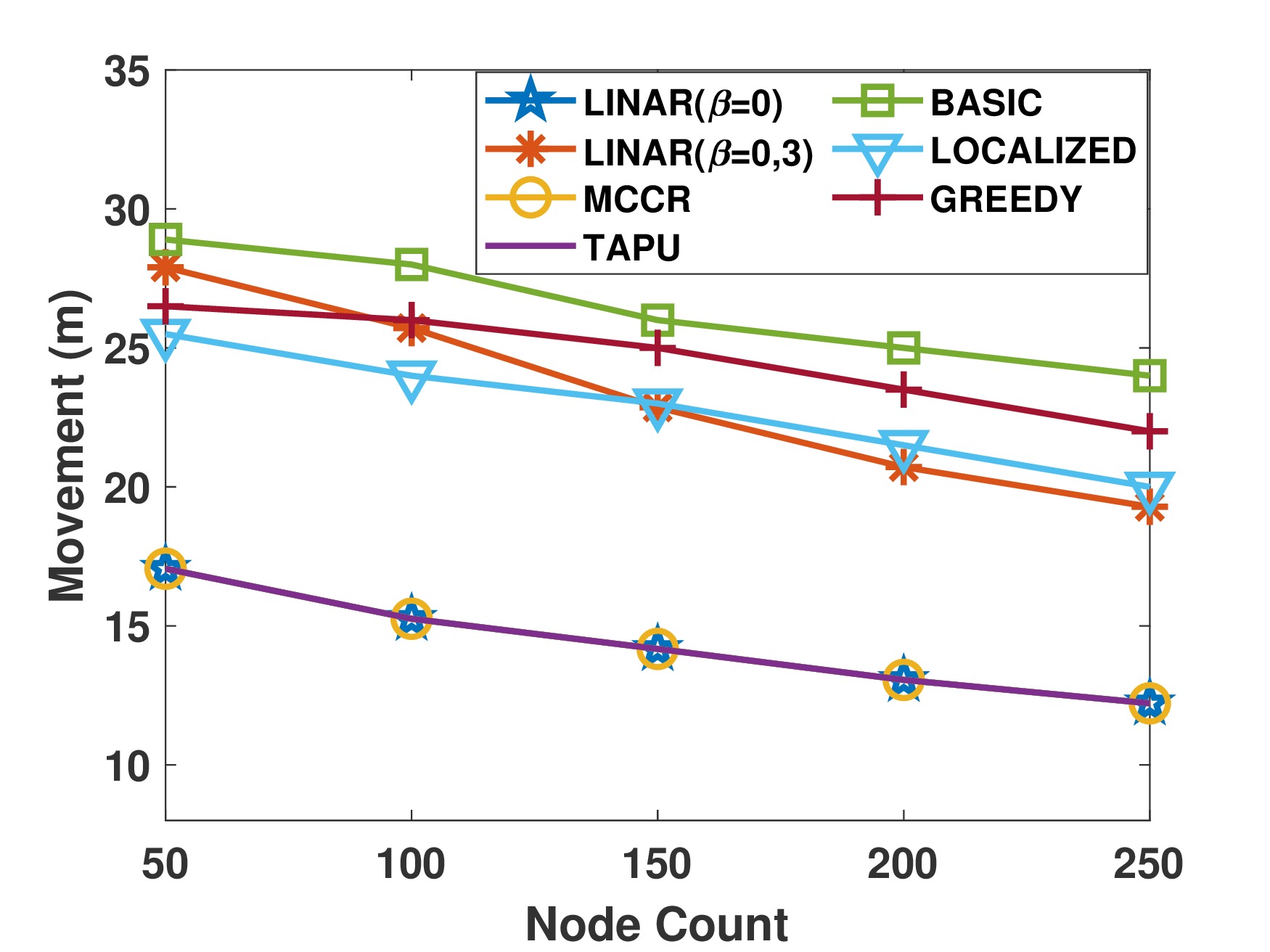}
\label{move_n_all}}
\hfil
\subfloat[]{\includegraphics[width=0.24\textwidth]{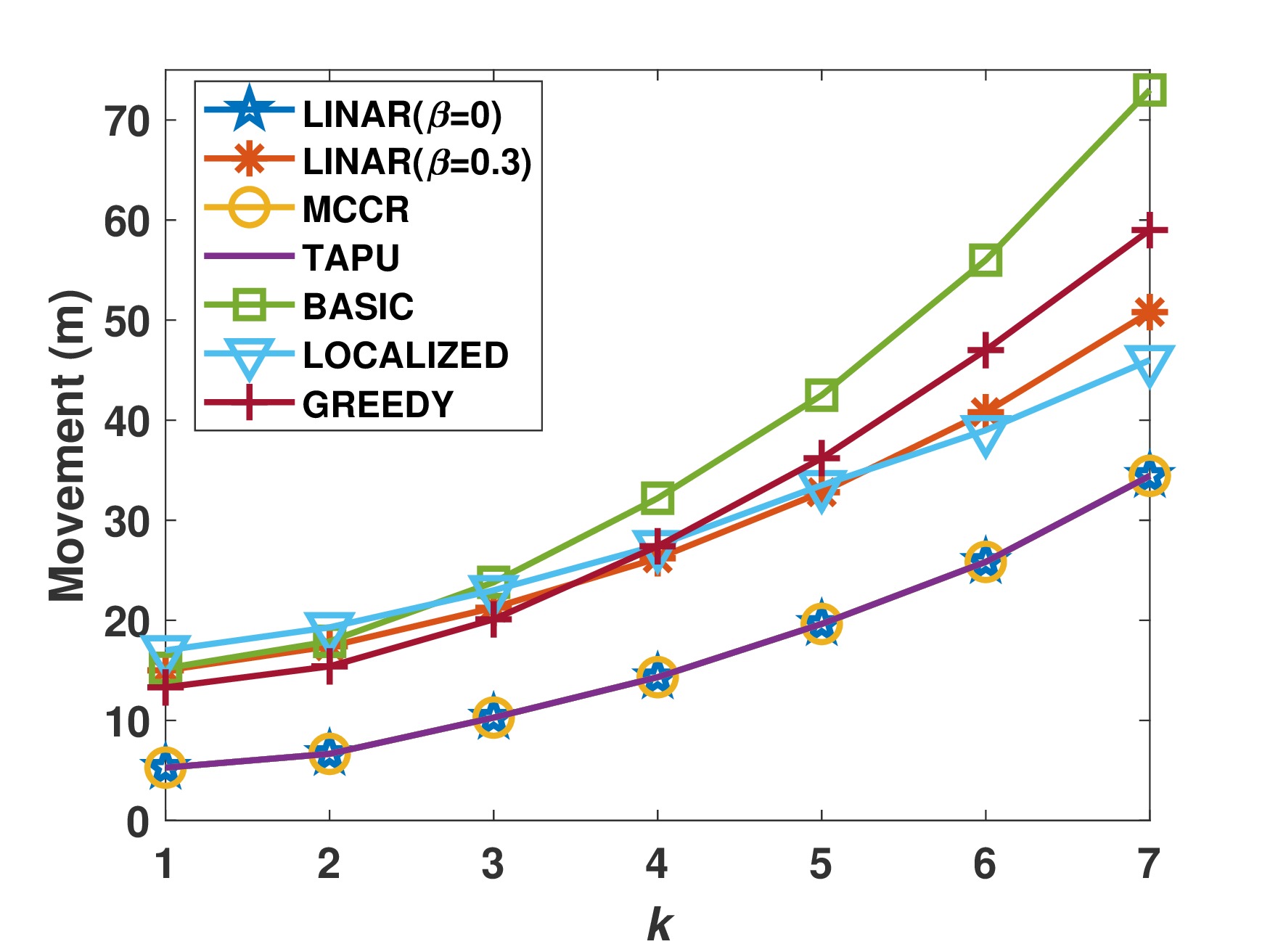}
\label{move_k_all}}
\hfil
\subfloat[]{\includegraphics[width=0.24\textwidth]{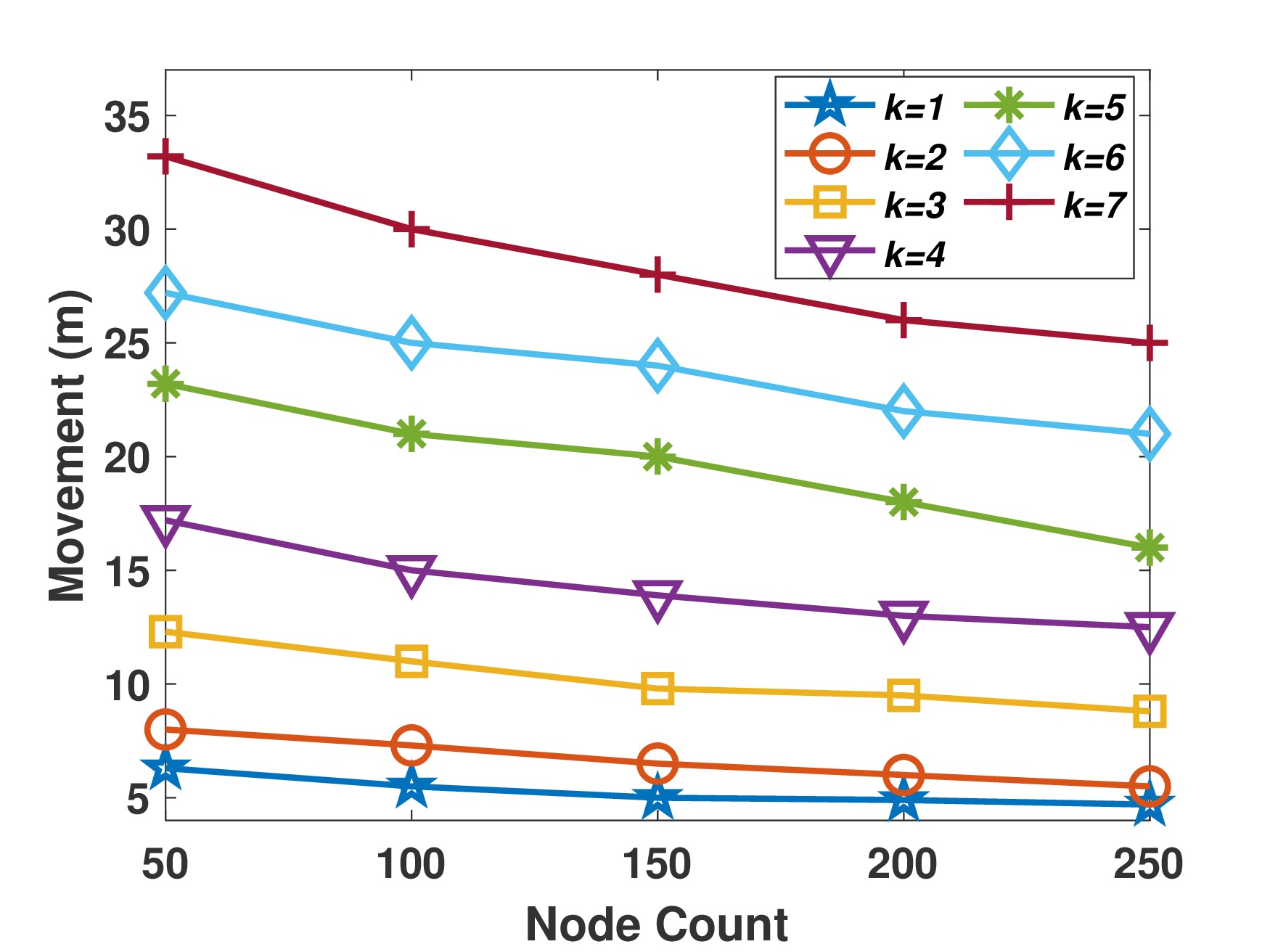}
\label{linar_k}}
\hfil
\subfloat[]{\includegraphics[width=0.25\textwidth]{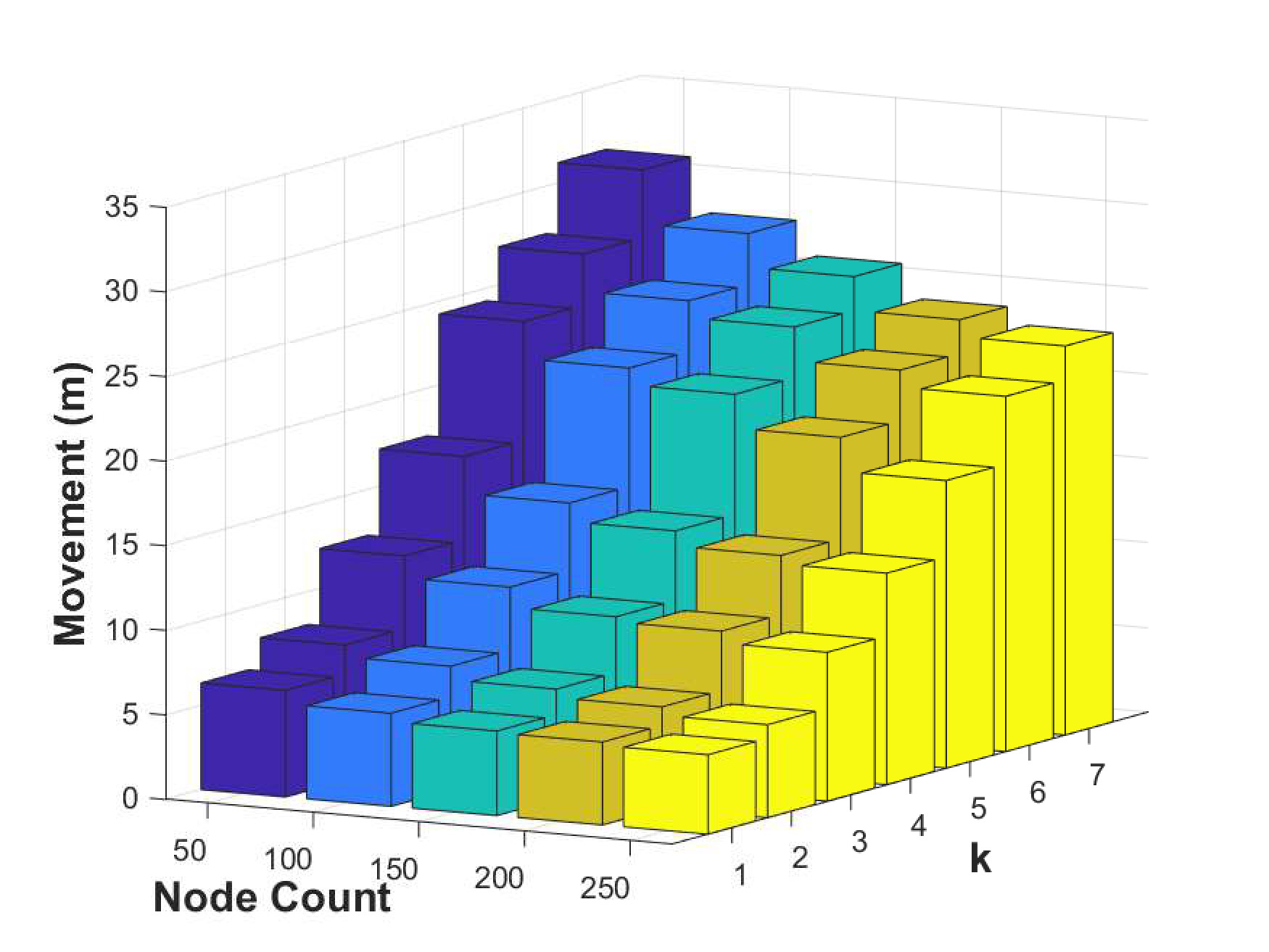}
\label{move_nk_linar}}
\hfil
\caption{Movement costs of: a) algorithms against node count, b) algorithms against $k$, c) LINAR ($\beta=0$) against node count, d) LINAR ($\beta=0$) against node count and $k$.}
\label{move}
\end{figure*}

To measure the performances of the algorithms in large-scale networks, we simulated the algorithms using Java on geometric bidirectional weighted graphs in a field of 1000 m $\times$ 1000 m area. In the generated random networks, the weight of each edge is the movement cost between the associated nodes. The transmission range of each node is set to 20 m and five classes of topologies with 50, 100, 150, 200, and 250 nodes  are generated. For each class, we generated random topologies with different $k$ values from 1 to 7. For each specific node count and $k$ value, we created 10 random topologies.

After starting the algorithms, we stopped 20\% of the nodes randomly and measured the total sent bytes, movement costs, wallclock times, and coverage area before and after the restoration for $\beta=0$, 0.3, 0.6, and 0.9. We saved each topology, before and after the restoration, in different text files and used the Shapely and Matplotlib libraries of Python to calculate the coverage area. Since the other $k$-connectivity restoration algorithms do not consider the coverage constraint, for the sake of fairness, we first compared the performances of LINAR with $\beta=0$ (no coverage constraint) and $\beta=0.3$ with the existing algorithms and then evaluated the performance of LINAR for different $\beta$ values, in detail.

MCCR has the highest wallclock time taking approximately 214 s in the networks with 250 nodes (Fig.~\ref{time_n_all}). After MCCR, LINAR with $\beta=0.3$ and $\beta=0$ exhibits the next highest wallclock times. 
Note that none of the algorithms have any coverage constraint except LINAR with $\beta=0.3$. In the networks with 50 nodes, the wallclock times of LINAR with $\beta=0$ are less than 10 s and increase up to 131 s in the networks with 250 nodes, which are approximately 38.7\% lower than MCCR and 19.1\% higher than the \textit{Greedy} algorithm. 
Increasing the $k$ value increases the wallclock time of MCCR higher than the other algorithms (Fig.~\ref{time_k_all}).
The \textit{Basic} algorithm is the fastest algorithm taking less than 45 s for all $k$ values. The wallclock time of LINAR with $\beta=0$ is up to 47.3\% lower than that of MCCR and grows almost linearly with increasing $k$ values. In the worst case, for $k$=7, the wallclock time of LINAR with $\beta=0$ is approximately 43.8\% higher than that of the fastest implemented algorithm. LINAR runs faster for small $k$ values, but the gaps between different $k$ values are limited (Fig.~\ref{time_n_linar}). In networks with 250 nodes, LINAR with $\beta=0$ takes approximately 89 s for $k$=1, whereas this value for $k$=7 is approximately 150 s, which indicates that LINAR exhibits a stable behavior in all topologies. The wallclock time of LINAR moderately increases when we increase the node count, whereas wallclock times grow slowly with increasing $k$ values (Fig.~\ref{time_nk_linar}).

Sending neighborhood information of all nodes to a single node increases the total sent bytes of the central approaches up to 34.3\% more than the sent bytes of LINAR with $\beta=0$ (Fig.~\ref{sentbytes_n_all}). In the \textit{Basic} algorithm, the sink node sends a 1-hop message after each failure, hence, its average sent bytes are lower than those of the other central algorithms. In the \textit{Localized}, \textit{Greedy}, MCCR, and TAPU algorithms, the sink can send more than one multi-hop messages after each node failure. The \textit{Basic}, MCCR, and TAPU algorithms send less bytes than the \textit{Greedy} and \textit{Localized} algorithms, but their sent bytes are at least 19.8\% higher than LINAR for all $k$ values (Fig.~\ref{sentbytes_k_all}). LINAR with $\beta=0.3$ always sends more bytes than LINAR with $\beta=0$ because LINAR with $\beta=0.3$ is the only algorithm that has a coverage constraint, unlike all the others. The sent bytes of LINAR with $\beta=0$ for small $k$ values are lower than the sent bytes for higher $k$ values because for smaller $k$ values more nodes can find their status with the proposed local relations (Fig.~\ref{sentbytes_n_linar}). In networks with 250 nodes, LINAR with $\beta=0$ sends approximately 286 kb on average for $k$=1, whereas this value is approximately 1252 kb for $k$=7. In networks with 50 nodes, LINAR sends less bytes and the effects of $k$ values on the sent bytes are not significant, yet, increasing the number of nodes increases the amount of sent bytes (Fig.~\ref{sentbytes_nk_linar}).

Fig. \ref{move_n_all} shows the average movement costs of all algorithms against the node count. In networks with 50 nodes, MCCR, TAPU, and LINAR (with $\beta=0$) algorithms generate approximately 17 m movement on average whereas for the same topologies the average movements of the \textit{Localized}, \textit{Greedy}, and \textit{Basic} algorithms are higher than 26 m (52.9\% higher than LINAR). By increasing the node count, the probability of a \textit{Joint} node failure and also the distance between the nodes decrease which leads to shorter movements. In networks with 250 nodes, the average movement costs of MCCR, TAPU, and LINAR (with $\beta=0$) are less than 12 m, whereas the average movement costs of the other three algorithm are higher than 20 m. Fig. \ref{move_k_all} compares the average movement costs of the algorithms against the $k$ values. The movement costs of MCCR, TAPU, and LINAR (with $\beta=0$) are up to 34\% lower than those of the other algorithms. LINAR with $\beta=0.3$ generates more movements than LINAR with $\beta=0$ because only this algorithm considers the coverage constraint. The average movement cost of LINAR for different $k$ values is presented in Fig. \ref{linar_k}. In general, for lower $k$ values the networks have fewer \textit{Joint} nodes than networks with higher $k$ values. Hence, the probability of a \textit{Joint} node failure for smaller $k$ values is lower. 
Fig.~\ref{linar_k} shows that for $k$=1, the average movement cost is approximately 7 m in networks with 50 nodes. This cost is reduced to approximately 5 m in networks with 250 nodes. For $k$=7, the average movement cost of LINAR  varies between 33 m and 25 m.  
Fig.~\ref{move_nk_linar} reveals that the $k$ value is more dominant for movement cost than for the node count. Increasing the node count slowly reduces the movement cost whereas increasing the $k$ value rapidly increases the 
movement 
cost. 

\begin{figure}[!htbp]
	\captionsetup[subfloat]{farskip=9pt}
	\centering
	\subfloat[]{\includegraphics[width=0.18\textwidth]{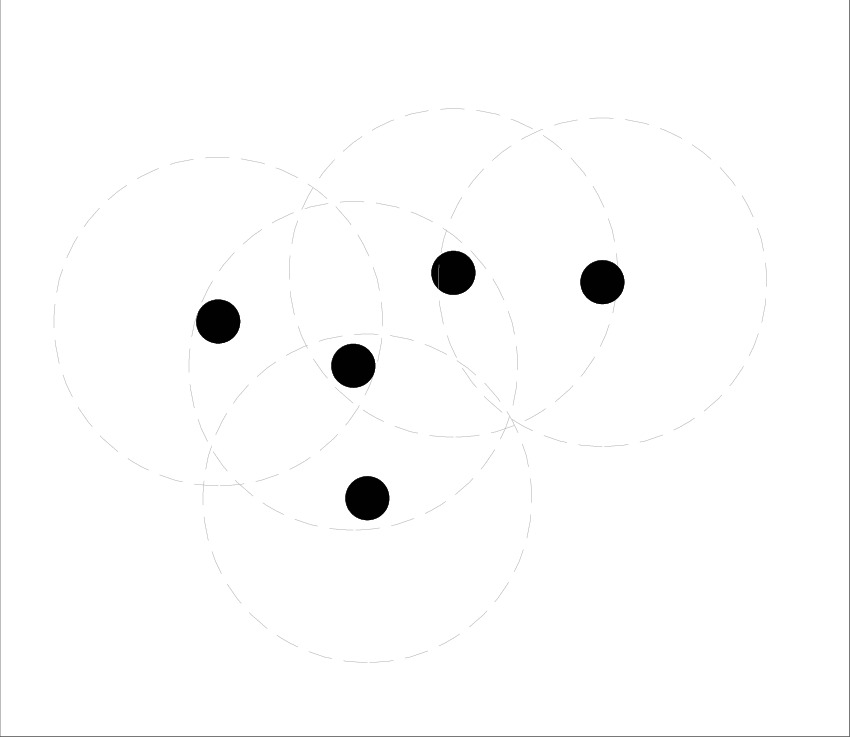}
		\label{pcov1}}
	\hfil
	\subfloat[]{\includegraphics[width=0.18\textwidth]{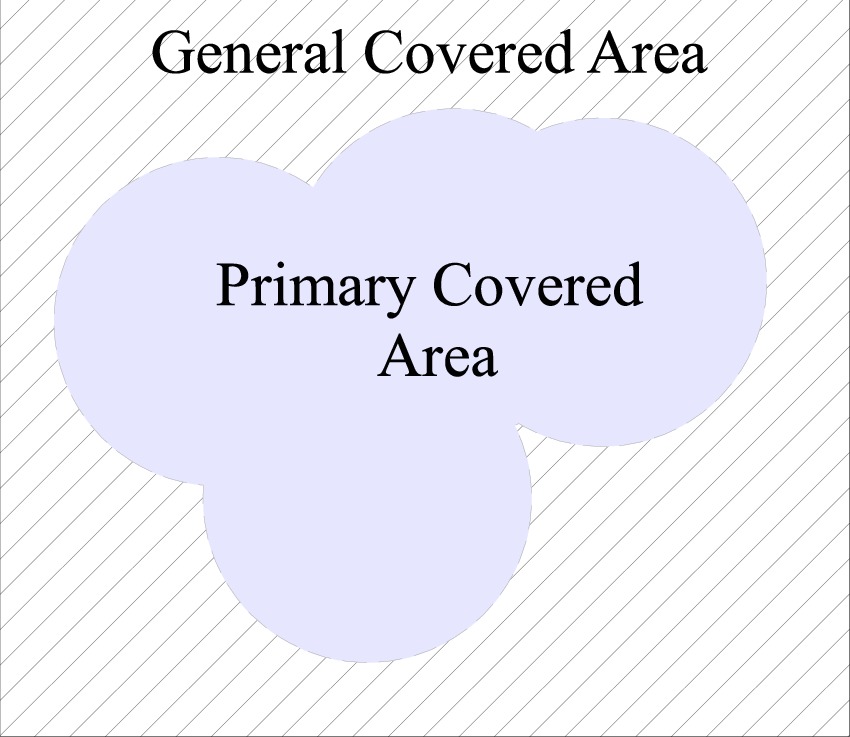}
		\label{pcov2}}	
	\caption{Primary versus general coverage. }
	\label{pcov}
\end{figure}

We measured the coverage loss of the network before and after the restoration for two cases (i.e., the primary coverage loss and the general coverage loss). Fig.~\ref{pcov1} shows a sample MSN and Fig.~\ref{pcov2} shows the difference between the primary and general coverage. We assumed that the sensing and communications ranges are equal~\cite{huang2005coverage}. We run LINAR with $\beta=0$, $\beta=0.3$, $\beta=0.6$, and $\beta=0.9$, however, because of the negligible difference between the results of $\beta=0.6$ and $\beta=0.9$, we present the results up to $\beta= 0.6$.

\begin{figure*}[!htbp]
	\captionsetup[subfloat]{farskip=9pt}
	\centering
	\subfloat[]{\includegraphics[width=0.25\textwidth]{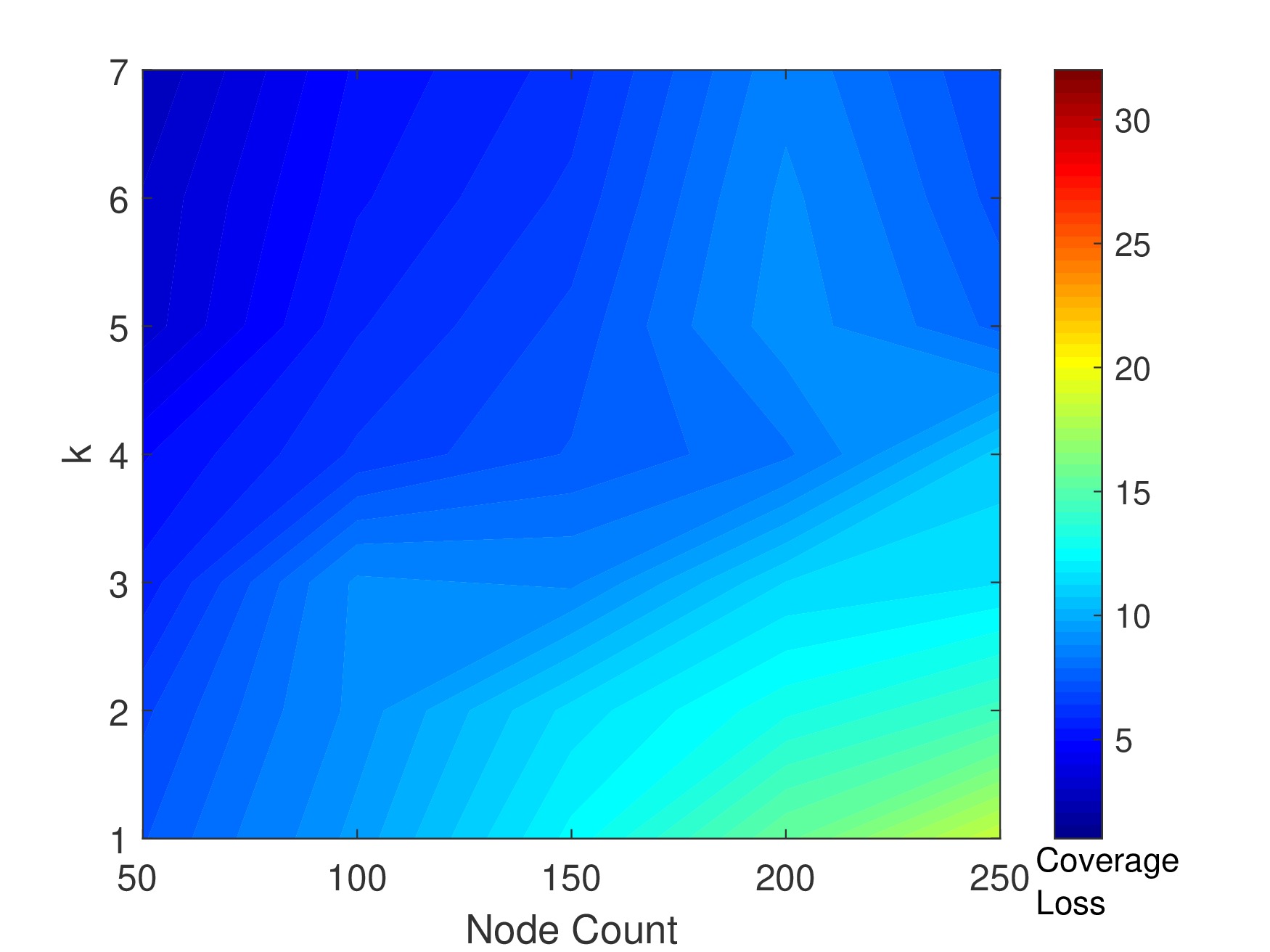}
		\label{p_cont}}
	\hfil
	\subfloat[]{\includegraphics[width=0.25\textwidth]{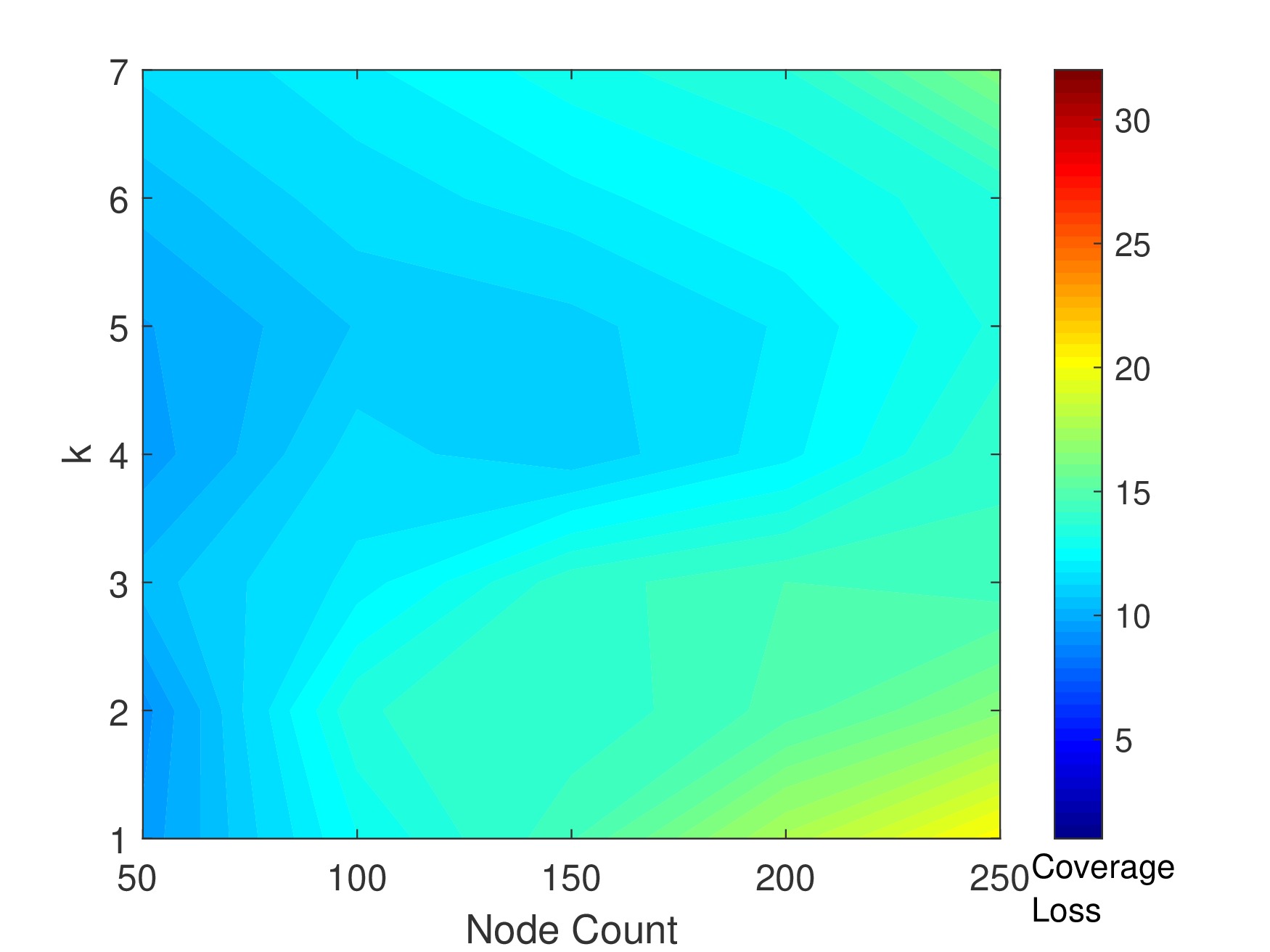}
	\label{p_b6_cont}}	
	\hfil
	\subfloat[]{\includegraphics[width=0.25\textwidth]{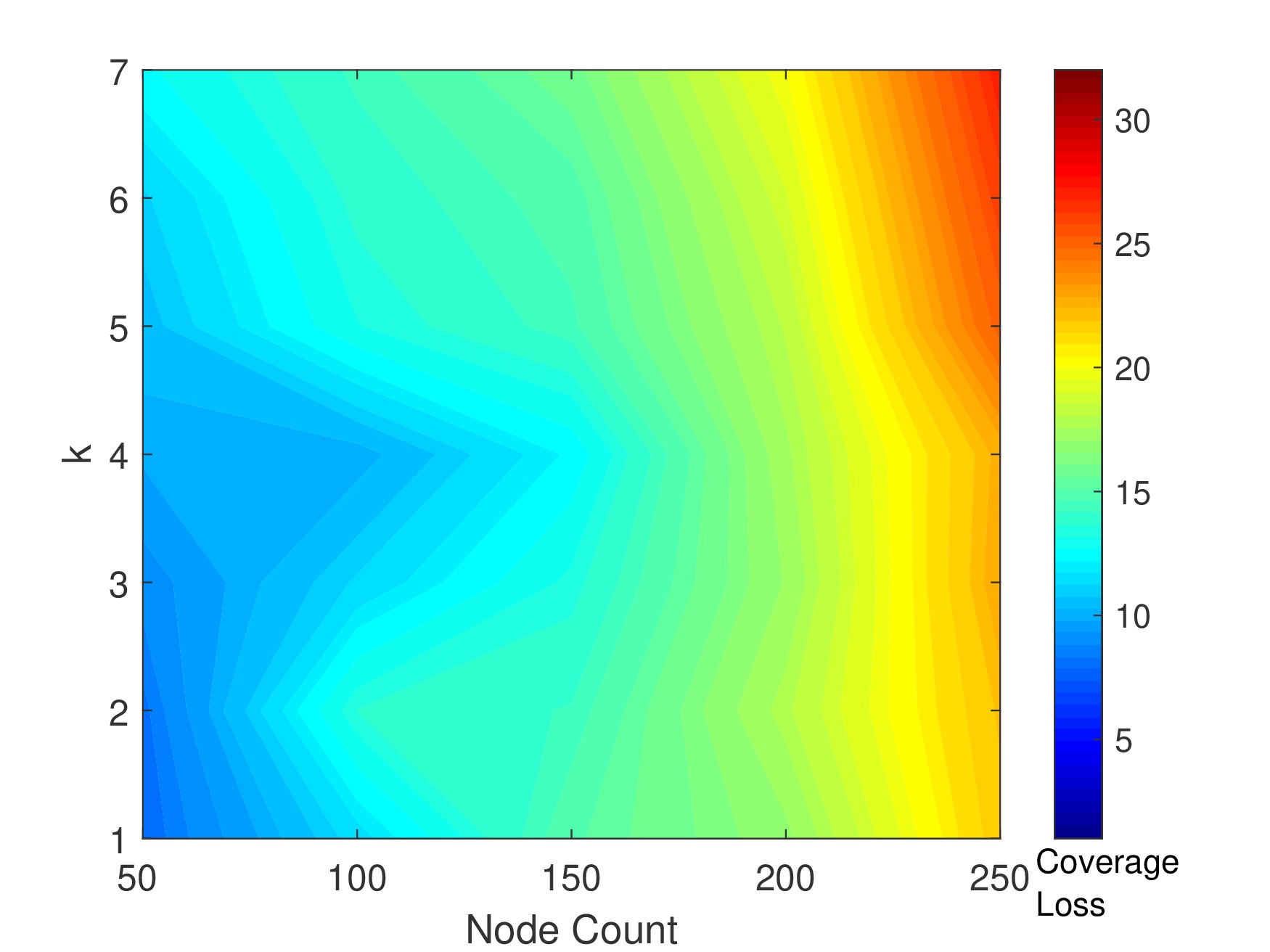}
		\label{p_b3_cont}}
	\hfil
\subfloat[]{\includegraphics[width=0.25\textwidth]{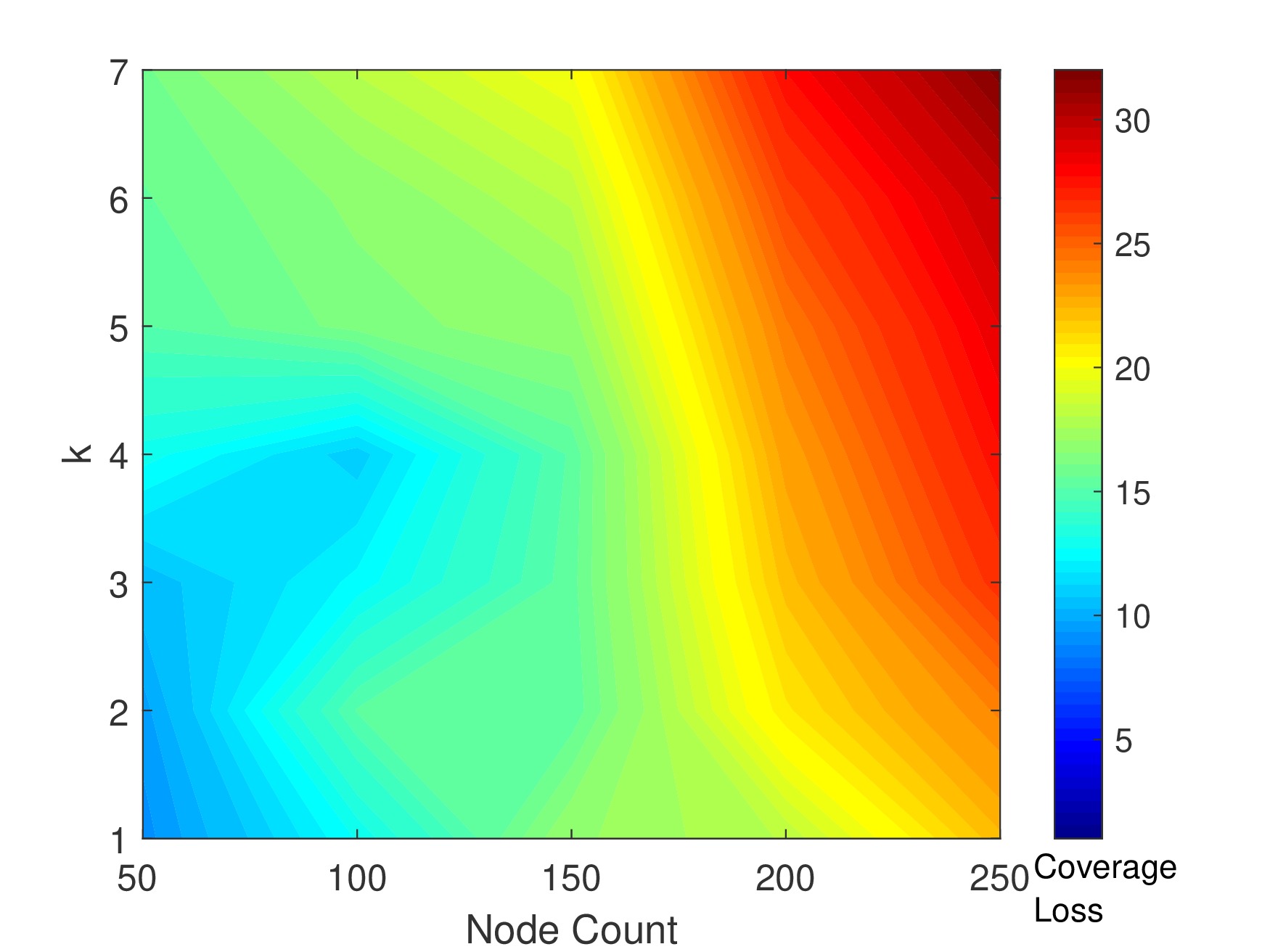}
	\label{p_b0_cont}}
	\hfil
	\caption{Primary coverage loss after node failures: a) without restoration, b) with $\beta=0.6$, c) with $\beta=0.3$, d) with $\beta=0$.}
	\label{primarycov}
\end{figure*}

Fig. \ref{primarycov} shows the primary coverage loss percentage after node failures. Fig.~\ref{p_cont} shows that we lose 3\% (for $n=50$ and $k=7$) up to 16\% (for $n=250$ and $k=1$) of the initial covered area  without performing any restoration after the failures of 20\% of the nodes. After $k$-connectivity restoration, we lose approximately 8\% (for $n=50$) up to 19\% (for $n=250$ and $k=1$) of the covered area with $\beta=0.6$ (Fig.~\ref{p_b6_cont}) and approximately 9\% (for $n=50$ and $k=1$) up to 28\% (for $n=250$ and $k>5$) with $\beta=0.3$ (Fig.~\ref{p_b3_cont}). For $\beta=0$ (Fig.~\ref{p_b0_cont}) the coverage loss varies between 10\% (for $n=50$ and $k=1$) and 33\% (for $n=250$ and $k>5$).

\begin{figure*}[!htbp]
	\captionsetup[subfloat]{farskip=9pt}
	\centering
	\subfloat[]{\includegraphics[width=0.25\textwidth]{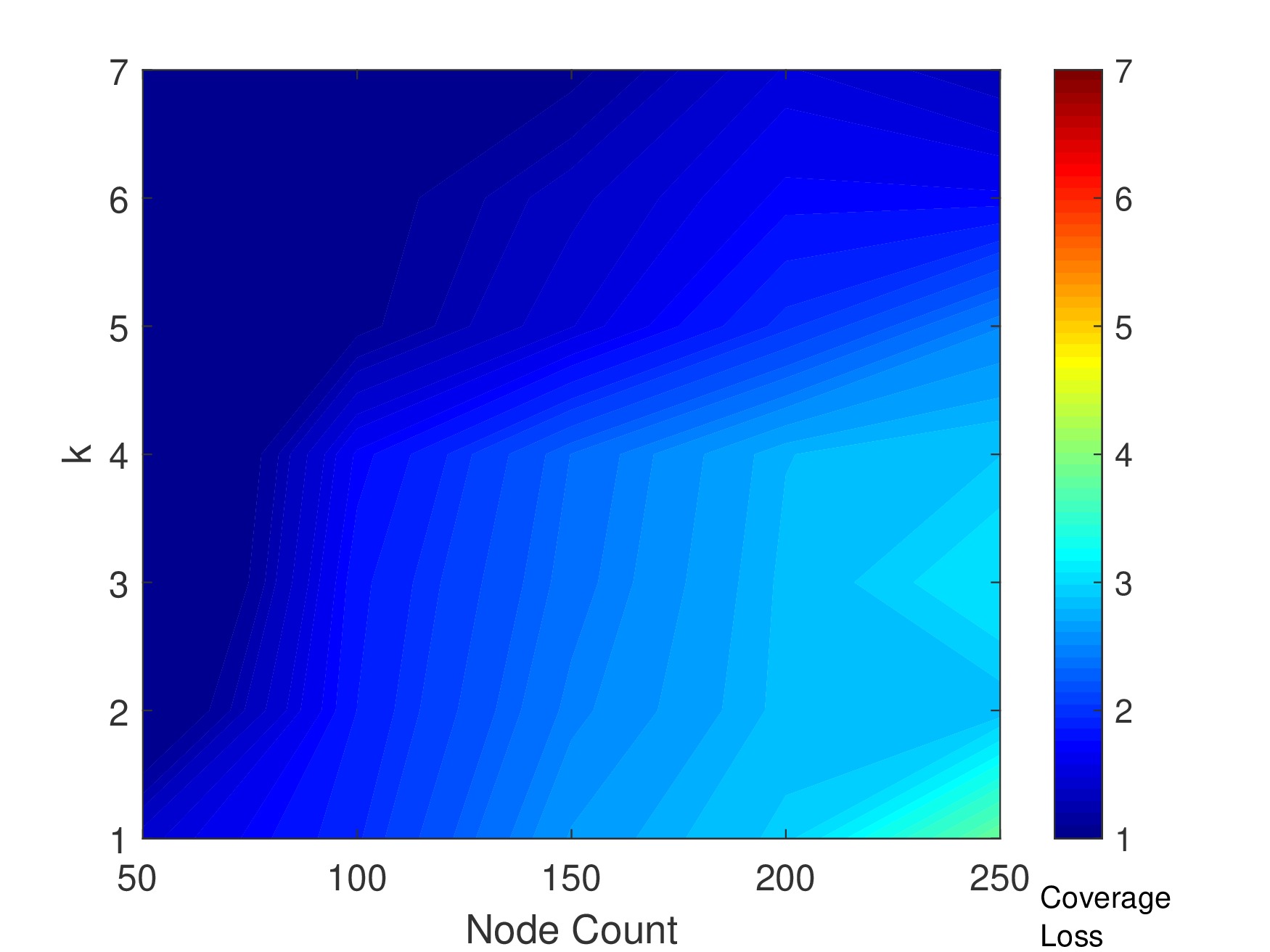}
		\label{gen_cont}}
	\hfil
	\subfloat[]{\includegraphics[width=0.25\textwidth]{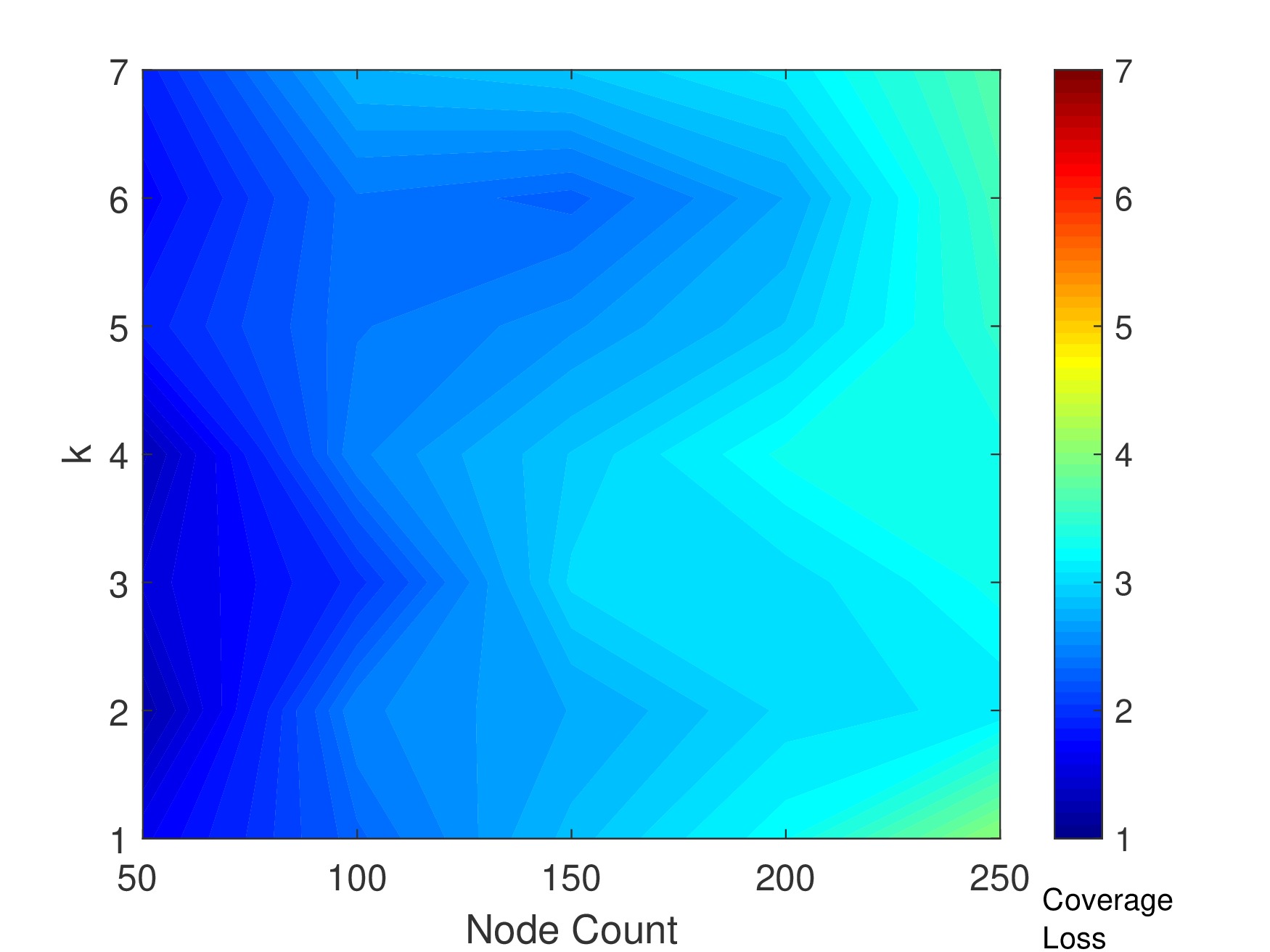}
		\label{gen_b6_cont}}
	\hfil
	\subfloat[]{\includegraphics[width=0.25\textwidth]{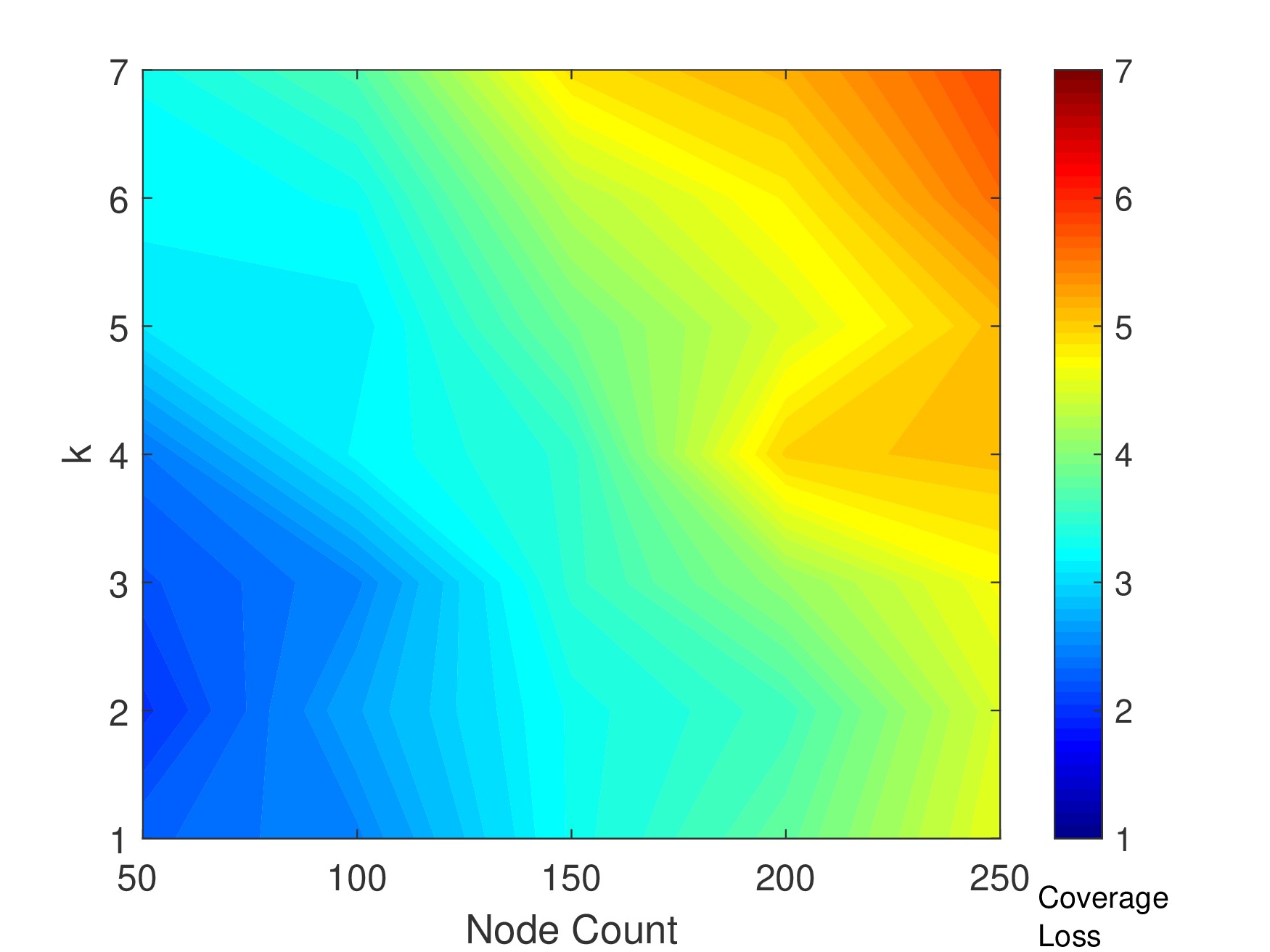}
		\label{gen_b3_cont}}
	\hfil
	\subfloat[]{\includegraphics[width=0.25\textwidth]{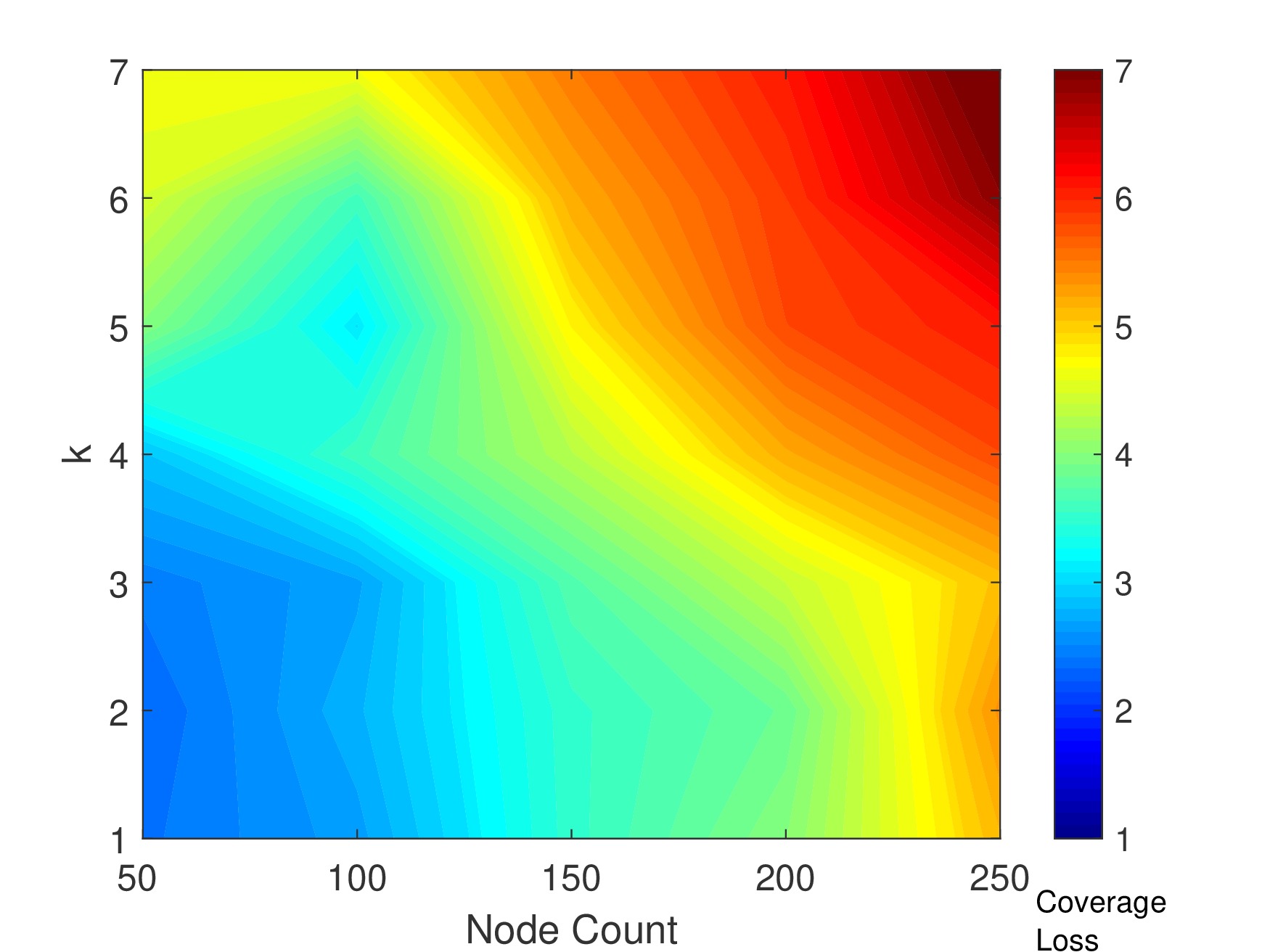}
		\label{gen_b0_cont}}
	\hfil
	\caption{General coverage loss after node failures: a) without restoration, b) with $\beta=0.6$, c) with $\beta=0.3$, d) with $\beta=0$.}
	\label{gen}
\end{figure*}

Fig.~\ref{gen} shows the general coverage loss after node failures. Fig.~\ref{gen_cont} shows that we lose 1\% (for $n=50$ and $k>2$) to 4\% (for $n=250$ and $k=1$) of the covered area without restoration after failures of 20\% of nodes. After $k$-connectivity restoration, we lose approximately 3\% (for $n=50$) to 5\% (for $n=250$ and $k=1$) of covered area with $\beta=0.6$ (Fig.~\ref{gen_b6_cont}) and about 3\% (for $n=50$ and $k<3$) up to 6\% (for $n=250$ and $k>4$) with $\beta=0.3$ (Fig.~\ref{gen_b3_cont}). For $\beta=0$ (Fig.~\ref{gen_b0_cont}), the coverage loss percentage varies between 3\% (for $n=50$ and $k=1$) and 8\% (for $n=250$ and $k>5$).



\begin{figure*}[!htbp]
	\captionsetup[subfloat]{farskip=9pt}
	\centering
	\subfloat[]{\includegraphics[width=0.25\textwidth]{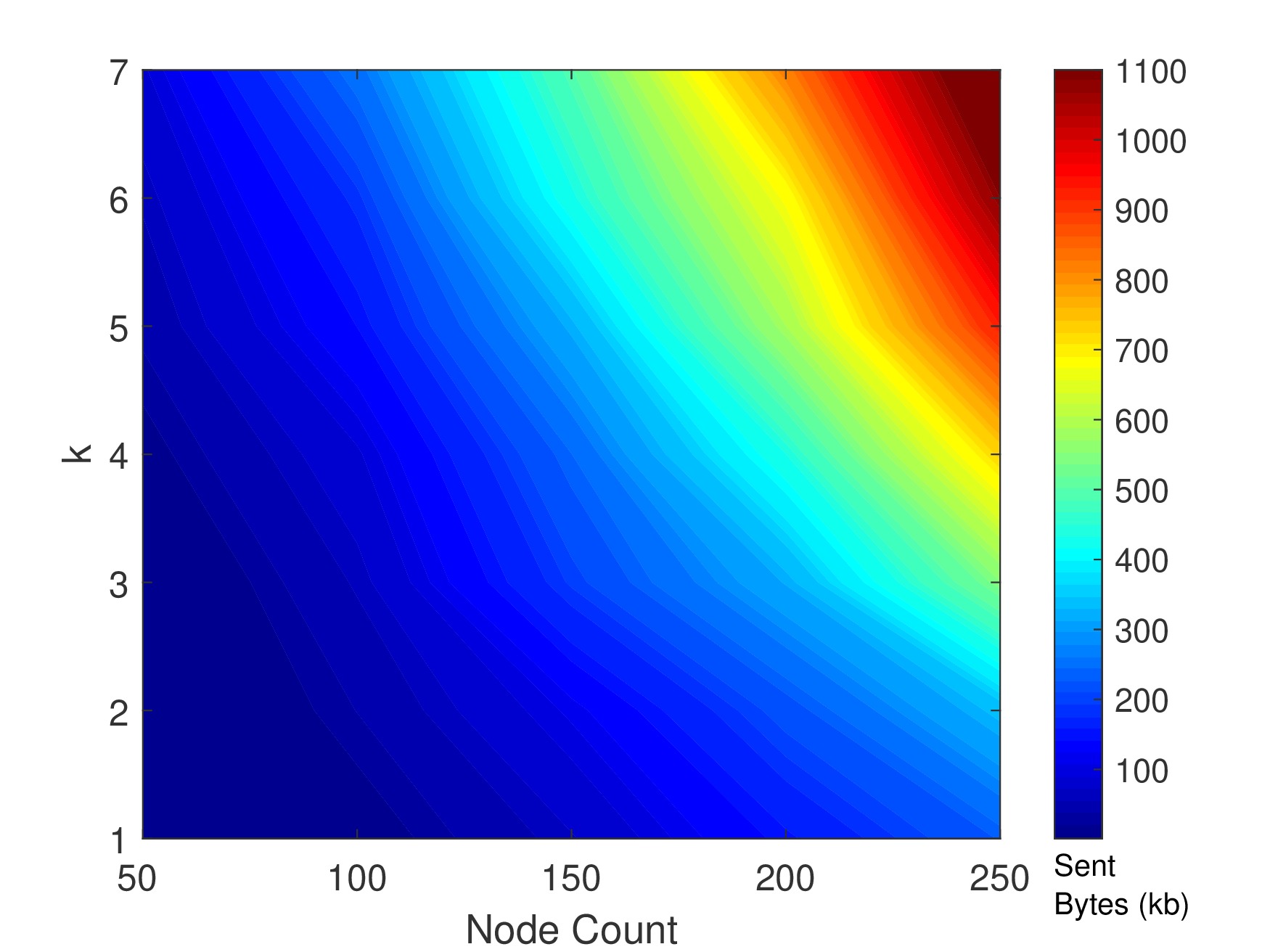}
		\label{sentb0}}
	\hfil
	\subfloat[]{\includegraphics[width=0.25\textwidth]{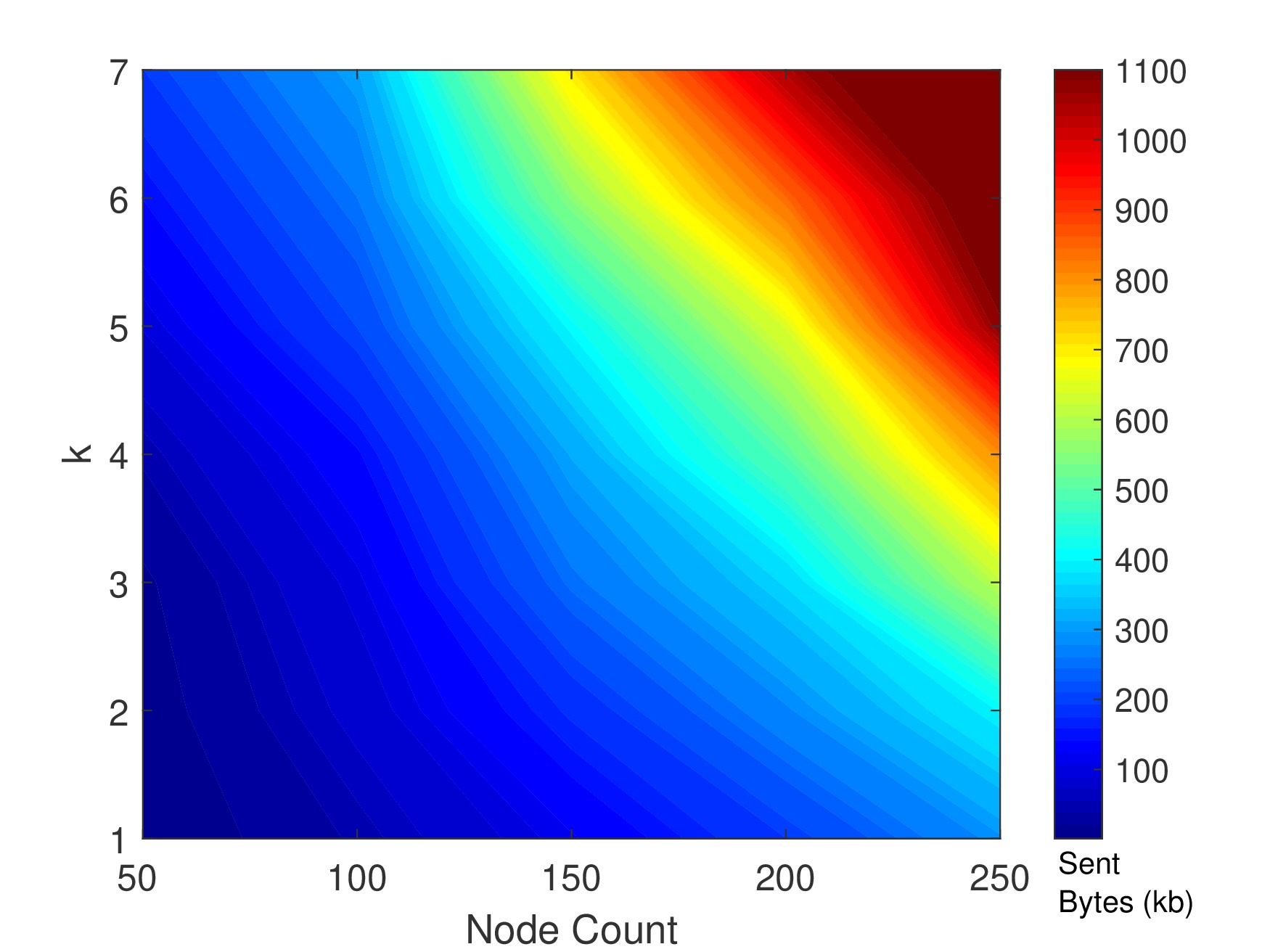}
		\label{sentb6}}
	\hfil
	\subfloat[]{\includegraphics[width=0.25\textwidth]{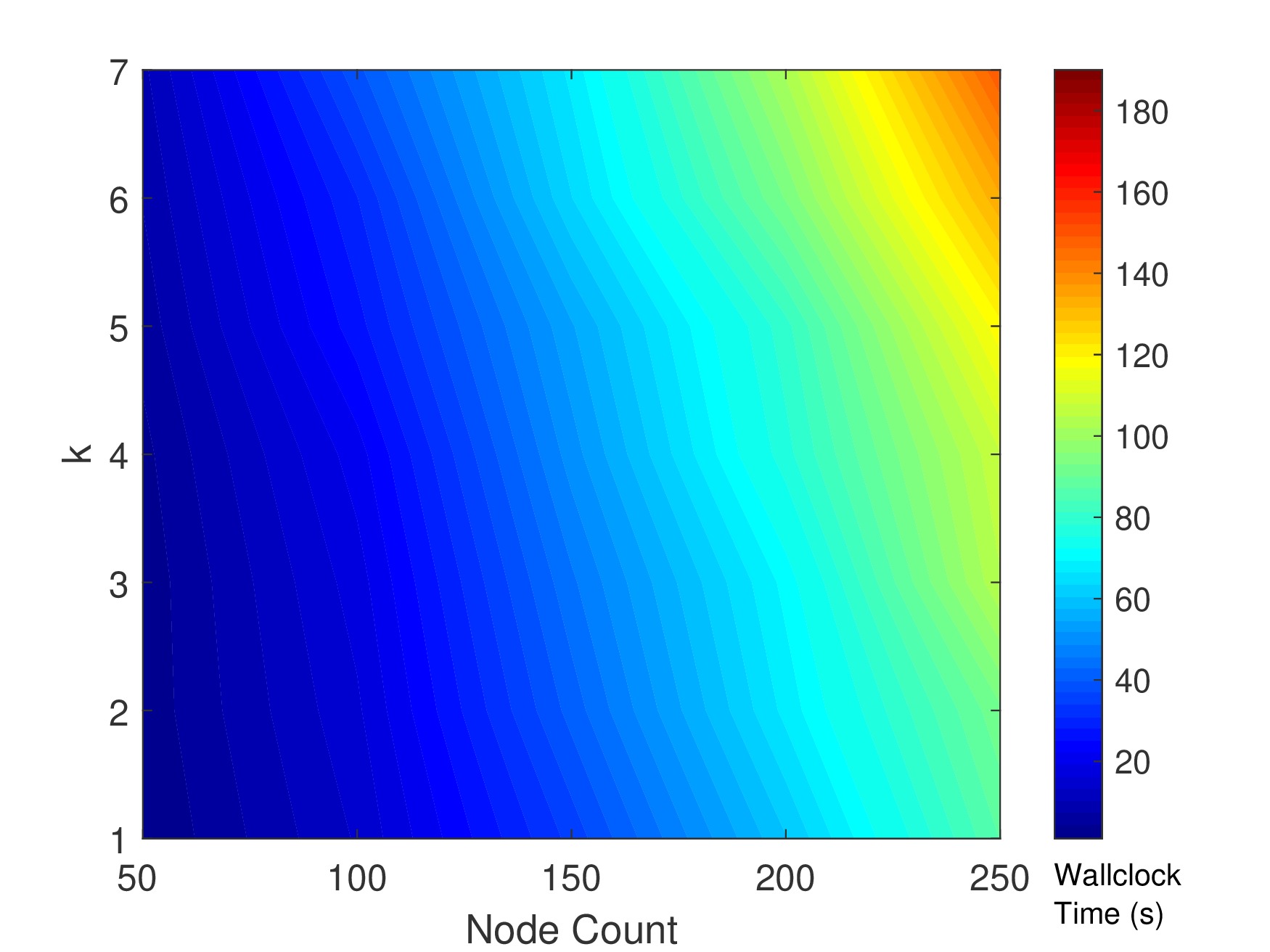}
		\label{timeb0}}
	\hfil
	\subfloat[]{\includegraphics[width=0.25\textwidth]{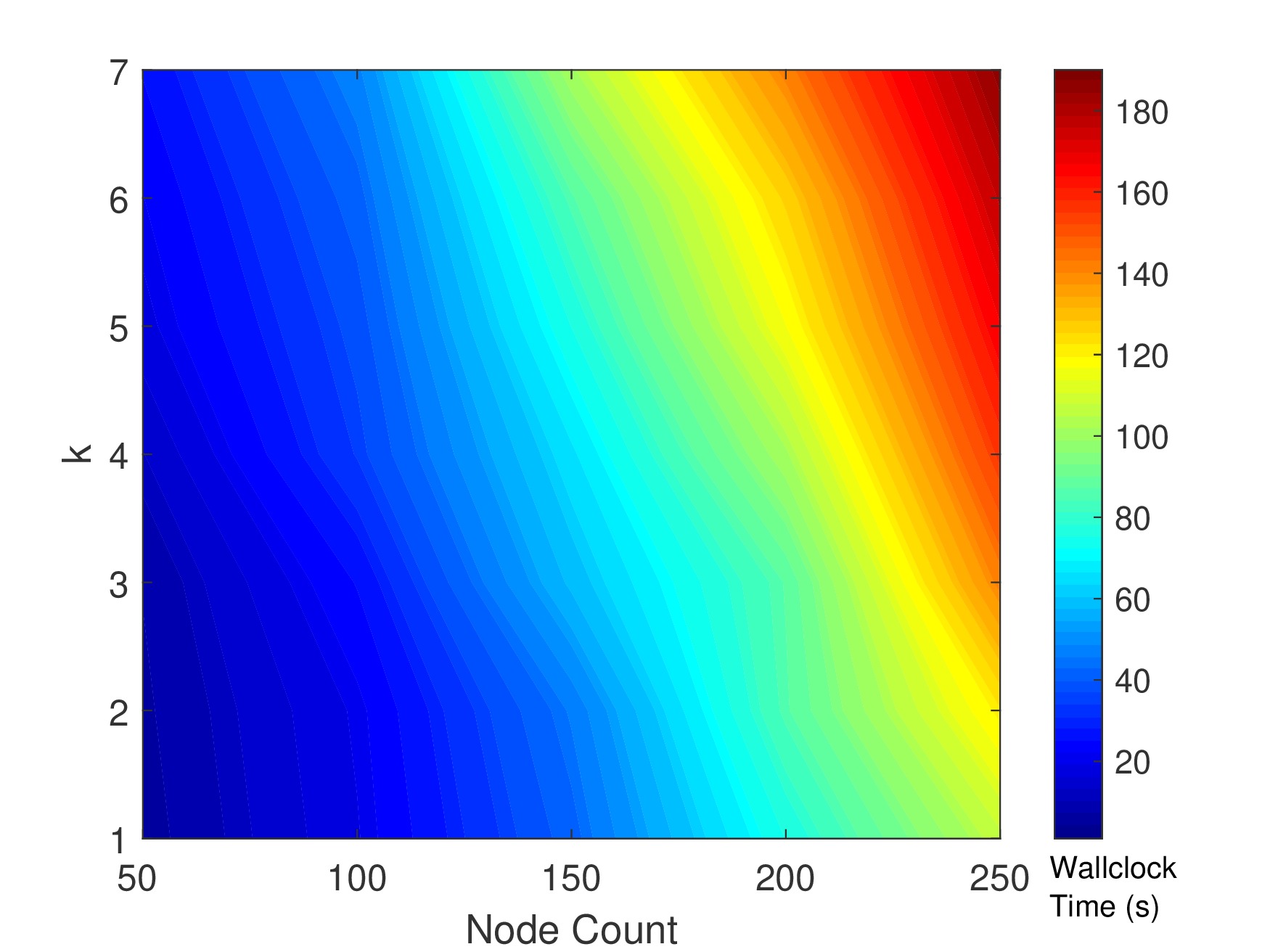}
		\label{timeb6}}
	\hfil
		\caption{Sent bytes of LINAR: a) with $\beta=0$, b) with $\beta=0.6$; wallclock times of LINAR: c) with $\beta=0$, d) with $\beta=0.6$.}
	\label{senttime}
\end{figure*}

Figs.~\ref{sentb0}~and~\ref{sentb6} show the total sent bytes of LINAR with $\beta=0$ and $\beta=0.6$, which reveal that increasing $\beta$ has a negligible effect on the sent bytes of LINAR. 
Figs.~\ref{timeb0}~and~\ref{timeb6} show the wallclock times of LINAR with $\beta=0$ and $\beta=0.6$, which indicate that the difference between $\beta=0$ and $\beta=0.6$ is negligible.

\begin{figure*}[!htbp]
	\captionsetup[subfloat]{farskip=9pt}
	\centering
	\subfloat[]{\includegraphics[width=0.25\textwidth]{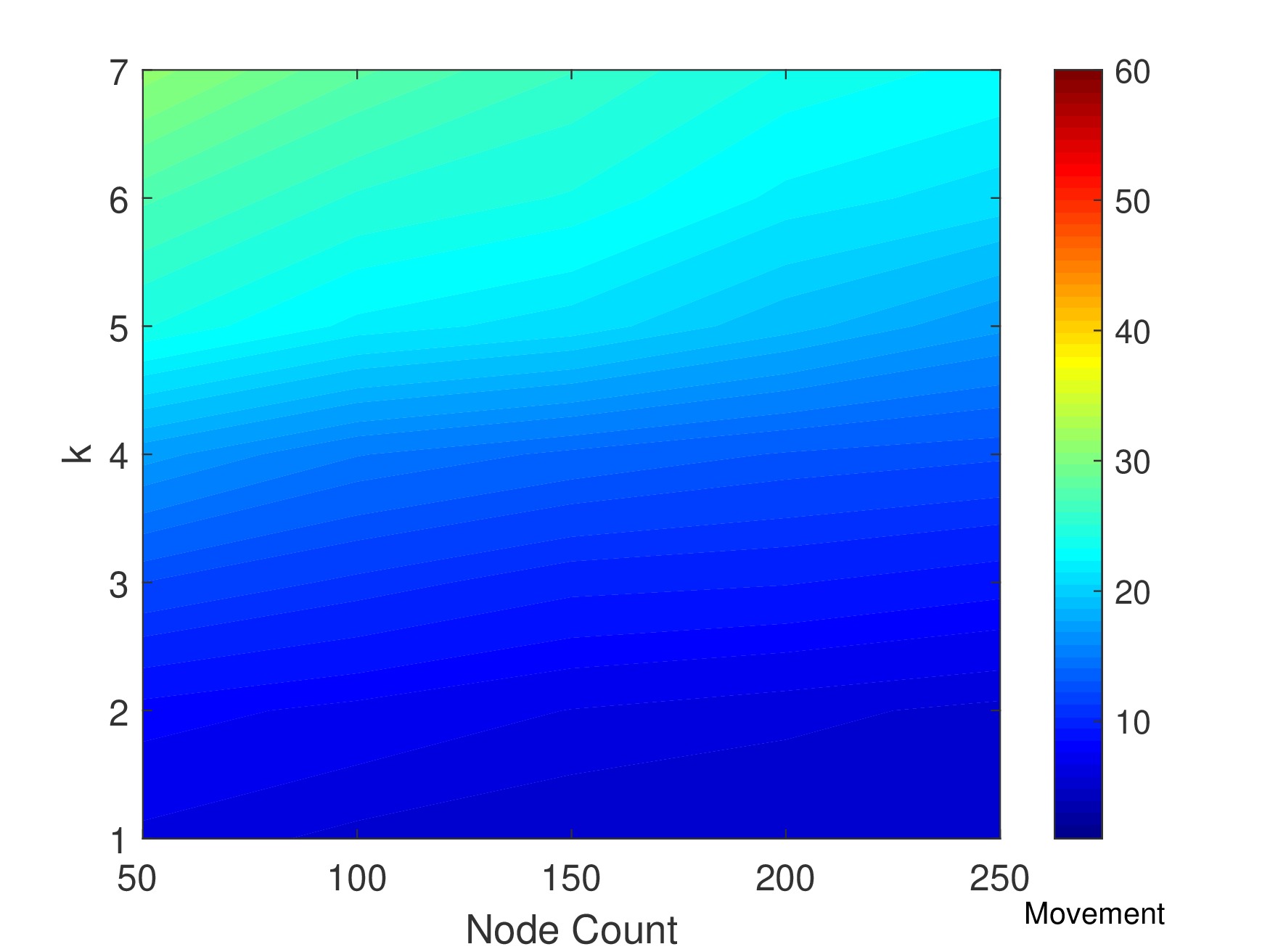}
		\label{moveb0}}
	\hfil
	\subfloat[]{\includegraphics[width=0.25\textwidth]{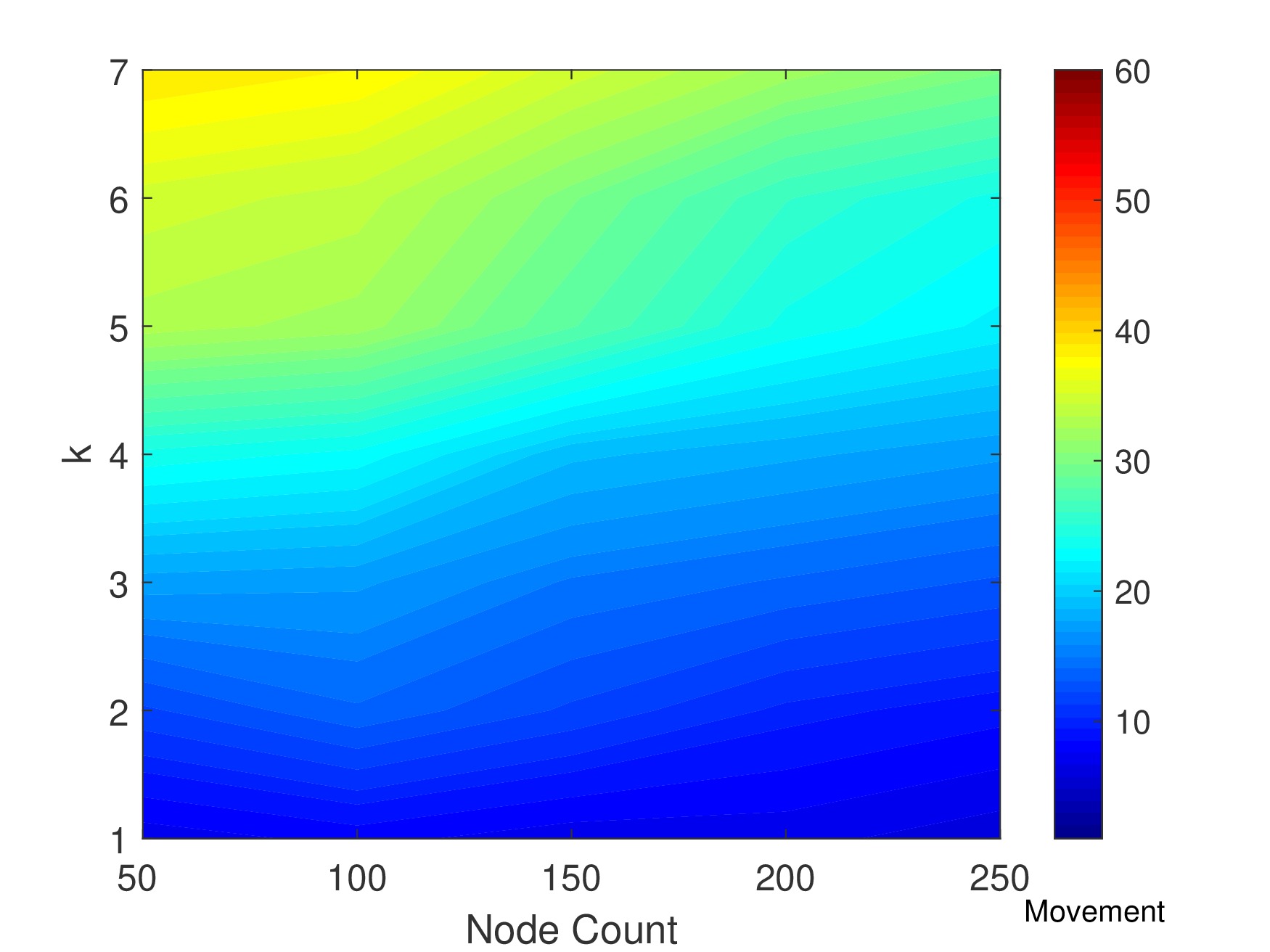}
		\label{moveb3}}
	\hfil
	\subfloat[]{\includegraphics[width=0.25\textwidth]{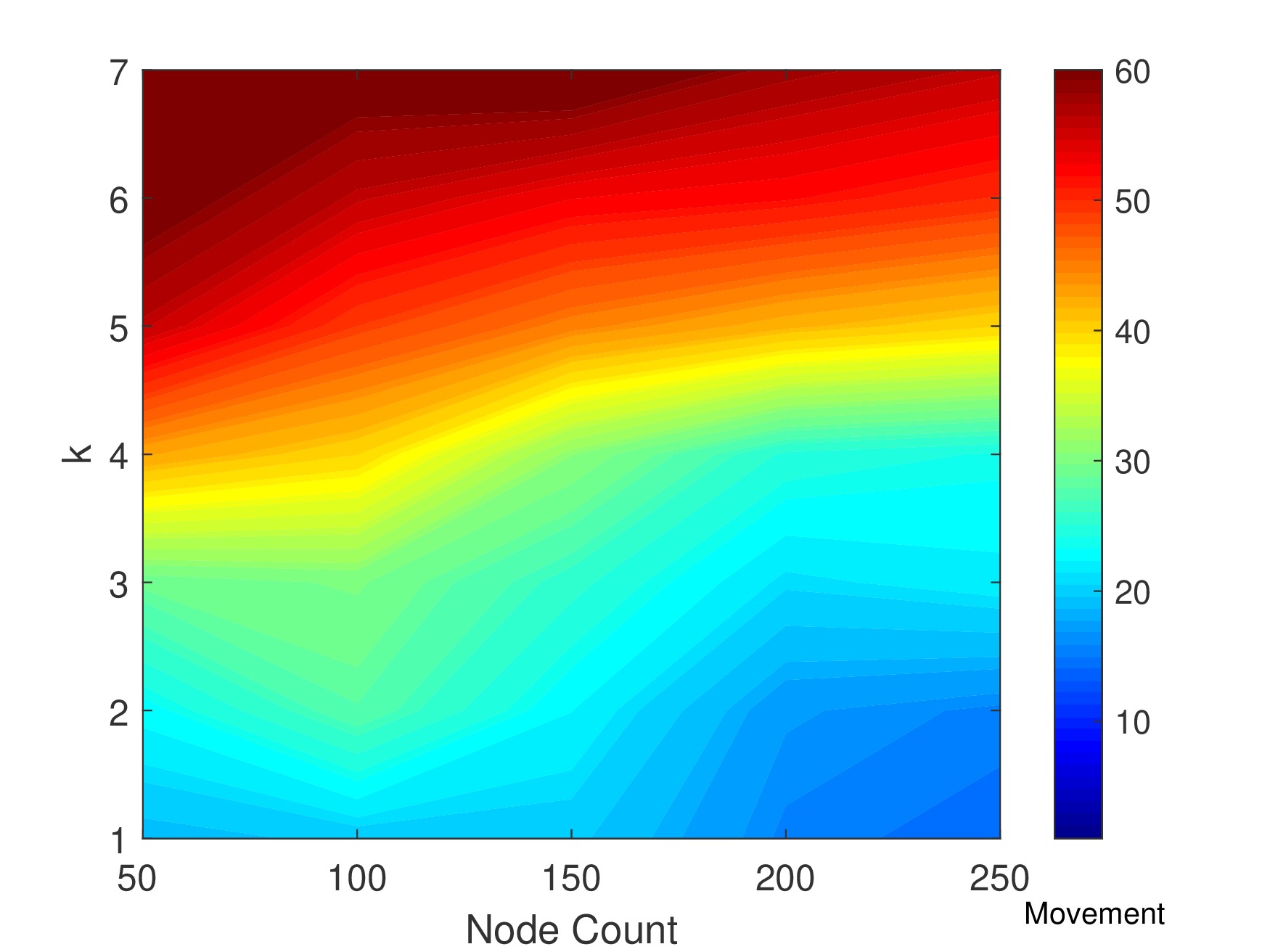}
		\label{moveb6}}
	\hfil
	\subfloat[]{\includegraphics[width=0.25\textwidth]{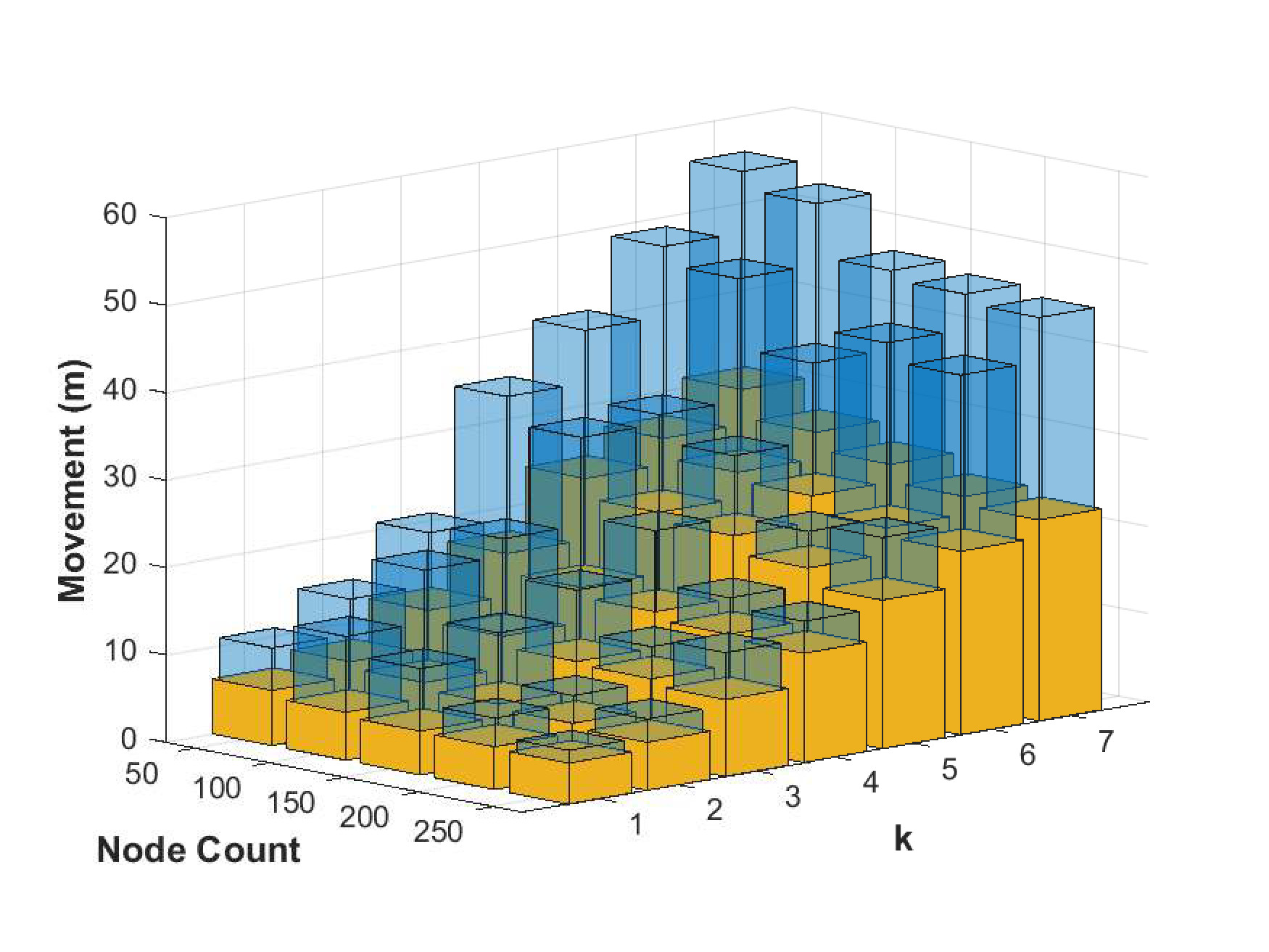}
		\label{move_nk_linar_beta}}
	\hfil
		\caption{LINAR movements: a) with $\beta=0$, b) with $\beta=0.3$, c) with $\beta=0.6$, d) with $\beta=0$ (yellow) and  $\beta=0.6$ (blue).}
	\label{movenew}
\end{figure*}

Fig.~\ref{movenew} presents the total generated movements by LINAR for different $\beta$ values. Fig.~\ref{moveb0} shows that the generated movements of the algorithm with $\beta=0$ vary between 4 m (for $n=250$ and $k=1$) and 32 m (for $n=50$ and $k=7$). The generated movements with $\beta=0.3$ (Fig. \ref{moveb3}) vary between 6 m (for $n=250$ and $k=1$) and 39 m (for $n=50$ and $k=7$). With $\beta=0.6$ (Fig.~\ref{moveb6}) the algorithm generates movements from 9 m (for $n=250$ and $k=1$) to 66 m (for $n=50$ and $k=7$). Fig.~\ref{move_nk_linar_beta} provides a comparison between the generated movements with $\beta=0$ and $\beta=0.6$, which shows that preserving the coverage and $k$-connectivity considerably increases the movement cost.

\section{Conclusion and Future Research Directions}
\label{conclusion}

In MSNs, connectivity and coverage should be maintained to achieve reliability and functionality. In this study, we present LINAR, which is a distributed algorithm for coverage-aware movement-based $k$-connectivity restoration in MSNs. Our algorithm is the first distributed movement-based $k$-connectivity restoration algorithm designed for arbitrarily large $k$ values, whereas all such algorithms in the literature are designed for $k\leq 2$. We introduce the theoretical foundations and show that most of the nodes can find their status using their local subgraphs. After a failure in a \textit{Joint} node, a \textit{Trusted} node or a chain of \textit{Joint} nodes change their position(s) to preserve the $k$ value with the minimum movement cost. LINAR is also capable of providing coverage preservation by sacrificing moderately from the movement minimization objective. In fact, the coverage conservation ratio ($\beta$), which is an input parameter, determines the extent of the tradeoff between coverage conservation and movement minimization. The comprehensive experimental and simulation results reveal that LINAR can restore $k$-connectivity with significantly better performance than the other algorithms in the literature. Furthermore, the obtained measurements also reveal that the difference between the primary coverage loss percentages before and after the restoration is, at most, 3\% for $\beta=0.6$.

Although we assume that paths and mobility costs between different locations are readily available to the nodes, in many real-word deployment scenarios, nodes have to discover paths and movement costs to other locations, which itself would not be available without significant exploration. In our algorithm, the movement paths are from one location to another established location, however, moving to a new location can result in better coverage and retain $k$-connectivity. In many real-life MSN deployments, accommodating sensing models other than a disk-shaped sensing abstraction (as in this study) is necessary. However, in this case, it is challenging to estimate what each node senses. In this study, our priority is to restore $k$-connectivity in a coverage-aware manner. However, it can also be argued that, for most applications, connectivity/coverage should be restored only after a part of the network becomes disconnected or after coverage of some area is lost. The proposed algorithm guarantees $k$-connectivity restoration after a node failure. 
Yet, our algorithm cannot guarantee the restoration of $k$-connectivity for all failure scenarios  (e.g., when all neighbors of the sink node fail simultaneously). Nevertheless, all of the aforementioned considerations warrant future research.

\bibliographystyle{ieeetran}
\bibliography{TNET-2019-00444.R3-references}

\end{document}